\documentclass[twocolumn,english,aps,pra,superscriptaddress,showkeys]{revtex4-2}
\usepackage[T1]{fontenc}
\usepackage[utf8]{inputenc}
\usepackage{babel}
\usepackage{array}
\usepackage{booktabs}
\usepackage{mathtools}
\usepackage{bm}
\usepackage{multirow}
\usepackage{amsmath}
\usepackage{amsthm}
\usepackage{graphicx}
\usepackage[unicode=true,pdfusetitle,
 bookmarks=true,bookmarksnumbered=false,bookmarksopen=false,
 breaklinks=false,pdfborder={0 0 1},backref=false,colorlinks=false]
 {hyperref}
\hypersetup{
 colorlinks=true}

\makeatletter

\providecommand{\tabularnewline}{\\}

\numberwithin{equation}{section}
\numberwithin{figure}{section}

\usepackage{units}
\usepackage{babel}
\usepackage{float}
\usepackage{graphicx}
\usepackage{babel}
\usepackage{enumitem}
\renewcommand{\theenumi}{\alph{enumi}}
\usepackage{tikz}
\usepackage{vmargin}
\setmargins{1.5cm}       
{0.5cm}                        
{17.5cm}                      
{25.42cm}                    
{14pt}                           
{1cm}                           
{0pt}                             
{0cm}                           
\usepackage{amsmath}
\usepackage{amssymb}
\usepackage{cancel}
\usepackage[nodayofweek,yyyymmdd]{datetime}
\usepackage{mdframed}
\definecolor{green}{RGB}{0,128,0}
\renewcommand{\fnum@figure}[1]{\textup{\textbf{FIG. \thefigure}}\textup{:} \upshape}  
\renewcommand{\fnum@table}[1]{\textup{\textbf{TABLE \thetable}}\textup{:} \upshape}  

\renewcommand\thefigure{\arabic{figure}}
\renewcommand\thetable{\arabic{table}}

\fontsize{9pt}{10pt}\selectfont 
\usepackage[utf8]{inputenc}

\makeatother

\begin{document}
\fontsize{9pt}{10pt}\selectfont 
\title{Alternant Hydrocarbon Diradicals as Optically Addressable Molecular
Qubits}
\author{\normalsize Yong Rui Poh}
\affiliation{Department of Chemistry and Biochemistry, University of California
San Diego, La Jolla, California 92093, USA}
\author{\normalsize Dmitry Morozov}
\affiliation{Terra Quantum AG, Kornhausstrasse 25, 9000 St. Gallen, Switzerland}
\author{\normalsize Nathanael P. Kazmierczak}
\affiliation{Division of Chemistry and Chemical Engineering, Arthur Amos Noyes
Laboratory of Chemical Physics, California Institute of Technology,
Pasadena, California 91125, USA}
\author{\normalsize Ryan G. Hadt}
\email{rghadt@caltech.edu}

\affiliation{Division of Chemistry and Chemical Engineering, Arthur Amos Noyes
Laboratory of Chemical Physics, California Institute of Technology,
Pasadena, California 91125, USA}
\author{\normalsize Gerrit Groenhof}
\email{gerrit.x.groenhof@jyu.fi}

\affiliation{Nanoscience Center and Department of Chemistry, University of Jyväskylä,
Jyväskylä, Finland}
\author{\normalsize Joel Yuen-Zhou}
\email{joelyuen@ucsd.edu}

\affiliation{Department of Chemistry and Biochemistry, University of California
San Diego, La Jolla, California 92093, USA}
\date{March 17, 2024}
\begin{abstract}
High-spin molecules allow for bottom-up qubit design and are promising
platforms for magnetic sensing and quantum information science. Optical
addressability of molecular electron spins has also been proposed
in first-row transition metal complexes via optically-detected magnetic
resonance (ODMR) mechanisms analogous to the diamond-NV colour centre.
However, significantly less progress has been made on the front of
metal-free molecules, which can deliver lower costs and milder environmental
impacts. At present, most luminescent open-shell organic molecules
are $\pi$-diradicals, but such systems often suffer from poor ground-state
open-shell characters necessary to realise a stable ground-state molecular
qubit. In this work, we use alternancy symmetry to selectively minimise
radical-radical interactions in the ground state, generating $\pi$-systems
with high diradical characters. We call them \emph{m}-dimers, referencing
the need to covalently link two benzylic radicals at their \emph{meta}
carbon atoms for the desired symmetry. Through a detailed electronic
structure analysis, we find that the excited states of alternant hydrocarbon
\emph{m}-diradicals contain important symmetries that can be used
to construct ODMR mechanisms leading to ground-state spin polarisation.
The molecular parameters are set in the context of a tris(2,4,6-trichlorophenyl)methyl
(TTM) radical dimer covalently tethered at the \emph{meta} position,
demonstrating the feasibility of alternant \emph{m}-diradicals as
molecular colour centres.
\end{abstract}

\maketitle
\global\long\def\blue#1{\textcolor{blue}{#1}}%
\global\long\def\red#1{\textcolor{red}{#1}}%
\global\long\def\green#1{\textcolor{green}{#1}}%
\global\long\def\purple#1{\textcolor{purple}{#1}}%
\global\long\def\orange#1{\textcolor{orange}{#1}}%

\global\long\def\it#1{\textit{\textrm{#1}}}%
\global\long\def\un#1{\underline{\textrm{#1}}}%
\global\long\def\br#1{\left( #1 \right)}%
\global\long\def\sqbr#1{\left[ #1 \right]}%
\global\long\def\curbr#1{\left\{  #1 \right\}  }%
\global\long\def\braket#1{\langle#1 \rangle}%
\global\long\def\bra#1{\langle#1 \vert}%
\global\long\def\ket#1{\vert#1 \rangle}%
\global\long\def\abs#1{\left|#1\right|}%
\global\long\def\mb#1{\mathbf{#1}}%
\global\long\def\doublebraket#1{\langle\langle#1 \rangle\rangle}%

\section{Introduction}

There has been sustained interest in designing optically addressable
electron spins that can be both initialised (polarised) and read out
by optical means \citep{Awschalom2018}. Known as optically-detected
magnetic resonance (ODMR), this technique offers precise control of
a single qubit and has found numerous applications in quantum sensing
\citep{Abobeih2019} and quantum information science \citep{Pfaff2014}.
The platform, referred to as a colour centre, is often a solid-state
spin defect such as a nitrogen-vacancy (NV) centre in diamond \citep{Taylor2008,Degen2017,Rose2018,Gottscholl2020,Chejanovsky2021,Mukherjee2023}.
However, because these defects are introduced post-synthesis and often
without control over their locations, defect-based colour centres
suffer from poor scalability and tunability.

These problems can be tackled with molecular spin systems because
they can be synthesised from bottom-up and extended into macromolecular
systems \citep{GaitaArino2019,Atzori2019,Wasielewski2020,Yu2021,Laorenza2022,Scholes2023,Wu2023}.
Metal complexes have seen the most advancements, particularly with
transition metals \citep{Wojnar2020,Bayliss2020,Fataftah2020,Mirzoyan2021,Kazmierczak2021,Laorenza2021,Amdur2022,Bayliss2022,Goh2022,Kazmierczak2022,Kazmierczak2023,Mullin2023}.
By contrast, little progress has been made with fully organic molecules
\citep{Smyser2020,Dill2023,Gorgon2023,Palmer2024,Mena2024,Singh2024}
which, being metal-free, will be more cost-efficient and sustainable
than their metal-based counterparts. Pioneering efforts were made
by Gorgon et al. using excited organic radicals \citep{Gorgon2023};
however, a drawback of microwave spin manipulation in an excited electronic
state is the limited lifetime of the latter. As for ground-state organic
systems, the same (and only) work \citep{Gorgon2023} also showed
light-induced selective preparation of a diradical in the triplet
ground state over its degenerate singlet counterpart. In this case,
it is unclear if the triplet magnetic sublevels were differentially
populated, which is key towards coherent spin control. To this end,
we present the first theoretical analysis for a general class of organic
spin-optical interfaces, focusing on attaining spin polarisation in
the ground-state magnetic sublevels.

\begin{figure*}
\emph{\includegraphics[width=1\textwidth]{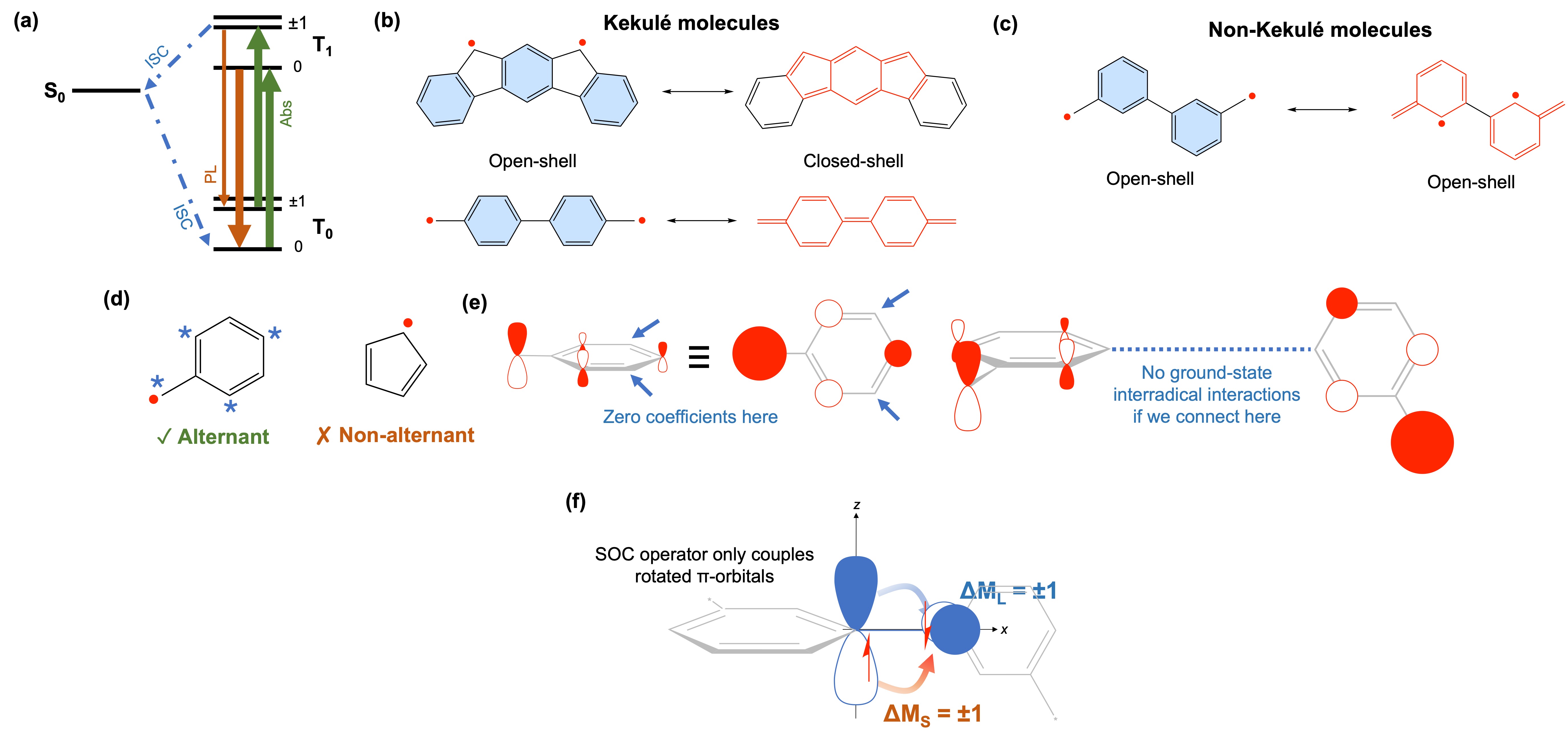}\caption{\label{fig:main}(a) Colour centres like diamond-NV defects can be
spin polarised by multiple photoexcitation (Abs) cycles, the intensity
of PL being an indicator of the ground-state spin polarisation. This
procedure is known as ODMR. (b) Kekulé hydrocarbons are conjugated
$\pi$-systems with at least one closed-shell Kekulé structure. These
molecules tend to have lower diradical characters. (c) Non-Kekulé
structures have no closed-shell Kekulé structures and are hence more
likely to have open-shell ground states. (d) Alternant hydrocarbons
are $\pi$-systems whereby the $\pi$-contributing atoms may be divided
into two classes, starred {[}{*}{]} and unstarred {[}~{]}, such that
no two atoms from the same class are adjacent. (e) Within the Hückel
framework, radicals constructed with alternacy symmetry (termed AHRs)
have their SOMOs localised on the starred atoms (the class with more
atoms). As such, two AHRs covalently bonded via the unstarred atoms
will have vanishing radical-radical interactions. These molecules
are non-Kekulé, as shown in (c). (f) In $\pi$-electron systems, SOC-mediated
ISC can only occur between rotated $\pi$-orbitals, in line with the
El-Sayed rules. Due to conservation of angular momentum, an electron
spin flip must occur concomitantly, making a triplet-singlet ISC process
spin-selective.}
}
\end{figure*}

In diamond-NV centres, ODMR is achieved via the following procedure
\citep{Doherty2013}: Upon photoexcitation of the triplet ground state
to the excited state (also a triplet), intersystem crossing (ISC)
to a singlet excited state occurs with a change in magnetic quantum
number $M_{S}$. This increases the relative population of the excited
triplet $M_{S}=0$ level, the photoluminescence (PL) of which transfers
the spin polarisation to the ground state (optical initialisation).
The PL intensity also indicates the ground-state polarisation (optical
readout). As for the singlet excited state, its population decays
non-radiatively to the triplet $M_{S}=0$ ground state, its irreversibility
being the key towards overall purification of the ground state spins
{[}Fig. \ref{fig:main}a{]}. There are thus three elements to engineer
in a metal-free molecular system: (1) a high-spin ground state, (2)
spin-selective interactions among the excited states, and (3) irreversible
relaxation to the ground state.

Element (1) suggests the use of an organic diradical. Thus far, most
luminescent diradicals have involved some degree of $\pi$-conjugation
\citep{Hattori2019,Kimura2021,Wonink2021,Feng2021,Huang2022,Abdurahman2023,Matsuoka2023,Abdurahman2023-2},
which has proven useful in the chemical assembly of spin arrays \citep{Kimura2021}.
However, large spatial separation between the two paramagnetic centres
is often necessary to minimise radical-radical interactions and achieve
a high-spin ground state {[}Fig. \ref{fig:main}b{]} \citep{Abe2013,Casado2017,Stuyver2019,Murto2022}.
In most cases, this separation extends to the excited states as well,
which is disadvantageous for element (2). Indeed, in Gorgon et al.'s
seminal work, interradical interaction was achieved via a charge transfer
to the spacer's triplet state, generating a complicated manifold of
12 coupled electronic levels \citep{Gorgon2023}. One solution would
be to reduce (improve) the stability of the molecule's closed-shell
(open-shell) Kekulé structures, with a crude measure being the number
of Clar sextets minus the number of unpaired electrons \citep{Prajapati2023}.
In the most extreme case, the $\pi$-diradical may have no closed-shell
Kekulé structures at all; such molecules are termed ``non-Kekulé''
\citep{Abe2013,Murto2022} and an example has been shown in Fig. \ref{fig:main}c.
However, this analysis is purely qualitative and there exist non-Kekulé
molecules with appreciable closed-shell characters \citep{Stuyver2019}.
(Interestingly, alkaline earth metal complexes \citep{Khvorost2024},
commonly used as optical cycling centres \citep{Dickerson2021,Zhu2022},
as well as nitrenes \citep{Gately2023} have also been investigated
for luminescent diradical properties and are useful candidates of
molecular colour centres as well.)

Alternant hydrocarbon radicals (AHRs) are a special type of hydrocarbon
$\pi$-electron radicals. In AHRs, atoms contributing to the $\pi$
framework may be divided into two classes such that no two atoms from
the same class are adjacent {[}Fig. \ref{fig:main}d{]}. Importantly,
within Hückel theory of nearest-neighbour hopping, the singly occupied
molecular orbital (SOMO) is localised on atoms from only one class
-- the class containing more atoms {[}Fig. \ref{fig:main}e{]}. As
such, by covalently tethering two AHRs via atoms of the \emph{other}
class, radical-radical interactions may be avoided in the ground state,
that is, the resulting diradical is non-Kekulé and quantitatively
predicted to be open-shell within the Hückel framework. Note that
interradical interactions are retained in the excited states because
the occupied and virtual molecular orbitals (MOs) are in general delocalised
across both atom classes; this is necessary for element (2).

Thus far, we have not addressed the spin selectivity component of
element (2). This was achieved in NV centres via an ISC process mediated
by spin-orbit coupling (SOC) \citep{Doherty2013}, but these rates
are negligible if the $\pi$-conjugated system is fully planar. The
reason may be found in the SOC operator, which in its mean-field Breit-Pauli
form has the structure of $\hat{{\bf L}}\cdot\hat{{\bf S}}=\hat{L}_{z}\hat{S}_{z}+\hat{L}_{+}\hat{S}_{-}/2+\hat{L}_{-}\hat{S}_{+}/2$,
where $\hat{{\bf L}}$ and $\hat{{\bf S}}$ are the orbital and spin
angular momentum vector operators respectively \citep{Penfold2018}.
The first operator vanishes when acting on $2p_{z}$ atomic orbitals
(AOs), while the other two operators rotate $2p_{z}$ orbitals into
linear combinations of $2p_{x}$ and $2p_{y}$ orbitals. Therefore,
the SOC operator only connects between rotated $2p$ (or $\pi$) orbitals
{[}Fig. \ref{fig:main}f{]}. This is essentially the statement of
El-Sayed \citep{El-Sayed1963} and implies that torsion between the
radical $\pi$-systems is important for element (2) \citep{Hong2001,Rybicki2009,Barford2010,Yu2012}.
Finally, since orbital rotation constitutes a $\Delta M_{L}=\pm1$
transition, to conserve total angular momentum the spin must also
change by $\Delta M_{S}=\pm1$, giving rise to the spin selectivity
of ISC in $\pi$-electron systems \citep{Hong2001,Barford2010,Yu2012}.

Therefore, through a qualitative analysis, we find potential in realising
an NV centre analogue by attaching two AHRs covalently at selected
atoms and with significant torsion. We call them AHR \emph{m}-dimers,
referencing the fact that most luminescent $\pi$-radicals have benzylic
structures and thus should be dimerised via atoms at the \emph{meta}
position. In this work, we make further progress by analysing the
electronic structure of AHR \emph{m}-dimers and the symmetries of
their excited wavefunctions. By adopting a mean-field approach to
the Pariser-Parr-Pople (PPP) model \citep{Pariser1953,Pople1953,Linderberg1968}
(also known as the extended Hubbard model), we find that not all excitations
are absorptive and only few within the subset can undergo triplet-singlet
ISC. This makes the excited state kinetics highly selective and useful
for generating robust ODMR mechanisms leading towards ground-state
spin polarisation. While general to any AHR \emph{m}-dimer, our observations
are also set in the context of two existing AHRs -- a methylated
benzylic radical and the tris(2,4,6-trichlorophenyl)methyl (TTM) radical
-- using parameters estimated from density functional theory (DFT).
For the methylated benzylic radical, our DFT-parametrised excited
states also matched the results from a multi-configurational level
of theory. These ab initio calculations further demonstrate the possibility
of designing a molecular colour center with AHR \emph{m}-dimers.

The paper is organised as such: In Sec. IIA, the electronic structure
of AHR \emph{m}-dimers is analysed using the PPP framework. This section
is technical and the reader may safely skip it at the first pass.
In Sec. IIB, ODMR mechanisms resulting in ground-state spin polarisation
are proposed based on results from the PPP model. In Sec. IIC, the
PPP predictions are verified by ab initio calculations. Finally, in
Sec. IID, a mapping between a polarised ground-state triplet and an
operational qubit is proposed.

\section{Results and Discussions}

\subsection{Electronic structure analysis}

The PPP model \citep{Pariser1953,Pople1953,Linderberg1968} is an
extension of the Hückel model to include additional Coulomb interactions.
It partially resolves the electron-electron correlation problem and
continues to be the reference model for $\pi$-conjugated hydrocarbons.
By considering only the minimal basis of $2p$ AOs that constitute
the $\pi$-electron system, the PPP Hamiltonian reads 
\begin{align}
\hat{H}_{\text{PPP}} & =\sum_{\mu}\alpha_{\mu}\hat{n}_{\mu}-\sum_{\braket{\nu>\mu}}\sum_{\sigma}\beta_{\mu\nu}\br{\hat{a}_{\mu\sigma}^{\dagger}\hat{a}_{\nu\sigma}+a_{\nu\sigma}^{\dagger}\hat{a}_{\mu\sigma}}\nonumber \\
 & \quad+\sum_{\mu}\gamma_{\mu\mu}\hat{n}_{\mu\alpha}\hat{n}_{\mu\beta}+\sum_{\nu>\mu}\gamma_{\mu\nu}\hat{n}_{\mu}\hat{n}_{\nu},\label{eq:H_PPP}
\end{align}
with $\hat{a}_{\mu\sigma}^{\dagger}$ ($\hat{a}_{\mu\sigma}$) being
the fermionic creation (annihilation) operator for an electron in
the $2p$ AO of atom $\mu$ with spin $\sigma$ ($=\alpha,\beta$),
$\hat{n}_{\mu\sigma}\equiv\hat{a}_{\mu\sigma}^{\dagger}\hat{a}_{\mu\sigma}$
being the number operator and $\hat{n}_{\mu}\equiv\sum_{\sigma=\alpha,\beta}\hat{n}_{\mu\sigma}$.
Here, we have used $\alpha$ ($\beta$) to denote spin-up (spin-down)
electrons. The parameters are defined as followed: $\alpha_{\mu}$
is the on-site energy of atom $\mu$, $\beta_{\mu\nu}$ is the hopping
amplitude between atoms $\mu$ and $\nu$, and $\gamma_{\mu\nu}$
represents the effective Coulombic repulsion between an electron on
atom $\mu$ and another electron on atom $\nu$. Finally, $\braket{\cdot}$
denotes nearest neighbours.

Its application towards AHRs was explored by Longuet-Higgins and Pople
using a mean-field approach similar to Hartree-Fock theory \citep{Longuet-Higgins1950,Dewar1954,Longuet-Higgins1955}.
The results remain qualitatively the same as the Hückel approach,
that is, if we label the alternacy class containing more atoms by
stars {[}{*}{]} {[}Fig. \ref{fig:main}d{]}, then each HOMO--$j$
has the same atomic coefficients as the corresponding LUMO+$j$, except
with opposite signs on the unstarred atoms. This is known as the pairing
theorem and, consequently, the SOMO has nodes on the unstarred atoms.
An excellent review on this topic has been provided by Hele \citep{Hele2021}
and we shall not rederive the results. Instead, we will briefly mention
that, within the aforementioned mean-field treatment, the AHR's electronic
ground state (GS) is a doublet described by the following (Slater)
determinants: 
\begin{align}
\ket{\Psi;+1/2} & =\ket{\cdots2\overline{2}1\overline{1}0}, & \ket{\Psi;-1/2} & =\ket{\cdots2\overline{2}1\overline{1}\overline{0}},\label{eq:GS-monomer}
\end{align}
that is, determinants with one electron in the SOMO ($j=0$) and two
electrons in each HOMO--$j$ ($j=1,2,\cdots$). In this work, we
shall follow standard electronic structure theory notation \citep{Szabo1989}.
In addition, the second state index labels the spin magnetic number
$M_{S}\in\curbr{+1/2,-1/2}$.

Yet another consequence of alternacy symmetry is the invariance of
$\hat{H}_{\text{PPP}}$ {[}Eq. (\ref{eq:H_PPP}){]} under particle-hole
transformation (PHT) $\hat{a}_{\mu\sigma}^{\dagger}\rightarrow f_{\mu}\hat{a}_{\mu\overline{\sigma}}$
with $f_{\mu}$ being $+1$ if $\mu$ is starred and $-1$ otherwise
($\overline{\sigma}$ is the spin complementary to $\sigma$). As
such, it is apt to label electronic states of AHRs by their symmetries
under PHT, be it odd or even. For instance, when considering an $\alpha$-spin
excitation from SOMO to LUMO+$j$, labelled $\ket{\Psi_{0}^{j'};+1/2}$,
one should take a linear combination with the $\beta$-spin HOMO--$j$
to SOMO transition $\ket{\Psi_{\overline{j}}^{\overline{0}};+1/2}$,
which is related to the latter by PHT, such that the resulting state
has a well-defined symmetry under PHT. Following the notation set
by Pariser \citep{Pariser1956,Hele2019}, we label these symmetry-adapted
states by ``$+$'' and ``$-$'' representing symmetric and antisymmetric
combinations respectively: 
\begin{align}
\ket{\Psi_{0j}^{\pm};+1/2} & \equiv\frac{\ket{\Psi_{\overline{j}}^{\overline{0}};+1/2}\pm\ket{\Psi_{0}^{j'};+1/2}}{\sqrt{2}}.\label{eq:PHT}
\end{align}
We note that due to differences in origin (Pariser was considering
configuration interactions), some ``$+$'' states may be odd (instead
of even) under PHT and the same can happen to ``$-$'' states too.
Despite that, there is a one-to-one mapping between the two notations,
justifying our choice of Pariser's (see Supplementary Information
S1).

Our \emph{m}-dimer system comprises two AHRs covalently tethered via
the unstarred atoms (the class containing fewer atoms) {[}Fig. \ref{fig:results}a{]}.
The entire $\pi$-electron system is modelled by the PPP Hamiltonian.
Because the monomers are rotated relative to each other by sterics,
any intermonomer couplings due to $\pi$-orbital overlaps may be treated
perturbatively. As such, we partition the Hamiltonian into monomeric
components: 
\begin{align}
\hat{H}_{r} & =\sum_{\mu\in\mathcal{N}_{r}}\alpha_{\mu}\hat{n}_{\mu}-\sum_{\braket{\nu>\mu\in\mathcal{N}_{r}}}\sum_{\sigma}\beta_{\mu\nu}\br{\hat{a}_{\mu\sigma}^{\dagger}\hat{a}_{\nu\sigma}+\hat{a}_{\nu\sigma}^{\dagger}\hat{a}_{\mu\sigma}}\nonumber \\
 & \quad+\sum_{\mu\in\mathcal{N}_{r}}\gamma_{\mu\mu}\hat{n}_{\mu\alpha}\hat{n}_{\mu\beta}+\sum_{\nu>\mu\in\mathcal{N}_{r}}\gamma_{\mu\nu}\hat{n}_{\mu}\hat{n}_{\nu},\label{eq:H_r}
\end{align}
and perturbative intermonomer couplings: 
\begin{align}
\hat{V}_{AB} & =-\sum_{\braket{\mu\in\mathcal{N}_{A},\nu\in\mathcal{N}_{B}}}\sum_{\sigma}\beta_{\mu\nu}\br{\hat{a}_{\mu\sigma}^{\dagger}\hat{a}_{\nu\sigma}+\hat{a}_{\nu\sigma}^{\dagger}\hat{a}_{\mu\sigma}}\nonumber \\
 & \quad+\sum_{\mu\in\mathcal{N}_{A},\nu\in\mathcal{N}_{B}}\gamma_{\mu\nu}\hat{n}_{\mu}\hat{n}_{\nu},\label{eq:V_AB}
\end{align}
such that the full Hamiltonian reads $\hat{H}=\hat{H}_{A}+\hat{H}_{B}+\hat{V}_{AB}$.
Here, $r\in\curbr{A,B}$ labels the two monomers and $\mathcal{N}_{r}$
denotes the set of atoms in monomer $r$.

Before solving the Hamiltonian $\hat{H}$, it is useful to understand
the types of electronic states to expect. We note that all dimers
have a $C_{2}$ rotational symmetry and both $\hat{H}_{A}+\hat{H}_{B}$
and $\hat{V}_{AB}$ are symmetric under this transformation. Furthermore,
the \emph{m}-dimer is also alternant so both $\hat{H}_{A}+\hat{H}_{B}$
and $\hat{V}_{AB}$ are symmetric under PHT as well. As such, focusing
on the zeroth-order term $\hat{H}_{A}+\hat{H}_{B}$ for now, the zeroth-order
electronic states may be organised by eigenvalues of the molecule's
spin and symmetry operators, namely (1) the spin numbers, $S$ and
$M_{S}$, (2) symmetry under PHT, $P\in\curbr{+,-}$, and (3) irreducible
representations (irreps) of the $C_{2}$ point group, $\Gamma\in\curbr{\mathtt{A},\mathtt{B}}$,
presented in typewriter font to avoid confusion with the monomer index.
In this work, we packed them into the following notation for electronic
states: 
\begin{align}
 & \ket{^{2S+1}\Phi_{\Gamma}^{jk,P};M_{S}},\label{eq:state_notation}
\end{align}
with $\Phi$ labelling the class of excitation and $\curbr{j,k}$
labelling the MOs involved in the excitation (more to follow).

\setcounter{figure}{1}
\begin{figure*}
\emph{\includegraphics[width=1\textwidth]{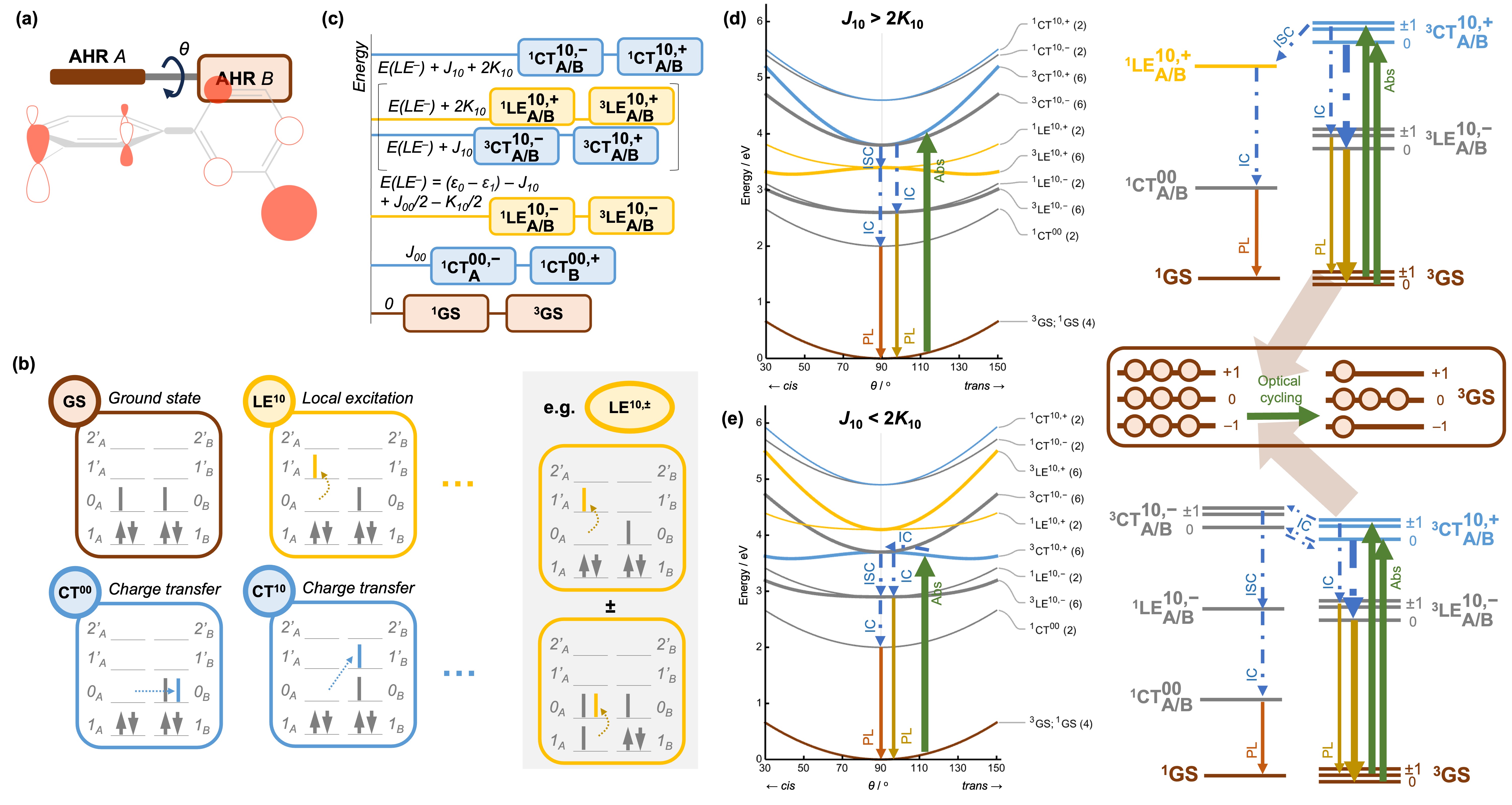}\caption{\label{fig:results}(a) Schematic of an AHR \emph{m}-dimer, which
contains two AHRs (indexed $r\in\protect\curbr{A,B}$) covalently
tethered at the unstarred atoms (the class of fewer atoms) with torsional
angle $\theta$. (b) Ground and excited state configurations of AHR
\emph{m}-dimers, expressed in the basis of MOs localised completely
on either monomer $A$, i.e. $\protect\curbr{\cdots,1_{A},0_{A},1_{A}',\cdots}$,
or monomer $B$, i.e. $\protect\curbr{\cdots,1_{B},0_{B},1_{B}',\cdots}$.
LEs keep the excited electron on the same monomer (e.g. $0_{A}\rightarrow1_{A}'$),
while CTs excite electrons across monomers (e.g. $0_{A}\rightarrow1_{B}'$).
Labels \textquotedblleft$\pm$\textquotedblright{} represent linear
combinations of SOMO-to-LUMO and HOMO-to-SOMO transitions, i.e. particle-hole
symmetries. (c) Energies of ground and excited electronic states at
$\hat{V}_{\text{NN}}=0$, presented in ascending order. The energy
ordering between the $\text{LE}^{10,+}$ states and the $^{3}\text{CT}^{10}$
states depends on the relative values of $J_{10}\equiv\protect\br{11\vert00}$
and $2K_{10}\equiv2\protect\br{10\vert01}$. ZFS is not being considered
by this work. (d,e) Plots of representative PESs based on perturbation
theory calculations for (d) $J_{10}>2K_{10}$ and (e) $J_{10}<2K_{10}$.
Non-absorptive states are drawn in grey and degeneracies are indicated
in parenthesis. Also illustrated are two possible ODMR mechanisms,
one for each case. In both cases, spin polarisation is attained in
the triplet GS by transferring its $M_{S}=\pm1$ populations into
the singlet GS. Parameters for (d): $E_{\text{steric}}=0.6\text{ eV}$;
$\theta_{\text{eq}}=90.0^{\text{o}}$; $\beta_{\mathcal{A}\mathcal{B}}^{0}=2.5\text{ eV}$;
$\gamma_{\mathcal{A}\mathcal{B}}^{0}=0\text{ eV}$; SOMO-LUMO gap
$=3.0\text{ eV}$; $c_{\mathcal{A}1_{A}}=c_{\mathcal{B}1_{B}}=0.5$;
$J_{00}=2.0\text{ eV}$; $J_{10}=1.2\text{ eV}$; $K_{10}=0.4\text{ eV}$.
Parameters for (e): Same as (d) except $J_{10}=0.8\text{ eV}$ and
$K_{10}=0.6\text{ eV}$. Abs: Photoexcitation.}
}
\end{figure*}

To solve the Hamiltonian, we note that the monoradical terms $\hat{H}_{A}$
and $\hat{H}_{B}$ each admit the same mean-field solutions as those
obtained by Longuet-Higgins and Pople, described by Eq. (\ref{eq:GS-monomer}).
As such, because the two monomers are non-interacting at $\hat{V}_{AB}=0$,
the approximate zeroth-order electronic ground states ($\Phi=\text{GS}$)
to the full zeroth-order Hamiltonian $\hat{H}_{A}+\hat{H}_{B}$ will
be (tensor) products of the monoradical's mean-field solutions. This
yields a singlet state $\ket{^{1}\text{GS};0}$ and a set of triplet
states $\curbr{\ket{^{3}\text{GS};M_{S}}\vert M_{S}=-1,0,+1}$, all
of which have ``$-$'' particle-hole symmetry. Note that we continue
to use the basis of MOs localised on each monomer; for instance, $\ket{^{3}\text{GS};+1}=\ket{\cdots1_{A}\overline{1_{A}}0_{A}\cdots1_{B}\overline{1_{B}}0_{B}}$,
with $j_{r}$ being the $j$-th MO of monomer $r$.

As for the zeroth-order low-lying excited states, they are approximated
by single excitations from the GS, just like the monoradical case
{[}Eq. (\ref{eq:PHT}){]}. With dimers, there can be two types of
excited states: local excitations ($\Phi=\text{LE}$), which move
electrons within the same monomer, and charge transfers ($\Phi=\text{CT}$),
which excite electrons to the other monomer. Fig. \ref{fig:results}b
provides a diagrammatic representation of these excitations. For simplicity,
we consider only the $1\rightarrow0$ and $0\rightarrow1'$ transitions
($j'$ labels LUMO+$j$), even though the subsequent analysis applies
to higher-lying excitations as well (such as $2\rightarrow0$ and
$0\rightarrow2'$). This totals up to 35 states and their expressions
have been compiled in Supplementary Information S2 after taking symmetry-
and spin-adapted linear combinations, as described by Eq. (\ref{eq:state_notation}).

We then compute the zeroth-order excitation energies by taking expectation
values over $\hat{H}_{A}+\hat{H}_{B}$, the results being compiled
in Fig. \ref{fig:results}c. We highlight that the singlet $\text{CT}^{P}$
state is higher in energy than the triplet ones because, unlike $\text{LE}^{P}$,
the unpaired spins in $\text{CT}^{P}$ experience intramonomer exchange
interactions that do not vanish with $\hat{V}_{AB}\rightarrow0$.
Also, the $\curbr{\ket{^{1}\text{LE}_{\Gamma}^{10,+};0}}$ and $\curbr{\ket{^{3}\text{CT}_{\Gamma}^{10,+};\pm1}}$
sets may have opposite energy rankings depending on the relative values
of $J_{10}\equiv\br{11\vert00}$ and $2K_{10}\equiv2\br{10\vert01}$
(in chemists' notation \citep{Szabo1989}). Finally, while the $\text{LE}^{+}$
and $\text{LE}^{-}$ states are separated in energy due to configuration
interaction between the $1\rightarrow0$ and $0\rightarrow1'$ transitions,
this separation is not present in the CT states because the two MO
transitions, being intermonomer in nature, are non-interacting at
$\hat{V}_{AB}=0$. These, as we shall see, are crucial to the ODMR
mechanism.

With regard to the linear absorption spectrum, we find that out of
all 35 zeroth-order low-lying excitations, only the $\text{LE}^{+}$
states are bright with non-vanishing transition dipole moments. The
reason lies in the form of the dipole operator within the neglect
of differential overlap assumption invoked by the PPP model \citep{Pariser1956}:
\begin{align}
\hat{\bm{\mu}} & =\sum_{\nu\in\mathcal{N}_{A}\cup\mathcal{N}_{B}}\bm{\mu}_{\nu}\hat{n}_{\nu},\label{eq:dipole}
\end{align}
which has ``$-$'' particle-hole symmetry and does not excite between
monomers. Hence, the GS, being of ``$-$'' symmetry, can only be
photoexcited locally and to ``$+$'' states. (An alternative interpretation
is to realise that the $j\rightarrow0$ and $0\rightarrow j'$ transitions
have identical transition dipole moments, so they cancel when taking
antisymmetric combinations in Eq. (\ref{eq:PHT}) \citep{Abdurahman2020,Hele2021}.)

Next, we consider perturbative effects on the approximate zeroth-order
eigenstates due to $\hat{V}_{AB}$. Because $\hat{V}_{AB}$ is a spin-preserving
intermonomer operator of ``$+$'' particle-hole symmetry and $\mathtt{A}$
irrep {[}Eq. (\ref{eq:V_AB}){]}, only CT-type mixing between states
of like symmetry can occur: 
\begin{align}
\ket{^{2S+1}\text{LE}_{\Gamma}^{10,P};M_{S}} & \xleftrightarrow{\hat{V}_{AB}}\ket{^{2S+1}\text{CT}_{\Gamma}^{10,P};M_{S}}.\label{eq:V_AB-rule}
\end{align}
This suggests that $\text{CT}^{+}$ excitations can also be partially
bright depending on the extent of $\text{LE}^{+}$ mixing. We note
that while $\hat{V}_{AB}$ can in principle connect GS to CT states
and also $\text{CT}^{00}$ to LE states, these matrix elements are
small because $\hat{V}_{AB}$ is largest near the dimer linkage where
nodes of the SOMOs reside, thus any excitations involving SOMO electrons
will be negligible. This is consistent with the notion that, in AHR
\emph{m}-dimers, $\pi$-bonding -- described by mixing of covalent
(LE) and ionic (CT) wavefunctions \citep{Salem1972} -- only occurs
in the excited states. Formally, we extend the nearest neighbour approximation
to the Coulomb interaction parameter $\gamma_{\mu\nu}$ (which scales
inversely with the distance between $\mu$ and $\nu$) and expand
$\hat{V}_{AB}$ in terms of MO fermionic operators using $\hat{a}_{\mu\sigma}^{\dagger}=\sum_{j_{r}}c_{\mu j_{r}}^{*}\hat{b}_{j_{r}\sigma}^{\dagger}$
($r$ indexes the monomer containing $\mu$). The result, labelled
$\hat{V}_{\text{NN}}$, reads {[}Eq. (\ref{eq:V_AB}){]} 
\begin{align}
\hat{V}_{AB} & \approx\hat{V}_{\text{NN}}\nonumber \\
 & \equiv-\beta_{\mathcal{A}\mathcal{B}}\sum_{\sigma}\br{\hat{a}_{\mathcal{A}\sigma}^{\dagger}\hat{a}_{\mathcal{B}\sigma}+\hat{a}_{\mathcal{B}\sigma}^{\dagger}\hat{a}_{\mathcal{A}\sigma}}\nonumber \\
 & \quad+\gamma_{\mathcal{A}\mathcal{B}}\hat{n}_{\mathcal{A}}\hat{n}_{\mathcal{B}}
\end{align}
in the AO basis and 
\begin{align}
\hat{V}_{AB} & \approx\hat{V}_{\text{NN}}\nonumber \\
 & \equiv-\beta_{\mathcal{A}\mathcal{B}}\sum_{j_{A},k_{B}}\sum_{\sigma}\left(c_{\mathcal{A}j_{A}}^{*}c_{\mathcal{B}k_{B}}\hat{b}_{j_{A}\sigma}^{\dagger}\hat{b}_{k_{B}\sigma}\right.\nonumber \\
 & \quad\left.+c_{\mathcal{B}k_{B}}^{*}c_{\mathcal{A}j_{A}}\hat{b}_{k_{B}\sigma}^{\dagger}\hat{b}_{j_{A}\sigma}\right)\nonumber \\
 & \quad+\gamma_{\mathcal{A}\mathcal{B}}\sum_{j_{A},l_{A},k_{B},m_{B}}\sum_{\sigma\tau}c_{\mathcal{A}j_{A}}^{*}c_{\mathcal{A}l_{A}}c_{\mathcal{B}k_{B}}^{*}c_{\mathcal{B}m_{B}}\nonumber \\
 & \quad\times\hat{b}_{j_{A}\sigma}^{\dagger}\hat{b}_{l_{A}\sigma}\hat{b}_{k_{B}\tau}^{\dagger}\hat{b}_{m_{B}\tau}\label{eq:V_NN}
\end{align}
in the MO basis, with $\mathcal{A}$ and $\mathcal{B}$ being the
unstarred atoms linking the two AHRs. Because unstarred atoms are
where the SOMO's nodes are located, we find that $c_{\mathcal{A}0_{A}}=c_{\mathcal{B}0_{B}}=0$,
restricting the sum over $\curbr{j_{A},l_{A},k_{B},m_{B}}$ in Eq.
(\ref{eq:V_NN}) to only non-zero values. This leaves $\hat{V}_{\text{NN}}$
with no excitations connected to the SOMO (index $j=0$), in line
with the above analysis. Importantly, all of these suggest a negligible
GS singlet-triplet gap despite $\hat{V}_{\text{NN}}\neq0$, a sign
of a true diradical.

Corrections to the states and energies are then evaluated to second
order in the approximate perturbation $\hat{V}_{\text{NN}}$, the
results being presented in Supplementary Information S3. Importantly,
because $\hat{V}_{\text{NN}}$ scales with the degree of $\pi$-orbital
overlap between the two AHRs, we expect its effects to be quantified
by an intermonomer torsional angle $\theta$ {[}Fig. \ref{fig:results}a{]}
such that the $\pi$-interactions are maximal when the two monomers
are coplanar at $\theta=0^{\text{o}}\text{ or }180^{\text{o}}$. To
this end, we parametrise $\hat{V}_{\text{NN}}=\hat{V}_{\text{NN}}\br{\theta}$
by substituting 
\begin{align}
\beta_{\mathcal{A}\mathcal{B}} & =\beta_{\mathcal{A}\mathcal{B}}\br{\theta}\approx\beta_{\mathcal{A}\mathcal{B}}^{0}\cos\theta,\label{eq:V_NN-torsion-1}\\
\gamma_{\mathcal{A}\mathcal{B}} & =\gamma_{\mathcal{A}\mathcal{B}}\br{\theta}\approx\gamma_{\mathcal{A}\mathcal{B}}^{0}\cos^{2}\theta.\label{eq:V_NN-torsion-2}
\end{align}
Here, $\cos\theta$ represents the projection of one AHR's local axis
of quantisation (the ``$z$-axis'') onto the other's, and two of
such terms are necessary for any two-electron integral. An additional
parabolic potential is also introduced to model steric effects in
the GS: 
\begin{align}
V_{\text{steric}}\br{\theta} & \approx E_{\text{steric}}\br{\theta-\theta_{\text{eq}}}^{2},\label{eq:V_steric}
\end{align}
which we assume to be the same in the excited states. In general,
we expect $\theta_{\text{eq}}\approx90^{\text{o}}$ due to the lack
of $\pi$-bonding in the GS.

Based on the perturbation theory results and the above $\theta$ parametrisation,
we plotted two representative potential energy surfaces (PESs) in
Fig. \ref{fig:results}d,e. Indeed, the singlet and triplet GSs are
degenerate over the entire range of $\theta$, consistent with the
lack of radical-radical interactions in the GS. The same holds for
LE states at $\theta=90^{\text{o}}$ (because $\hat{V}_{AB}=0$).
However, away from this angle, mixing with the CT states opens up
a gap between the two spin multiplicites -- a sign of $\pi$-bonding
\citep{Salem1972}.

The ODMR cycle is only complete with a spin-selective ISC, the key
towards spin polarisation. At the heart of this effect is the SOC
operator that controls the ISC rate and, as mentioned earlier, only
links rotated $\pi$-orbitals \citep{Hong2001,Rybicki2009,Barford2010,Yu2012}.
Because AHRs are mostly planar, SOC will have to be an intermonomer
effect that is largest at the maximum torsion of $\theta=90^{\text{o}}$.
Furthermore, we expect this operator to act locally at the dimer connection
since its matrix elements scale inversely with the cube of the interatomic
distance. Given these restrictions, we further examine the nature
of SOC by adapting the single-electron Breit-Pauli SOC operator in
PPP theory \citep{Barford2010} to our system. We obtain 
\begin{align}
\hat{V}_{\text{SOC}}\br{\theta} & =\sum_{\sigma}B\sin\br{\theta}\br{\hat{a}_{\mathcal{A}\sigma}^{\dagger}\hat{a}_{\mathcal{B}\overline{\sigma}}-\hat{a}_{\mathcal{B}\overline{\sigma}}^{\dagger}\hat{a}_{\mathcal{A}\sigma}},
\end{align}
or, in terms of MO operators, 
\begin{align}
\hat{V}_{\text{SOC}}\br{\theta} & =\sum_{j_{A}\neq0_{A},k_{B}\neq0_{B}}\sum_{\sigma}B\sin\br{\theta}\left(c_{\mathcal{A}j_{A}}^{*}c_{\mathcal{B}k_{B}}\hat{b}_{j_{A}\sigma}^{\dagger}\hat{b}_{k_{B}\overline{\sigma}}\right.\nonumber \\
 & \quad\left.-c_{\mathcal{B}k_{B}}^{*}c_{\mathcal{A}j_{A}}\hat{b}_{k_{B}\overline{\sigma}}^{\dagger}\hat{b}_{j_{A}\sigma}\right),\label{eq:V_SOC}
\end{align}
where $B$ is a purely imaginary value defined in Supplementary Information
S4. Similar to the analysis of $\hat{V}_{AB}$, we find that $\hat{V}_{\text{SOC}}$
only couples LE and CT excitations and leaves the GS and $\text{CT}^{00}$
states unperturbed due to the SOMO's nodal structure. It creates spin
flips, has ``$+$'' particle-hole symmetry and is odd under $C_{2}$
rotation. Hence, we find that the only allowed ISC processes are 
\begin{align}
\ket{^{1}\text{LE}_{\Gamma}^{10,P};0} & \xleftrightarrow{\hat{V}_{\text{SOC}}}\ket{^{3}\text{CT}_{\overline{\Gamma}}^{10,P};\pm1},\label{eq:V_SOC-rule-1}\\
\ket{^{3}\text{LE}_{\Gamma}^{10,P};\pm1} & \xleftrightarrow{\hat{V}_{\text{SOC}}}\ket{^{1}\text{CT}_{\overline{\Gamma}}^{10,P};0},\label{eq:V_SOC-rule-2}
\end{align}
with $\overline{\Gamma}$ being the irrep opposite to $\Gamma$. As
an aside, in deriving Eq. (\ref{eq:V_SOC}) we have assumed the dimer
linkage to be perpendicular to the spin quantisation axis (Supplementary
Note S4). When no magnetic field is applied, the latter is determined
by the spin-spin interactions intrinsic to the molecule \citep{Atherton1993}.
This implies spin quantisation along the high-symmetry axis of the
molecule, making the earlier assumption valid in AHR \emph{m}-dimers
given the orientation of the $C_{2}$ axis.

\subsection{Proposed ODMR mechanisms}

To sum it up, a detailed electronic structure analysis of AHR \emph{m}-dimers
revealed important symmetries that affect the ODMR mechanism. Of particular
importance are the following:
\begin{enumerate}
\item Excited states may be classified based on their character at torsion
$\theta=90^{\text{o}}$ {[}Fig. \ref{fig:results}a{]}. This can be
either local excitations (LEs), which promote electrons in an intramonomer
fashion, or charge transfers (CTs), which move electrons across monomers
{[}Fig. \ref{fig:results}b{]}. Also labelled are the state's particle-hole
symmetry (a consequence of alternacy symmetry), which can be either
``$+$'' or ``$-$'' {[}Eq. (\ref{eq:state_notation}){]}. These
states are ordered by energy in Fig. \ref{fig:results}c. (\emph{Aside:}
States are also labelled by their irrep $\Gamma\in\curbr{\mathtt{A},\mathtt{B}}$,
but this is inconsequential to the ODMR mechanism.)
\item Near the equilibrium torsion of $\theta\approx90^{\text{o}}$ {[}Fig.
\ref{fig:results}a{]}, intermonomer $\pi$-bonding is negligible.
Therefore, only the CT states have appreciable singlet-triplet gaps
due to intramonomer exchange interactions {[}Fig. \ref{fig:results}c{]}.
This implies that selective photoexcitation of the triplets (over
the singlets) can only occur to CT-dominated states. 
\item Of all lowest-lying LE and CT excitations, only LE states of ``$+$''
particle-hole symmetry are bright at $\theta=90^{\text{o}}$ {[}Eq.
(\ref{eq:dipole}){]}. However, CT states of ``$+$'' particle-hole
symmetry can borrow intensity from these LE states at $\theta$ values
away from $90^{\text{o}}$ due to intermonomer $\pi$-bonding {[}Eqs.
(\ref{eq:V_NN}), (\ref{eq:V_NN-torsion-1}) and (\ref{eq:V_NN-torsion-2}){]}.
\item ISC is fastest at $\theta=90^{\text{o}}$, between CT and LE states,
between states of like particle-hole symmetry, and when the triplet
$M_{S}=\pm1$ levels are involved -- this aligns with El-Sayed rules
\citep{El-Sayed1963}. In addition, ISC processes involving the $\ket{^{1}\text{CT}^{00,-}}$
and $\ket{^{1}\text{CT}^{00,+}}$ states are slow because the SOMO
has a node at the bond connecting the two monomers {[}Eqs. (\ref{eq:V_SOC}),
(\ref{eq:V_SOC-rule-1}) and (\ref{eq:V_SOC-rule-2}){]}.
\end{enumerate}
These observations are consistent with those made in the introduction.
From here, we report two possible pathways for ODMR according to the
excitation energies {[}Fig. \ref{fig:results}c{]}: $
\global\long\def\theenumi{\arabic{enumi}}%
$
\begin{enumerate}
\item In the first case where $J_{10}>2K_{10}$ {[}Fig. \ref{fig:results}d{]},
we propose to directly photoexcite the $\ket{^{3}\text{CT}^{10,+}}$
states at $\theta$ values away from $90^{\text{o}}$ (via thermal
effects) where $\pi$-bonding with the $\ket{^{3}\text{LE}^{10,+}}$
states is appreciable. Relaxation to the equilibrium torsion of $\theta\approx90^{\text{o}}$
is expected. At this point, ISC effects are strongest and couple the
$M_{S}=\pm1$ sublevels of the $\ket{^{3}\text{CT}^{10,+}}$ state
to the $\ket{^{1}\text{LE}^{10,+}}$ state, selectively depopulating
the triplet $M_{S}=\pm1$ levels. Following Kasha's rule, the resulting
$\ket{^{1}\text{LE}^{10,+}}$ population should emit from the $\ket{^{1}\text{CT}^{00,-}}$
and $\ket{^{1}\text{CT}^{00,+}}$ states, relaxing to the ground singlet
state. As for the remaining $\ket{^{3}\text{CT}^{10,+}}$ molecules
in the $M_{S}=0$ level, emission to the ground triplet state should
occur through the $\ket{^{3}\text{LE}^{10,-}}$ states ($M_{S}=0$)
at a different wavelength from the singlets, offering optical readout
of the triplet $M_{S}=0$ levels. We note that both non-radiative
internal conversions (ICs) and emissions are spin-preserving. Therefore,
overall, population from the triplet $M_{S}=\pm1$ ground levels have
been transferred away to the singlet ground state, creating ground-state
spin polarisation. Since ISC rates decay exponentially with the energy
gap \citep{Englman1970}, this mechanism works best when the $\ket{^{3}\text{CT}^{10,+}}$
and $\ket{^{1}\text{LE}^{10,+}}$ states lie close in energy, achieved
when $J_{10}\approx2K_{10}$.
\item In the second case where $J_{10}<2K_{10}$ {[}Fig. \ref{fig:results}e{]},
the photoexcited $\ket{^{3}\text{CT}^{10,+}}$ molecules can, through
a single-phonon process, rotate to $\theta\approx90^{\text{o}}$,
at which point the $\ket{^{3}\text{CT}^{10,-}}$ states are near-degenerate
and directly accessible via spin-preserving ICs \citep{Englman1970}.
This is also the geometry with the best ISC, taking $\ket{^{3}\text{CT}^{10,-}}$
to $\ket{^{1}\text{LE}^{10,-}}$ through the triplet $M_{S}=\pm1$
sublevels, which then relaxes to the ground singlet state. Meanwhile,
the remaining triplet $M_{S}=0$ population in the $\ket{^{3}\text{CT}^{10}}$
states have no ISC channel and will return to the ground triplet state,
with optical readout being possible because emissions from both spin
multiplicities are distinguishable by wavelength (just like in case
1). This mechanism works best with smaller $J_{10}$ values because
not only is the rotation barrier on the $\ket{^{3}\text{CT}^{10,+}}$
surfaces smaller, but also ISC can occur over a smaller energy gap.
\end{enumerate}
A few points are now in order: Firstly, photoexcitation away from
the equilibrium torsion of $\theta\approx90^{\text{o}}$ is possible
at room temperature, verified by ab initio calculations (more to follow)
and with the trade-off of faster spin decoherence. Next, while the
$\ket{^{1}\text{GS}}\rightarrow\ket{^{3}\text{GS}}$ ISC through the
triplet $M_{S}=\pm1$ sublevels will lead to spin relaxation, such
couplings are not mediated by the single-electron Breit-Pauli SOC
due to the operator's CT nature {[}Eq. (\ref{eq:V_SOC}){]}. Thirdly,
while $\text{LE}^{-}$ and CT states of AHR dimers are, at $\theta\approx90^{\text{o}}$
torsion, less emissive than their $\text{LE}^{+}$ counterparts {[}Eq.
(\ref{eq:dipole}){]}, PL can still be detected; for instance, TTM
radicals have quantum yields of around $0.02$ with emission from
$\text{LE}^{-}$ states \citep{Gamero2006,Abdurahman2020,Abdurahman2023-2}.
On a related note, we did not explicitly model IC in the PPP framework
but only expected them to occur without preserving particle-hole symmetry,
as with most AHRs \citep{Abdurahman2020,Hele2021}. Finally, because
spins are polarised by shelving triplet $M_{S}=\pm1$ population into
the singlet state, at most 25\% of the ensemble will be a triplet
$M_{S}=0$, unlike the NV centre where this value is 100\%.

Note that even though we have described only the $1\rightarrow0$
and $0\rightarrow1'$ excitations for simplicity, the same selection
rules apply to any general $j\rightarrow0$ and $0\rightarrow j'$
transitions ($j>0$), which may be low-lying in some AHR \emph{m}-dimers.

\subsection{Validation by ab initio calculations}

\begin{figure*}
\emph{\includegraphics[width=1\textwidth]{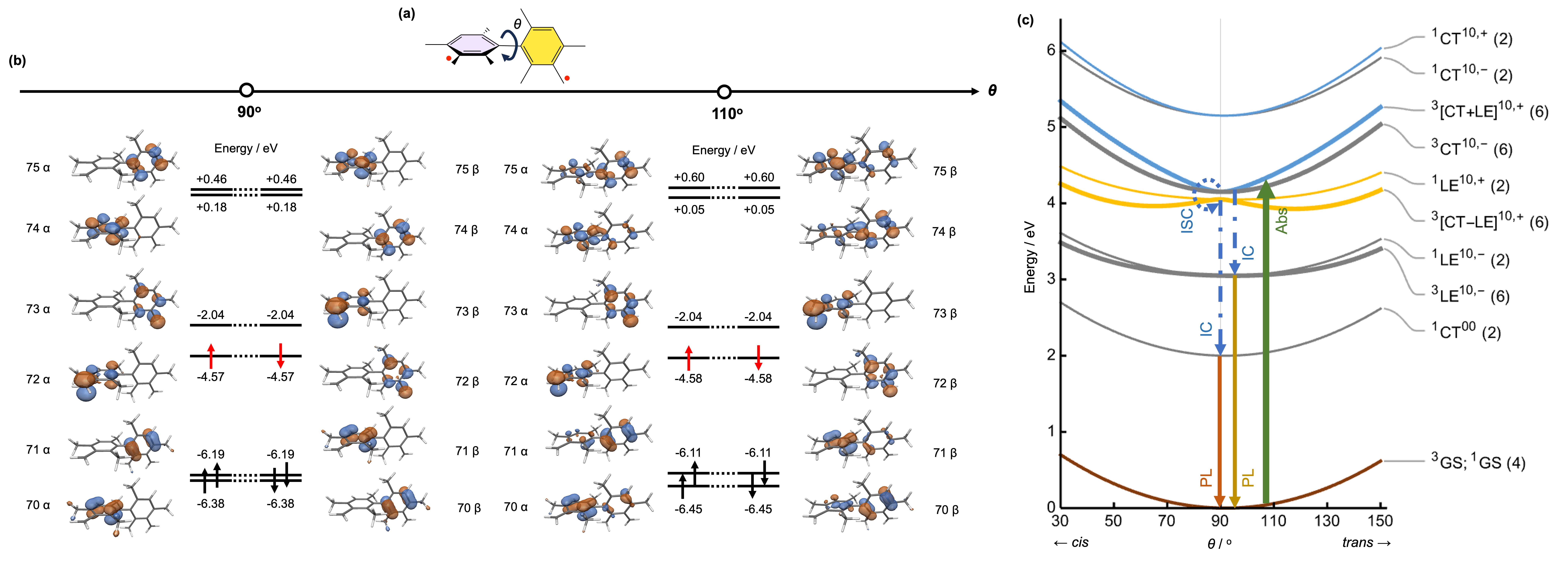}\caption{\label{fig:benzyl-Me}(a) Structure of an AHR \emph{m}-dimer constructed
using two methylated benzylic radicals. (b) Frontier MOs (isovalue
$=0.060$) of the dimer in the open-shell BS ground state at two different
conformations ($\theta=90^{\text{o}}$ and $110^{\text{o}}$), calculated
using DFT (UB3LYP/6--31G(d,p)). (c) Plots of PESs using parameters
from a TDDFT/TDA calculation (UB3LYP/6--31G(d,p)) of the monoradical.
Parameters: $E_{\text{steric}}=0.6\text{ eV}$; $\theta_{\text{eq}}=91.9^{\text{o}}$;
$\beta_{\mathcal{A}\mathcal{B}}^{0}=2.5\text{ eV}$; $\gamma_{\mathcal{A}\mathcal{B}}^{0}=0\text{ eV}$;
SOMO-LUMO gap $=3.4\text{ eV}$; $c_{\mathcal{A}1_{A}}=c_{\mathcal{B}1_{B}}=0.5$;
$J_{00}=2.0\text{ eV}$; $J_{10}=1.1\text{ eV}$; $K_{10}=0.5\text{ eV}$.}
}
\end{figure*}

\begin{figure*}
\emph{\includegraphics[width=1\textwidth]{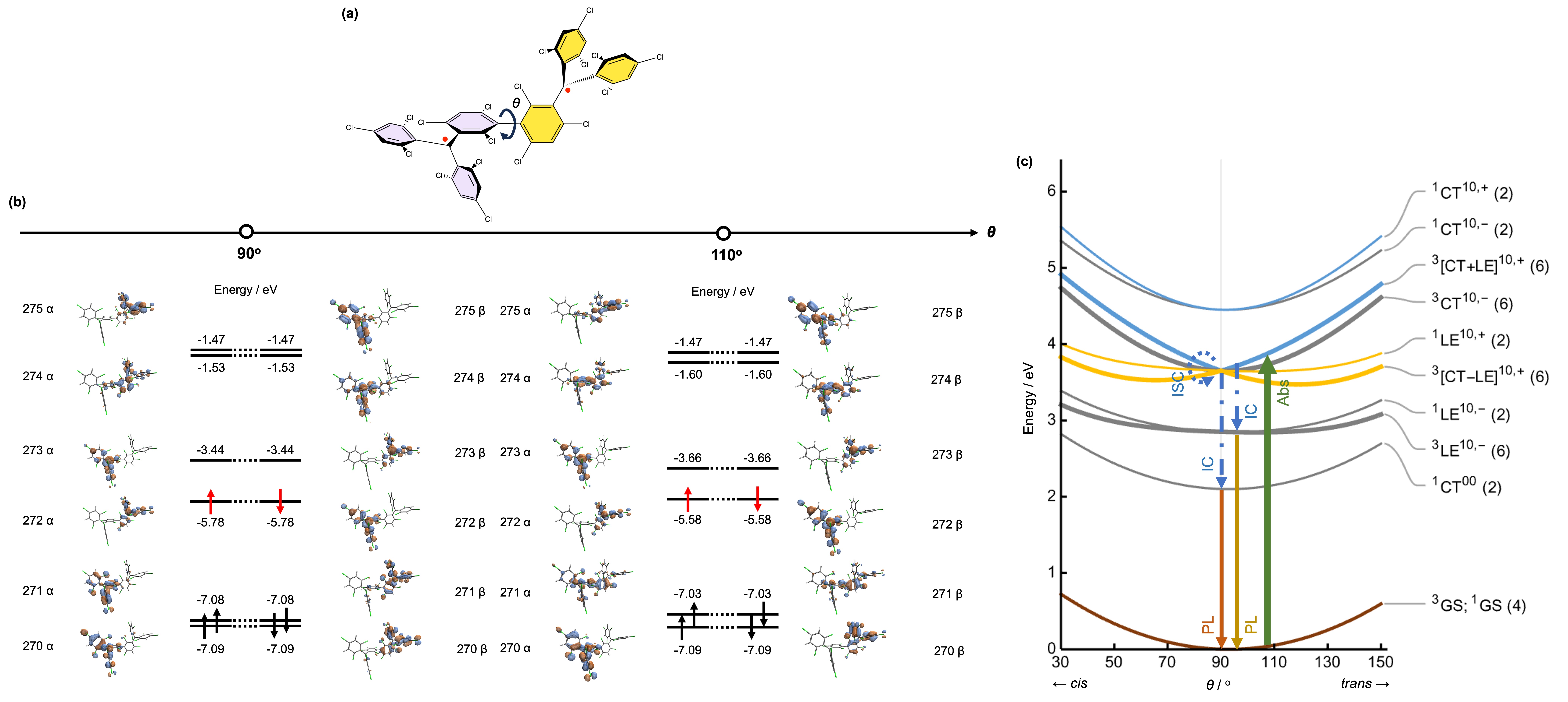}\caption{\label{fig:TTM}(a) Structure of an AHR \emph{m}-dimer constructed
using two TTM radicals. (b) Frontier MOs (isovalue $=0.030$) of the
dimer in the open-shell BS ground state at two different conformations
($\theta=90^{\text{o}}$ and $110^{\text{o}}$), calculated using
DFT (UB3LYP/6--31G(d,p)). (c) Plots of PESs using parameters from
a TDDFT/TDA calculation (UB3LYP/6--31G(d,p)) of the monoradical.
Parameters: $E_{\text{steric}}=0.6\text{ eV}$; $\theta_{\text{eq}}=92.9^{\text{o}}$;
$\beta_{\mathcal{A}\mathcal{B}}^{0}=2.5\text{ eV}$; $\gamma_{\mathcal{A}\mathcal{B}}^{0}=0\text{ eV}$;
SOMO-LUMO gap $=2.8\text{ eV}$; $c_{\mathcal{A}1_{A}}=c_{\mathcal{B}1_{B}}=0.5$;
$J_{00}=2.1\text{ eV}$; $J_{10}=0.8\text{ eV}$; $K_{10}=0.4\text{ eV}$.}
}
\end{figure*}

To set the results in the context of existing systems, we consider
two specific AHRs: A methylated benzylic radical {[}Fig. \ref{fig:benzyl-Me}a{]}
and TTM {[}Fig. \ref{fig:TTM}a{]}. Ground-state properties of the
AHR \emph{m}-dimers were calculated using unrestricted DFT at the
B3LYP/6--31G(d,p) level. In both cases, significant torsion was found
at the equilibrium geometry, with $\theta_{\text{eq}}$ values of
$91.9^{\text{o}}$ and $92.9^{\text{o}}$ respectively. Through calculations
converging to the open-shell broken-symmetry (BS) state, we also found
small singlet-triplet energy gaps ($\Delta E_{\text{ST}}$) at both
$\theta=90^{\text{o}}$ ($-0.003\text{ eV}$ and $-0.053\text{ eV}$
respectively) and $\theta=110^{\text{o}}$ ($-0.006\text{ eV}$ and
$-0.031\text{ eV}$), indicating weak exchange interactions between
the two monoradicals \footnote{These values were calculated using the Q-Chem package (version 6.0.2)
\citep{Epifanovsky2021}. For the same computation on the Gaussian16
software \citep{g16}, the respective values were $-0.005\text{ eV}$
and $+0.001\text{ eV}$ at $\theta=90^{\text{o}}$, and $-0.009\text{ eV}$
and $+0.001\text{ eV}$ at $\theta=110^{\text{o}}$.}. Indeed, the two SOMOs of the BS state were degenerate, each localised
on separate monomers {[}Fig. \ref{fig:benzyl-Me},\ref{fig:TTM}b{]}.
Closed-shell singlet (CS) states were also found, with energies that
were at least $1.0\text{ eV}$ higher than the open-shell triplet
states (respective values: $1.021\text{ eV}$ and $1.031\text{ eV}$
at $\theta=90^{\text{o}}$; $0.974\text{ eV}$ and $1.019\text{ eV}$
at $\theta=110^{\text{o}}$). This suggests that AHR \emph{m}-dimers
are likely to be open-shell diradicals in the ground state across
different torsional angles $\theta$. Finally, rotating the molecules
away from $\theta=90^{\text{o}}$ distributed the HOMOs and LUMOs
(but not the SOMOs) of the BS state across the dimer, a sign of interradical
interactions in the excited states {[}Fig. \ref{fig:benzyl-Me},\ref{fig:TTM}b{]}.
These findings are consistent with the analytical results from the
PPP Hamiltonian.

For the methylated benzylic radical \emph{m}-dimer, multi-configurational
self-consistent field (MCSCF) and configuration interaction (CI) methods
were also used to verify the DFT calculations (MCSCF approach: complete
active space self-consistent field (CASSCF) method with 10 electrons
and 10 orbitals optimised for the ground state; see Methods). $\Delta E_{\text{ST}}$
was computed to be $-0.004\text{ eV}$ at $\theta=90^{\text{o}}$
and $-0.009\text{ eV}$ at $\theta=110^{\text{o}}$ with the first
excited singlet energies being $2.227\text{ eV}$ and $2.207\text{ eV}$
respectively relative to the triplet ground state. These results agree
with those from ground-state DFT (note that the CS energy is approximately
half of the first singlet excitation energy; see Methods). 

We also approximated the width of the Boltzmann distribution of $\theta$
by $\Delta\theta\equiv\sqrt{k_{\text{B}}T/E_{\text{steric}}}$ {[}Eq.
(\ref{eq:V_steric}){]} with $k_{\text{B}}$ and $T$ representing
the Boltzmann constant and temperature respectively. For both AHR
\emph{m}-dimers, $\Delta\theta=12^{\text{o}}$ at $T=298\text{ K}$,
indicating that excitation away from $\theta=90^{\text{o}}$, required
for ODMR {[}Fig. \ref{fig:results}d,e{]}, may be achieved by operating
at room temperature (at the expense of enhancing deleterious decoherence
processes). Importantly, this value can be tuned by modifying the
degree of steric hindrance at the dimer linkage. As an example, to
the methylated benzylic radical \emph{m}-dimer, we removed the methyl
substituents closest to the dimer linkage that are \emph{ortho} to
the methylene group. This widened the ground-state PES, yielding the
same $\Delta\theta$ value at a lower temperature of $T=108\text{ K}$
(Supplementary Information S7).

Coulomb and exchange integrals among the MOs were then estimated from
the excited-state properties of the monoradicals using unrestricted
time-dependent DFT (TD-DFT) calculations, conducted within the Tamm--Dancoff
approximation (TDA) and with the same functional and basis set as
the dimers. We note that these excitation energies have poorer accuracies
of $\sim0.3\text{ eV}$ than a rigorous multi-reference electron correlation
approach \citep{Laurent2013} but are nevertheless sufficient for
the present goal of setting our analytical results in the correct
parameter range. With the methylated benzylic radical, an additional
$\ket{\Psi_{02}^{-};\pm1/2}$ state (following the notation of Eq.
(\ref{eq:PHT})) was found to lie between the $\ket{\Psi_{01}^{-};\pm1/2}$
and $\ket{\Psi_{01}^{+};\pm1/2}$ excitations. Because states of ``$-$''
particle-hole symmetry are dark, the $\ket{\Psi_{02}^{-};\pm1/2}$
excitation only contributes an additional intermediate state for non-radiative
decay and does not affect the overall ODMR process. Other than that,
the methylated benzylic radical is an example of case 1, with $J_{10}\approx1.1\text{ eV}$
and $K_{10}\approx0.5\text{ eV}$. We note that given the small energy
difference between $J_{10}$ and $2K_{10}$, perturbation theory in
$\hat{V}_{\text{NN}}$ between the $\ket{^{3}\text{LE}_{\Gamma}^{10,+};M_{S}}$
and $\ket{^{3}\text{CT}_{\Gamma}^{10,+};M_{S}}$ states breaks down
and it is necessary to exactly diagonalise the matrix representation
of $\hat{V}_{\text{NN}}$ within this subspace; these eigenstates
are labelled $\ket{^{3}\sqbr{\text{CT}\pm\text{LE}}_{\Gamma}^{10,+};M_{S}}$
(see Supplementary Information S5). Despite that, the ODMR mechanism
remains unchanged {[}Fig. \ref{fig:benzyl-Me}c{]}.

A smaller $J_{10}$ value is desirable for case 1 because the energy
gap for ISC will be reduced. This may be achieved by extending the
degree of $\pi$-conjugation, which spreads the electron clouds of
the valence MOs out, thereby reducing Coulombic repulsion between
the occupying electrons. Indeed, the TTM radical, which has a (chlorinated)
benzylic radical $\pi$-conjugated to two more (chlorinated) phenyl
groups, was estimated to have a lower $J_{10}$ value of $\approx0.8\text{ eV}$
($K_{10}\approx0.4\text{ eV}$) when averaged over its five near-degenerate
lowest energy LUMOs. This places the TTM \emph{m}-dimer in the most
ideal scenario for case 1 of ODMR in AHR \emph{m}-dimers {[}Fig. \ref{fig:TTM}c{]}.

\begin{figure*}
\emph{\includegraphics[width=1\textwidth]{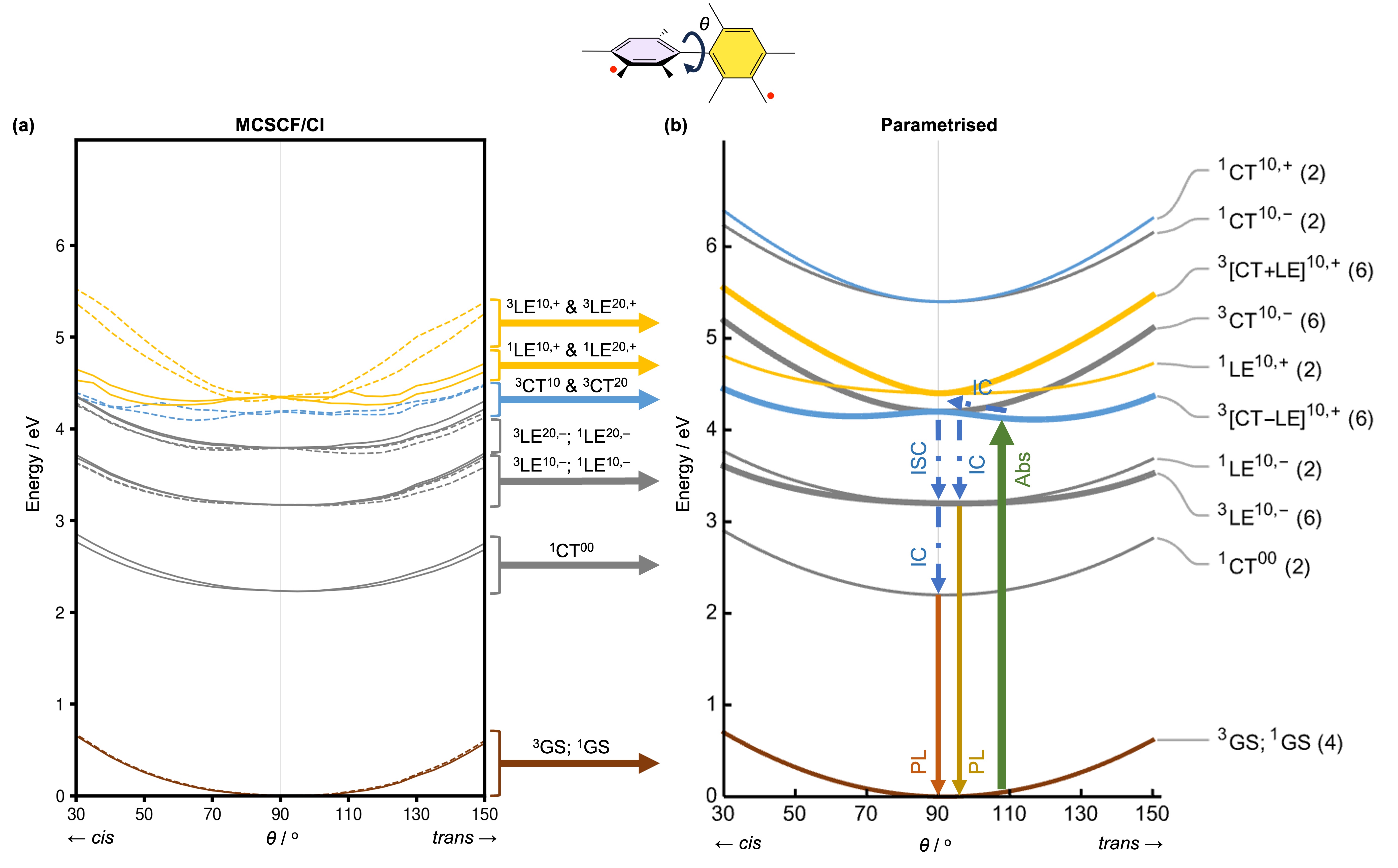}\caption{\label{fig:benzyl-Me-CAS}(a) PESs of the nine lowest-lying triplet
$M_{S}=+1$ states (dashed lines) and the nine lowest-lying singlet
states (solid lines) of the methylated benzylic radical \emph{m}-dimer,
obtained with MCSCF/CI/6--31G(d,p) calculations (theory level: CASSCF(10,10)/CASCI/QD-SC-NEVPT2;
see Methods). States were assigned based on their characters at $\theta\approx90^{\text{o}}$.
(b) Plots of PESs using parameters from MCSCF/CI/6--31G(d,p) calculations
of the \emph{m}-dimer. Parameters: $E_{\text{steric}}=0.6\text{ eV}$;
$\theta_{\text{eq}}=91.9^{\text{o}}$; $\beta_{\mathcal{A}\mathcal{B}}^{0}=2.5\text{ eV}$;
$\gamma_{\mathcal{A}\mathcal{B}}^{0}=0\text{ eV}$; SOMO-LUMO gap
$=3.4\text{ eV}$; $c_{\mathcal{A}1_{A}}=c_{\mathcal{B}1_{B}}=0.5$;
$J_{00}=2.2\text{ eV}$; $J_{10}=1.0\text{ eV}$; $K_{10}=0.6\text{ eV}$.}
}
\end{figure*}

Because the MCSCF/CI approach fully describes the multi-configurational
nature of the methylated benzylic radical \emph{m}-dimer, we used
that to verify the PPP model parameters, which we had obtained earlier
from a single-reference TDDFT/TDA calculation of the monoradical.
At this level of theory, we estimated $\text{SOMO-LUMO gap}\approx3.4\text{ eV}$,
$J_{00}\approx2.2\text{ eV}$, $J_{10}\approx1.0\text{ eV}$, and
$K_{10}\approx0.6\text{ eV}$. PESs plotted using these parameters
agree well with the MCSCF/CI calculations {[}Fig. \ref{fig:benzyl-Me-CAS}a,b{]}.
By placing $J_{10}$ and $2K_{10}$ at slightly different values from
TDDFT/TDA (with differences of $\approx0.1\text{ eV}$), MCSCF/CI
changed the ODMR mechanism to that of case 2 {[}Fig. \ref{fig:benzyl-Me-CAS}b{]}.
Nevertheless, we believe both mechanisms to be possible given the
uncertainty in TDDFT/TDA energies and the sensitivity of MCSCF/CI
results to input parameters. Another advantage of the MCSCF/CI method
is the possibility for ab initio SOC calculations. Indeed, only CT-type
ISCs had appreciable SOC matrix elements with $\Delta M_{S}=\pm1$
selectivity, in agreement with Eqs. (\ref{eq:V_SOC-rule-1}) and (\ref{eq:V_SOC-rule-2})
(Supplementary Information S7). All these lend credence to the semiempirical
analysis described earlier.

\subsection{Mapping the triplet ground state to a controllable qubit}

Finally, we note that the triplet ground state of AHR \emph{m}-dimers
is likely to have minimal zero-field splittings (ZFSs) due to weak
spin dipolar interactions between spatially separated radicals \citep{Atherton1993}.
As such, a two-state qubit cannot be isolated by simply applying an
external magnetic field, in contrast to the diamond-NV centre. Nevertheless,
we propose for an \emph{effective} qubit with the $M_{S}=0$ sublevel
being one of the qubit states and the $M_{S}=\pm1$ sublevels functioning
collectively as an \emph{effective} second qubit state. This is possible
because for systems of two non-interacting electron spins experiencing
an external magnetic field along its quantisation axis ($C_{2}$ axis
for AHR \emph{m}-dimers), circularly-polarised microwave light will
produce unitary operations of the 3D rotation group SO(3), that is,
it rotates the $\ket{M_{S}=0}$ or $\ket z$ state into linear combinations
of $\ket x$ and $\ket y$ states, defined as 
\begin{align}
\ket x & \equiv\frac{\ket{M_{S}=+1}-\ket{M_{S}=-1}}{\sqrt{2}},\label{eq:spin-1-x}\\
\ket y & \equiv i\frac{\ket{M_{S}=+1}+\ket{M_{S}=-1}}{\sqrt{2}}.\label{eq:spin-1-y}
\end{align}
For instance, if an AHR \emph{m}-dimer were to be spin polarised into
an initial quantum state of $\ket{\psi}=\ket z$, then a microwave
pulse will drive it into 
\begin{align}
\ket{\psi} & \rightarrow\ket{\psi'}=a\ket x+b\ket y+c\ket z,
\end{align}
keeping $a,b,c\in\mathbb{R}$ and preserving the norm of $a^{2}+b^{2}+c^{2}=1$.
In spherical coordinates, this reads 
\begin{align}
\ket{\psi'} & =\sin\theta\cos\phi\ket x+\sin\theta\sin\phi\ket y+\cos\theta\ket z,
\end{align}
with $\theta\in\sqbr{0,\pi}$ and $\phi\in\left[0,2\pi\right)$. Since
two real numbers are sufficient to describe a qubit state, an effective
qubit arises here because the normalised state $\ket{\psi'}$ is also
characterised by two real-valued coefficients $\theta$ and $\phi$.
In this way, the $M_{S}=\pm1$ sublevels have become the second qubit
state through either $\ket x$ or $\ket y$ (with the other state
being redundant). This is essentially a statement of the isomorphism
between SO(3) and SU(2), the latter of which describes a qubit. Here,
our accessible pure state space is a unit 2-sphere with the three
Cartesian axes corresponding to $\ket x$, $\ket y$ and $\ket z$
states, analogous to the Bloch sphere for a pure qubit state {[}Table
\ref{tab:qubit}{]}.

\begin{table*}
\emph{\caption{\label{tab:qubit}Mapping between a standard qubit and our effective
qubit. A pure qubit state \textquotedblleft lives\textquotedblright{}
on a Bloch sphere spanned by eigenstates of the Pauli vector operator
$\hat{\bm{\sigma}}$, denoted here by $\protect\curbr{\protect\ket{\uparrow},\protect\ket{\downarrow}}$.
The general form of a single qubit operator is thus a rotation on
the Bloch sphere, generated by ${\bf n}\cdot\hat{\bm{\sigma}}/2$
for some unit vector ${\bf n}$. Because our effective qubit state,
if pure, can only access a unit sphere in 3D real space, the corresponding
single qubit operator is a 3D rotation, generated by ${\bf n}\cdot\hat{{\bf S}}$.
Here, $\hat{{\bf S}}$ is the spin-1 operator with eigenstates $\protect\curbr{\protect\ket{M_{S}}\vert M_{S}=-1,0,+1}$,
where $\protect\ket z\equiv\protect\ket{M_{S}=0}$ and $\protect\curbr{\protect\ket x,\protect\ket y}$
are linear combinations of $\protect\curbr{\protect\ket{M_{S}=\pm1}}$
states {[}Eqs. (\ref{eq:spin-1-x}) and (\ref{eq:spin-1-y}){]}.}
}

\begin{tabular*}{1\textwidth}{@{\extracolsep{\fill}}ccc}
\toprule 
 & \textbf{Standard qubit} & \textbf{Our effective qubit}\tabularnewline
\midrule
\multirow{2}{*}{Pure state} & $\ket{\psi}=\cos\br{\theta/2}\ket{\uparrow}+e^{i\phi}\sin\br{\theta/2}\ket{\downarrow}$ & $\ket{\psi}=\sin\theta\cos\phi\ket x+\sin\theta\sin\phi\ket y+\cos\theta\ket z$\tabularnewline
 & with $\theta\in\sqbr{0,\pi}$ and $\phi\in\left[0,2\pi\right)$ & with $\theta\in\sqbr{0,\pi}$ and $\phi\in\left[0,2\pi\right)$\tabularnewline
 &  & \tabularnewline
Single qubit operator & $\hat{R}_{{\bf n}}\br{\theta}=e^{-i\theta{\bf n}\cdot\hat{\bm{\sigma}}/2}$ & $\hat{R}_{{\bf n}}\br{\theta}=e^{-i\theta{\bf n}\cdot\hat{{\bf S}}}$\tabularnewline
\bottomrule
\end{tabular*}
\end{table*}

To see how circularly-polarised microwave pulses result in unitary
operations of the SO(3) group, we consider a Hamiltonian $\mathcal{H}$
of two electron spins $\hat{{\bf S}}_{j}$ ($j\in\curbr{A,B}$) with
negligible exchange and spin-dipolar couplings. Both spins are precessing
under the influence of both a static magnetic field $B_{0}{\bf z}$
and an oscillating circularly-polarised microwave field $B_{1}\sqbr{\cos\br{\omega t}{\bf x}+\sin\br{\omega t}{\bf y}}$
(${\bf x}$, ${\bf y}$ and ${\bf z}$ are unit vectors of the Cartesian
axes). This Hamiltonian reads 
\begin{align}
\mathcal{H}\br t & =\sum_{j=A,B}g_{j}\mu_{\text{B}}\left[B_{0}\hat{S}_{j,z}\right.\nonumber \\
 & \quad\left.+B_{1}\hat{S}_{j,x}\cos\br{\omega t}+B_{1}\hat{S}_{j,y}\sin\br{\omega t}\right],
\end{align}
where $g_{j}$ is the \emph{g}-factor of electron $j$ and $\mu_{\text{B}}$
is the Bohr magneton. By substituting $g_{A}=g_{B}\equiv g$ for dimers
and $\hat{{\bf S}}\equiv\hat{{\bf S}}_{A}+\hat{{\bf S}}_{B}$ as the
total spin operator, we find $\mathcal{H}$ in the rotating frame
of the oscillating field to be 
\begin{align}
\mathcal{H}_{\text{r.f.}} & =\br{g\mu_{\text{B}}B_{0}-\hbar\omega}\hat{S}_{z}+g\mu_{\text{B}}B_{1}\hat{S}_{x}.
\end{align}
Thus, the time evolution operator at resonance is $\hat{U}\br t=\exp\br{-i\omega_{1}\hat{S}_{x}t}$,
the rotation operator about the \emph{x}-axis by angle $\omega_{1}t$
($\omega_{1}\equiv g\mu_{\text{B}}B_{1}/\hbar$). Similarly, a \emph{y}-axis
rotation may be obtained by shining light of the opposite polarisation.
Thus, through a series of microwave pulses we can achieve any arbitrary
single qubit operation in our effective qubit (for instance, through
the \emph{X}--\emph{Y} decomposition \citep{Nielsen2010}), satisfying
the generality criteria.

We note that general rotations of a three-level system (qutrit) belong
to the SU(3) group, which contains the SO(3) group, so the operations
proposed above are only a subset of all possible rotations in a qutrit,
an observation consistent with our mapping of a qutrit to a qubit.

\section{Conclusion}

In conclusion, we demonstrate AHR \emph{m}-dimers to be promising
sources of fully-organic molecular colour centres. By analysing their
electronic structure using the PPP framework, we propose different
ODMR mechanisms for different energetic orderings of the excited states.
The general principle is the same in all cases: Because the SOC operator,
responsible for spin filtering through its $\Delta M_{S}=\pm1$ rule,
only mediates transitions across two rotated AHRs, charge transfer
must be engineered between a triplet and a singlet in the excited
states. Moreover, only CT-type excited states have energetically well-separated
singlet and triplet levels because an unpaired electron from one monoradical
has been excited into the other monomer and thus experiences stronger
exchange interactions. Therefore, one may begin with a CT-type photoexcitation
of \emph{only} the triplet molecules, which will then see their $M_{S}=\pm1$
population reverse the charge transfer via SOC to the singlet excited
states, i.e. undergo spin-selective ISC. The excited-state spin polarisation
may then be transferred to the ground state via any series of spin-selective
relaxation, such as IC or emission. Since the emission probabilities
and wavelengths of the triplet $M_{S}$ levels will be different due
to different decay channels ($\pm1$ through the singlets and $0$
through the triplets), one can achieve optical readout of the ground-state
spin polarisation by observing changes to the emission spectrum. The
specific method for optical readout will differ based on the molecular
parameters.

To the best of our knowledge, this is the first theoretical analysis
on carbon-based ground-state optically-addressable molecular qubits
and we hope that our findings can guide future efforts in this field.
To this end, we have outlined the possibility of driving the ZFS-free
triplet sublevels using microwave pulses, an example of a single qubit
operation. Ongoing efforts include further structural optimisation
of AHR \emph{m}-dimers and generalisation of the qubit pulse protocol
using a density matrix formalism.

\section{Methods}

Unless otherwise stated, all DFT-based calculations were performed
using the Q-Chem package (version 6.0.2) \citep{Epifanovsky2021}.
In the next two paragraphs, we describe our DFT methods, conducted
at the UB3LYP/6--31G(d,p) level and in vacuo. Ground-state geometries
of the AHR \emph{m}-dimers were optimised as triplets and PES scans
were performed as triplets over $20^{o}\le\theta\le160^{\text{o}}$
(interval $=5^{\text{o}}$), of which the resulting energies were
fitted into $V_{\text{steric}}$ {[}Eq. (\ref{eq:V_steric}){]}. Using
geometries obtained for $\theta=90^{\text{o}}$ and $110^{\text{o}}$
(\emph{trans}), energies of the open-shell singlet were found by the
spin-unrestricted BS approach and used to compute $\Delta E_{\text{ST}}$
following Yamaguchi et al. \citep{Yamaguchi1988}: 
\begin{align}
\Delta E_{\text{ST}} & \equiv E_{\text{S}}-E_{\text{T}}=\frac{\braket{{\bf S}^{2}}_{\text{T}}}{\braket{{\bf S}^{2}}_{\text{T}}-\braket{{\bf S}^{2}}_{\text{BS}}}\br{E_{\text{BS}}-E_{\text{T}}}.
\end{align}
Energies of the CS states were also computed at the same conformations.
Due to $C_{2}$ symmetry, the CS states were described by $\ket{\cdots1_{+}\overline{1_{+}}1_{-}\overline{1_{-}}0_{+}\overline{0_{+}}}$
with $j_{\pm}\approx2^{-1/2}\br{j_{A}\pm j_{B}}$ ($j=\cdots,1,0,1',\cdots$).
Hence, for AHR \emph{m}-dimers described by the PPP Hamiltonian $\hat{H}$
{[}Eqs. (\ref{eq:H_r}) and (\ref{eq:V_AB}){]}, the CS states will
have energies of 
\begin{align}
E_{\text{CS}} & \approx E_{\text{T}}+\frac{J_{00}}{2}\label{eq:CS-energy}
\end{align}
(Supplementary Information S6). From that, we estimated the values
of $J_{00}$.

Geometries of AHR monomers were optimised as doublets, following which
unrestricted TDDFT/TDA was applied without further structural changes.
For the methylated benzylic radical, the SOMO-LUMO gap ($\varepsilon_{01}$)
was estimated as half of the HOMO-LUMO gap. Within the mean-field
approach set by Longuet-Higgins and Pople \citep{Longuet-Higgins1950,Dewar1954,Longuet-Higgins1955},
we find $J_{10}$ and $K_{10}$ to have the following expressions:
\begin{align}
K_{10} & =\frac{E_{01}^{+}-E_{01}^{-}}{2},\\
J_{10} & =-E_{01}^{-}+\varepsilon_{01}+\frac{J_{00}}{2}-\frac{1}{2}K_{10},
\end{align}
where $E_{01}^{\pm}$ are the excitation energies of $\ket{\Psi_{01}^{\pm};+1/2}$,
estimated using TDDFT/TDA. A similar approach was taken for the TTM
radical. However, because the highest-energy HOMOs and lowest-energy
LUMOs were (almost) five-fold degenerate each, for simplicity the
$\curbr{\br{j\rightarrow0}\vert1\le j\le5}$ and $\curbr{\br{0\rightarrow j'}\vert1\le j\le5}$
sets of lowest-lying excitations were each modelled as effective single
$1\rightarrow0$ and $0\rightarrow1'$ excitations with parameters
averaged over the five MOs. This is sufficient to place the PPP model
parameters in the correct energy range.

For the methylated benzylic radical \emph{m}-dimer, MCSCF calculations
were performed on the UB3LYP-optimised triplet ground-state geometry
using the CASSCF method. We conducted state-specific CASSCF(10,10)
calculations to obtain the ground triplet and singlet states, following
which the excited states were estimated using the complete active
space configuration interaction (CASCI) method. All states and energies
were corrected by the van Vleck quasi-degenerate (QD) extension to
the strongly contracted second-order N-electron valence state perturbation
theory (SC-NEVPT2). We were able to classify all lowest-lying excited
states into an excitation from Fig. \ref{fig:results}c by observing
the CI vector characters and oscillator strengths. This allowed us
to estimate the parameters $\varepsilon_{01}$, $J_{00}$, $J_{10}$,
and $K_{10}$ by comparing the computed energies with the expressions
in Fig. \ref{fig:results}c. In particular, the first singlet excitation
had $\text{CT}^{00}$ character, placing its energy at approximately
$J_{00}$ or twice that of the DFT CS state {[}Eq. (\ref{eq:CS-energy}){]}.
Meanwhile, SOC matrix elements were computed between singlet and triplet
wavefunctions of a CASCI/QD-SC-NEVPT2 calculation using state-specific
orbitals optimised by CASSCF(10,10) for the triplet ground state.
Only for this calculation were the orbitals shared between singlets
and triplets. Also, the $\hat{S}_{z}$ eigenbasis was used to express
the electron spin states with the $z$-axis aligned to the molecular
$C_{2}$ axis -- these are the most-likely spin eigenstates of the
molecule (see discussion following Eq. (\ref{eq:V_SOC-rule-2})).
Finally, ground and excited PESs were estimated via the same state-specific
CASSCF(10,10)/CASCI/QD-SC-NEVPT2 strategy with separate orbitals for
singlets and triplets and using the UB3LYP-scanned triplet conformations
mentioned earlier. These multi-configurational calculations were performed
with the ORCA 5.0 code \citep{Neese2022} and the basis set remains
unchanged at 6--31G(d,p).

\section*{Acknowledgements}

Y.R.P., N.P.K., R.G.H., and J.Y.-Z. were supported through the U.S.
Department of Energy (DOE) under 2019030-SP DOE CalTech Sub S532207
(DE-SC0022089). D.M. and G.G. were supported by the Academy of Finland,
grants 323996 and 332743. N.P.K. acknowledges support by the Hertz
Fellowship and the National Science Foundation Graduate Research Fellowship
Program under Grant No. DGE-1745301. We also thank Arghadip Koner
and Kai Schwennicke for helpful discussions.


\begin{thebibliography}{79}%
\makeatletter
\providecommand \@ifxundefined [1]{%
 \@ifx{#1\undefined}
}%
\providecommand \@ifnum [1]{%
 \ifnum #1\expandafter \@firstoftwo
 \else \expandafter \@secondoftwo
 \fi
}%
\providecommand \@ifx [1]{%
 \ifx #1\expandafter \@firstoftwo
 \else \expandafter \@secondoftwo
 \fi
}%
\providecommand \natexlab [1]{#1}%
\providecommand \enquote  [1]{``#1''}%
\providecommand \bibnamefont  [1]{#1}%
\providecommand \bibfnamefont [1]{#1}%
\providecommand \citenamefont [1]{#1}%
\providecommand \href@noop [0]{\@secondoftwo}%
\providecommand \href [0]{\begingroup \@sanitize@url \@href}%
\providecommand \@href[1]{\@@startlink{#1}\@@href}%
\providecommand \@@href[1]{\endgroup#1\@@endlink}%
\providecommand \@sanitize@url [0]{\catcode `\\12\catcode `\$12\catcode
  `\&12\catcode `\#12\catcode `\^12\catcode `\_12\catcode `\%12\relax}%
\providecommand \@@startlink[1]{}%
\providecommand \@@endlink[0]{}%
\providecommand \url  [0]{\begingroup\@sanitize@url \@url }%
\providecommand \@url [1]{\endgroup\@href {#1}{\urlprefix }}%
\providecommand \urlprefix  [0]{URL }%
\providecommand \Eprint [0]{\href }%
\providecommand \doibase [0]{https://doi.org/}%
\providecommand \selectlanguage [0]{\@gobble}%
\providecommand \bibinfo  [0]{\@secondoftwo}%
\providecommand \bibfield  [0]{\@secondoftwo}%
\providecommand \translation [1]{[#1]}%
\providecommand \BibitemOpen [0]{}%
\providecommand \bibitemStop [0]{}%
\providecommand \bibitemNoStop [0]{.\EOS\space}%
\providecommand \EOS [0]{\spacefactor3000\relax}%
\providecommand \BibitemShut  [1]{\csname bibitem#1\endcsname}%
\let\auto@bib@innerbib\@empty
\bibitem [{\citenamefont {Awschalom}\ \emph {et~al.}(2018)\citenamefont
  {Awschalom}, \citenamefont {Hanson}, \citenamefont {Wrachtrup},\ and\
  \citenamefont {Zhou}}]{Awschalom2018}%
  \BibitemOpen
  \bibfield  {author} {\bibinfo {author} {\bibfnamefont {D.~D.}\ \bibnamefont
  {Awschalom}}, \bibinfo {author} {\bibfnamefont {R.}~\bibnamefont {Hanson}},
  \bibinfo {author} {\bibfnamefont {J.}~\bibnamefont {Wrachtrup}},\ and\
  \bibinfo {author} {\bibfnamefont {B.~B.}\ \bibnamefont {Zhou}},\ }\href
  {https://doi.org/10.1038/s41566-018-0232-2} {\bibfield  {journal} {\bibinfo
  {journal} {Nature Photonics}\ }\textbf {\bibinfo {volume} {12}},\ \bibinfo
  {pages} {516} (\bibinfo {year} {2018})}\BibitemShut {NoStop}%
\bibitem [{\citenamefont {Abobeih}\ \emph {et~al.}(2019)\citenamefont
  {Abobeih}, \citenamefont {Randall}, \citenamefont {Bradley}, \citenamefont
  {Bartling}, \citenamefont {Bakker}, \citenamefont {Degen}, \citenamefont
  {Markham}, \citenamefont {Twitchen},\ and\ \citenamefont
  {Taminiau}}]{Abobeih2019}%
  \BibitemOpen
  \bibfield  {author} {\bibinfo {author} {\bibfnamefont {M.~H.}\ \bibnamefont
  {Abobeih}}, \bibinfo {author} {\bibfnamefont {J.}~\bibnamefont {Randall}},
  \bibinfo {author} {\bibfnamefont {C.~E.}\ \bibnamefont {Bradley}}, \bibinfo
  {author} {\bibfnamefont {H.~P.}\ \bibnamefont {Bartling}}, \bibinfo {author}
  {\bibfnamefont {M.~A.}\ \bibnamefont {Bakker}}, \bibinfo {author}
  {\bibfnamefont {M.~J.}\ \bibnamefont {Degen}}, \bibinfo {author}
  {\bibfnamefont {M.}~\bibnamefont {Markham}}, \bibinfo {author} {\bibfnamefont
  {D.~J.}\ \bibnamefont {Twitchen}},\ and\ \bibinfo {author} {\bibfnamefont
  {T.~H.}\ \bibnamefont {Taminiau}},\ }\href
  {https://doi.org/10.1038/s41586-019-1834-7} {\bibfield  {journal} {\bibinfo
  {journal} {Nature}\ }\textbf {\bibinfo {volume} {576}},\ \bibinfo {pages}
  {411} (\bibinfo {year} {2019})}\BibitemShut {NoStop}%
\bibitem [{\citenamefont {Pfaff}\ \emph {et~al.}(2014)\citenamefont {Pfaff},
  \citenamefont {Hensen}, \citenamefont {Bernien}, \citenamefont {van Dam},
  \citenamefont {Blok}, \citenamefont {Taminiau}, \citenamefont {Tiggelman},
  \citenamefont {Schouten}, \citenamefont {Markham}, \citenamefont {Twitchen},\
  and\ \citenamefont {Hanson}}]{Pfaff2014}%
  \BibitemOpen
  \bibfield  {author} {\bibinfo {author} {\bibfnamefont {W.}~\bibnamefont
  {Pfaff}}, \bibinfo {author} {\bibfnamefont {B.~J.}\ \bibnamefont {Hensen}},
  \bibinfo {author} {\bibfnamefont {H.}~\bibnamefont {Bernien}}, \bibinfo
  {author} {\bibfnamefont {S.~B.}\ \bibnamefont {van Dam}}, \bibinfo {author}
  {\bibfnamefont {M.~S.}\ \bibnamefont {Blok}}, \bibinfo {author}
  {\bibfnamefont {T.~H.}\ \bibnamefont {Taminiau}}, \bibinfo {author}
  {\bibfnamefont {M.~J.}\ \bibnamefont {Tiggelman}}, \bibinfo {author}
  {\bibfnamefont {R.~N.}\ \bibnamefont {Schouten}}, \bibinfo {author}
  {\bibfnamefont {M.}~\bibnamefont {Markham}}, \bibinfo {author} {\bibfnamefont
  {D.~J.}\ \bibnamefont {Twitchen}},\ and\ \bibinfo {author} {\bibfnamefont
  {R.}~\bibnamefont {Hanson}},\ }\href
  {https://doi.org/10.1126/science.1253512} {\bibfield  {journal} {\bibinfo
  {journal} {Science}\ }\textbf {\bibinfo {volume} {345}},\ \bibinfo {pages}
  {532} (\bibinfo {year} {2014})}\BibitemShut {NoStop}%
\bibitem [{\citenamefont {Taylor}\ \emph {et~al.}(2008)\citenamefont {Taylor},
  \citenamefont {Cappellaro}, \citenamefont {Childress}, \citenamefont {Jiang},
  \citenamefont {Budker}, \citenamefont {Hemmer}, \citenamefont {Yacoby},
  \citenamefont {Walsworth},\ and\ \citenamefont {Lukin}}]{Taylor2008}%
  \BibitemOpen
  \bibfield  {author} {\bibinfo {author} {\bibfnamefont {J.~M.}\ \bibnamefont
  {Taylor}}, \bibinfo {author} {\bibfnamefont {P.}~\bibnamefont {Cappellaro}},
  \bibinfo {author} {\bibfnamefont {L.}~\bibnamefont {Childress}}, \bibinfo
  {author} {\bibfnamefont {L.}~\bibnamefont {Jiang}}, \bibinfo {author}
  {\bibfnamefont {D.}~\bibnamefont {Budker}}, \bibinfo {author} {\bibfnamefont
  {P.~R.}\ \bibnamefont {Hemmer}}, \bibinfo {author} {\bibfnamefont
  {A.}~\bibnamefont {Yacoby}}, \bibinfo {author} {\bibfnamefont
  {R.}~\bibnamefont {Walsworth}},\ and\ \bibinfo {author} {\bibfnamefont
  {M.~D.}\ \bibnamefont {Lukin}},\ }\href {https://doi.org/10.1038/nphys1075}
  {\bibfield  {journal} {\bibinfo  {journal} {Nature Physics}\ }\textbf
  {\bibinfo {volume} {4}},\ \bibinfo {pages} {810} (\bibinfo {year}
  {2008})}\BibitemShut {NoStop}%
\bibitem [{\citenamefont {Degen}\ \emph {et~al.}(2017)\citenamefont {Degen},
  \citenamefont {Reinhard},\ and\ \citenamefont {Cappellaro}}]{Degen2017}%
  \BibitemOpen
  \bibfield  {author} {\bibinfo {author} {\bibfnamefont {C.}~\bibnamefont
  {Degen}}, \bibinfo {author} {\bibfnamefont {F.}~\bibnamefont {Reinhard}},\
  and\ \bibinfo {author} {\bibfnamefont {P.}~\bibnamefont {Cappellaro}},\
  }\href {https://doi.org/10.1103/RevModPhys.89.035002} {\bibfield  {journal}
  {\bibinfo  {journal} {Reviews of Modern Physics}\ }\textbf {\bibinfo {volume}
  {89}},\ \bibinfo {pages} {035002} (\bibinfo {year} {2017})}\BibitemShut
  {NoStop}%
\bibitem [{\citenamefont {Rose}\ \emph {et~al.}(2018)\citenamefont {Rose},
  \citenamefont {Huang}, \citenamefont {Zhang}, \citenamefont {Stevenson},
  \citenamefont {Tyryshkin}, \citenamefont {Sangtawesin}, \citenamefont
  {Srinivasan}, \citenamefont {Loudin}, \citenamefont {Markham}, \citenamefont
  {Edmonds}, \citenamefont {Twitchen}, \citenamefont {Lyon},\ and\
  \citenamefont {de~Leon}}]{Rose2018}%
  \BibitemOpen
  \bibfield  {author} {\bibinfo {author} {\bibfnamefont {B.~C.}\ \bibnamefont
  {Rose}}, \bibinfo {author} {\bibfnamefont {D.}~\bibnamefont {Huang}},
  \bibinfo {author} {\bibfnamefont {Z.-H.}\ \bibnamefont {Zhang}}, \bibinfo
  {author} {\bibfnamefont {P.}~\bibnamefont {Stevenson}}, \bibinfo {author}
  {\bibfnamefont {A.~M.}\ \bibnamefont {Tyryshkin}}, \bibinfo {author}
  {\bibfnamefont {S.}~\bibnamefont {Sangtawesin}}, \bibinfo {author}
  {\bibfnamefont {S.}~\bibnamefont {Srinivasan}}, \bibinfo {author}
  {\bibfnamefont {L.}~\bibnamefont {Loudin}}, \bibinfo {author} {\bibfnamefont
  {M.~L.}\ \bibnamefont {Markham}}, \bibinfo {author} {\bibfnamefont {A.~M.}\
  \bibnamefont {Edmonds}}, \bibinfo {author} {\bibfnamefont {D.~J.}\
  \bibnamefont {Twitchen}}, \bibinfo {author} {\bibfnamefont {S.~A.}\
  \bibnamefont {Lyon}},\ and\ \bibinfo {author} {\bibfnamefont {N.~P.}\
  \bibnamefont {de~Leon}},\ }\href {https://doi.org/10.1126/science.aao0290}
  {\bibfield  {journal} {\bibinfo  {journal} {Science}\ }\textbf {\bibinfo
  {volume} {361}},\ \bibinfo {pages} {60} (\bibinfo {year} {2018})}\BibitemShut
  {NoStop}%
\bibitem [{\citenamefont {Gottscholl}\ \emph {et~al.}(2020)\citenamefont
  {Gottscholl}, \citenamefont {Kianinia}, \citenamefont {Soltamov},
  \citenamefont {Orlinskii}, \citenamefont {Mamin}, \citenamefont {Bradac},
  \citenamefont {Kasper}, \citenamefont {Krambrock}, \citenamefont {Sperlich},
  \citenamefont {Toth}, \citenamefont {Aharonovich},\ and\ \citenamefont
  {Dyakonov}}]{Gottscholl2020}%
  \BibitemOpen
  \bibfield  {author} {\bibinfo {author} {\bibfnamefont {A.}~\bibnamefont
  {Gottscholl}}, \bibinfo {author} {\bibfnamefont {M.}~\bibnamefont
  {Kianinia}}, \bibinfo {author} {\bibfnamefont {V.}~\bibnamefont {Soltamov}},
  \bibinfo {author} {\bibfnamefont {S.}~\bibnamefont {Orlinskii}}, \bibinfo
  {author} {\bibfnamefont {G.}~\bibnamefont {Mamin}}, \bibinfo {author}
  {\bibfnamefont {C.}~\bibnamefont {Bradac}}, \bibinfo {author} {\bibfnamefont
  {C.}~\bibnamefont {Kasper}}, \bibinfo {author} {\bibfnamefont
  {K.}~\bibnamefont {Krambrock}}, \bibinfo {author} {\bibfnamefont
  {A.}~\bibnamefont {Sperlich}}, \bibinfo {author} {\bibfnamefont
  {M.}~\bibnamefont {Toth}}, \bibinfo {author} {\bibfnamefont {I.}~\bibnamefont
  {Aharonovich}},\ and\ \bibinfo {author} {\bibfnamefont {V.}~\bibnamefont
  {Dyakonov}},\ }\href {https://doi.org/10.1038/s41563-020-0619-6} {\bibfield
  {journal} {\bibinfo  {journal} {Nature Materials}\ }\textbf {\bibinfo
  {volume} {19}},\ \bibinfo {pages} {540} (\bibinfo {year} {2020})}\BibitemShut
  {NoStop}%
\bibitem [{\citenamefont {Chejanovsky}\ \emph {et~al.}(2021)\citenamefont
  {Chejanovsky}, \citenamefont {Mukherjee}, \citenamefont {Geng}, \citenamefont
  {Chen}, \citenamefont {Kim}, \citenamefont {Denisenko}, \citenamefont
  {Finkler}, \citenamefont {Taniguchi}, \citenamefont {Watanabe}, \citenamefont
  {Dasari}, \citenamefont {Auburger}, \citenamefont {Gali}, \citenamefont
  {Smet},\ and\ \citenamefont {Wrachtrup}}]{Chejanovsky2021}%
  \BibitemOpen
  \bibfield  {author} {\bibinfo {author} {\bibfnamefont {N.}~\bibnamefont
  {Chejanovsky}}, \bibinfo {author} {\bibfnamefont {A.}~\bibnamefont
  {Mukherjee}}, \bibinfo {author} {\bibfnamefont {J.}~\bibnamefont {Geng}},
  \bibinfo {author} {\bibfnamefont {Y.-C.}\ \bibnamefont {Chen}}, \bibinfo
  {author} {\bibfnamefont {Y.}~\bibnamefont {Kim}}, \bibinfo {author}
  {\bibfnamefont {A.}~\bibnamefont {Denisenko}}, \bibinfo {author}
  {\bibfnamefont {A.}~\bibnamefont {Finkler}}, \bibinfo {author} {\bibfnamefont
  {T.}~\bibnamefont {Taniguchi}}, \bibinfo {author} {\bibfnamefont
  {K.}~\bibnamefont {Watanabe}}, \bibinfo {author} {\bibfnamefont {D.~B.~R.}\
  \bibnamefont {Dasari}}, \bibinfo {author} {\bibfnamefont {P.}~\bibnamefont
  {Auburger}}, \bibinfo {author} {\bibfnamefont {A.}~\bibnamefont {Gali}},
  \bibinfo {author} {\bibfnamefont {J.~H.}\ \bibnamefont {Smet}},\ and\
  \bibinfo {author} {\bibfnamefont {J.}~\bibnamefont {Wrachtrup}},\ }\href
  {https://doi.org/10.1038/s41563-021-00979-4} {\bibfield  {journal} {\bibinfo
  {journal} {Nature Materials}\ }\textbf {\bibinfo {volume} {20}},\ \bibinfo
  {pages} {1079} (\bibinfo {year} {2021})}\BibitemShut {NoStop}%
\bibitem [{\citenamefont {Mukherjee}\ \emph {et~al.}(2023)\citenamefont
  {Mukherjee}, \citenamefont {Zhang}, \citenamefont {Oblinsky}, \citenamefont
  {de~Vries}, \citenamefont {Johnson}, \citenamefont {Gibson}, \citenamefont
  {Mayes}, \citenamefont {Edmonds}, \citenamefont {Palmer}, \citenamefont
  {Markham}, \citenamefont {Gali}, \citenamefont {Thiering}, \citenamefont
  {Dalis}, \citenamefont {Dumm}, \citenamefont {Scholes}, \citenamefont
  {Stacey}, \citenamefont {Reineck},\ and\ \citenamefont
  {de~Leon}}]{Mukherjee2023}%
  \BibitemOpen
  \bibfield  {author} {\bibinfo {author} {\bibfnamefont {S.}~\bibnamefont
  {Mukherjee}}, \bibinfo {author} {\bibfnamefont {Z.-H.}\ \bibnamefont
  {Zhang}}, \bibinfo {author} {\bibfnamefont {D.~G.}\ \bibnamefont {Oblinsky}},
  \bibinfo {author} {\bibfnamefont {M.~O.}\ \bibnamefont {de~Vries}}, \bibinfo
  {author} {\bibfnamefont {B.~C.}\ \bibnamefont {Johnson}}, \bibinfo {author}
  {\bibfnamefont {B.~C.}\ \bibnamefont {Gibson}}, \bibinfo {author}
  {\bibfnamefont {E.~L.~H.}\ \bibnamefont {Mayes}}, \bibinfo {author}
  {\bibfnamefont {A.~M.}\ \bibnamefont {Edmonds}}, \bibinfo {author}
  {\bibfnamefont {N.}~\bibnamefont {Palmer}}, \bibinfo {author} {\bibfnamefont
  {M.~L.}\ \bibnamefont {Markham}}, \bibinfo {author} {\bibfnamefont
  {{\'A}.}~\bibnamefont {Gali}}, \bibinfo {author} {\bibfnamefont
  {G.}~\bibnamefont {Thiering}}, \bibinfo {author} {\bibfnamefont
  {A.}~\bibnamefont {Dalis}}, \bibinfo {author} {\bibfnamefont
  {T.}~\bibnamefont {Dumm}}, \bibinfo {author} {\bibfnamefont {G.~D.}\
  \bibnamefont {Scholes}}, \bibinfo {author} {\bibfnamefont {A.}~\bibnamefont
  {Stacey}}, \bibinfo {author} {\bibfnamefont {P.}~\bibnamefont {Reineck}},\
  and\ \bibinfo {author} {\bibfnamefont {N.~P.}\ \bibnamefont {de~Leon}},\
  }\href {https://doi.org/10.1021/acs.nanolett.2c04608} {\bibfield  {journal}
  {\bibinfo  {journal} {Nano Letters}\ }\textbf {\bibinfo {volume} {23}},\
  \bibinfo {pages} {2557} (\bibinfo {year} {2023})}\BibitemShut {NoStop}%
\bibitem [{\citenamefont {Gaita-Ariño}\ \emph {et~al.}(2019)\citenamefont
  {Gaita-Ariño}, \citenamefont {Luis}, \citenamefont {Hill},\ and\
  \citenamefont {Coronado}}]{GaitaArino2019}%
  \BibitemOpen
  \bibfield  {author} {\bibinfo {author} {\bibfnamefont {A.}~\bibnamefont
  {Gaita-Ariño}}, \bibinfo {author} {\bibfnamefont {F.}~\bibnamefont {Luis}},
  \bibinfo {author} {\bibfnamefont {S.}~\bibnamefont {Hill}},\ and\ \bibinfo
  {author} {\bibfnamefont {E.}~\bibnamefont {Coronado}},\ }\href
  {https://doi.org/10.1038/s41557-019-0232-y} {\bibfield  {journal} {\bibinfo
  {journal} {Nature Chemistry}\ }\textbf {\bibinfo {volume} {11}},\ \bibinfo
  {pages} {301} (\bibinfo {year} {2019})}\BibitemShut {NoStop}%
\bibitem [{\citenamefont {Atzori}\ and\ \citenamefont
  {Sessoli}(2019)}]{Atzori2019}%
  \BibitemOpen
  \bibfield  {author} {\bibinfo {author} {\bibfnamefont {M.}~\bibnamefont
  {Atzori}}\ and\ \bibinfo {author} {\bibfnamefont {R.}~\bibnamefont
  {Sessoli}},\ }\href {https://doi.org/10.1021/jacs.9b00984} {\bibfield
  {journal} {\bibinfo  {journal} {Journal of the American Chemical Society}\
  }\textbf {\bibinfo {volume} {141}},\ \bibinfo {pages} {11339} (\bibinfo
  {year} {2019})}\BibitemShut {NoStop}%
\bibitem [{\citenamefont {Wasielewski}\ \emph {et~al.}(2020)\citenamefont
  {Wasielewski}, \citenamefont {Forbes}, \citenamefont {Frank}, \citenamefont
  {Kowalski}, \citenamefont {Scholes}, \citenamefont {Yuen-Zhou}, \citenamefont
  {Baldo}, \citenamefont {Freedman}, \citenamefont {Goldsmith}, \citenamefont
  {Goodson}, \citenamefont {Kirk}, \citenamefont {McCusker}, \citenamefont
  {Ogilvie}, \citenamefont {Shultz}, \citenamefont {Stoll},\ and\ \citenamefont
  {Whaley}}]{Wasielewski2020}%
  \BibitemOpen
  \bibfield  {author} {\bibinfo {author} {\bibfnamefont {M.~R.}\ \bibnamefont
  {Wasielewski}}, \bibinfo {author} {\bibfnamefont {M.~D.~E.}\ \bibnamefont
  {Forbes}}, \bibinfo {author} {\bibfnamefont {N.~L.}\ \bibnamefont {Frank}},
  \bibinfo {author} {\bibfnamefont {K.}~\bibnamefont {Kowalski}}, \bibinfo
  {author} {\bibfnamefont {G.~D.}\ \bibnamefont {Scholes}}, \bibinfo {author}
  {\bibfnamefont {J.}~\bibnamefont {Yuen-Zhou}}, \bibinfo {author}
  {\bibfnamefont {M.~A.}\ \bibnamefont {Baldo}}, \bibinfo {author}
  {\bibfnamefont {D.~E.}\ \bibnamefont {Freedman}}, \bibinfo {author}
  {\bibfnamefont {R.~H.}\ \bibnamefont {Goldsmith}}, \bibinfo {author}
  {\bibfnamefont {T.}~\bibnamefont {Goodson}}, \bibinfo {author} {\bibfnamefont
  {M.~L.}\ \bibnamefont {Kirk}}, \bibinfo {author} {\bibfnamefont {J.~K.}\
  \bibnamefont {McCusker}}, \bibinfo {author} {\bibfnamefont {J.~P.}\
  \bibnamefont {Ogilvie}}, \bibinfo {author} {\bibfnamefont {D.~A.}\
  \bibnamefont {Shultz}}, \bibinfo {author} {\bibfnamefont {S.}~\bibnamefont
  {Stoll}},\ and\ \bibinfo {author} {\bibfnamefont {K.~B.}\ \bibnamefont
  {Whaley}},\ }\href {https://doi.org/10.1038/s41570-020-0200-5} {\bibfield
  {journal} {\bibinfo  {journal} {Nature Reviews Chemistry}\ }\textbf {\bibinfo
  {volume} {4}},\ \bibinfo {pages} {490} (\bibinfo {year} {2020})}\BibitemShut
  {NoStop}%
\bibitem [{\citenamefont {Yu}\ \emph {et~al.}(2021)\citenamefont {Yu},
  \citenamefont {von Kugelgen}, \citenamefont {Laorenza},\ and\ \citenamefont
  {Freedman}}]{Yu2021}%
  \BibitemOpen
  \bibfield  {author} {\bibinfo {author} {\bibfnamefont {C.-J.}\ \bibnamefont
  {Yu}}, \bibinfo {author} {\bibfnamefont {S.}~\bibnamefont {von Kugelgen}},
  \bibinfo {author} {\bibfnamefont {D.~W.}\ \bibnamefont {Laorenza}},\ and\
  \bibinfo {author} {\bibfnamefont {D.~E.}\ \bibnamefont {Freedman}},\ }\href
  {https://doi.org/10.1021/acscentsci.0c00737} {\bibfield  {journal} {\bibinfo
  {journal} {ACS Central Science}\ }\textbf {\bibinfo {volume} {7}},\ \bibinfo
  {pages} {712} (\bibinfo {year} {2021})}\BibitemShut {NoStop}%
\bibitem [{\citenamefont {Laorenza}\ and\ \citenamefont
  {Freedman}(2022)}]{Laorenza2022}%
  \BibitemOpen
  \bibfield  {author} {\bibinfo {author} {\bibfnamefont {D.~W.}\ \bibnamefont
  {Laorenza}}\ and\ \bibinfo {author} {\bibfnamefont {D.~E.}\ \bibnamefont
  {Freedman}},\ }\href {https://doi.org/10.1021/jacs.2c07775} {\bibfield
  {journal} {\bibinfo  {journal} {Journal of the American Chemical Society}\
  }\textbf {\bibinfo {volume} {144}},\ \bibinfo {pages} {21810} (\bibinfo
  {year} {2022})}\BibitemShut {NoStop}%
\bibitem [{\citenamefont {Scholes}(2023)}]{Scholes2023}%
  \BibitemOpen
  \bibfield  {author} {\bibinfo {author} {\bibfnamefont {G.~D.}\ \bibnamefont
  {Scholes}},\ }\href {https://doi.org/10.1098/rspa.2023.0599} {\bibfield
  {journal} {\bibinfo  {journal} {Proceedings of the Royal Society A:
  Mathematical, Physical and Engineering Sciences}\ }\textbf {\bibinfo {volume}
  {479}},\ \bibinfo {pages} {20230599} (\bibinfo {year} {2023})}\BibitemShut
  {NoStop}%
\bibitem [{\citenamefont {Wu}\ and\ \citenamefont {Scholes}(2023)}]{Wu2023}%
  \BibitemOpen
  \bibfield  {author} {\bibinfo {author} {\bibfnamefont {W.}~\bibnamefont
  {Wu}}\ and\ \bibinfo {author} {\bibfnamefont {G.~D.}\ \bibnamefont
  {Scholes}},\ }\href@noop {} {\bibinfo {title} {Foundations of quantum
  information for physical chemistry}} (\bibinfo {year} {2023}),\ \Eprint
  {https://arxiv.org/abs/2311.12238} {arXiv:2311.12238 [quant-ph]} \BibitemShut
  {NoStop}%
\bibitem [{\citenamefont {Wojnar}\ \emph {et~al.}(2020)\citenamefont {Wojnar},
  \citenamefont {Laorenza}, \citenamefont {Schaller},\ and\ \citenamefont
  {Freedman}}]{Wojnar2020}%
  \BibitemOpen
  \bibfield  {author} {\bibinfo {author} {\bibfnamefont {M.~K.}\ \bibnamefont
  {Wojnar}}, \bibinfo {author} {\bibfnamefont {D.~W.}\ \bibnamefont
  {Laorenza}}, \bibinfo {author} {\bibfnamefont {R.~D.}\ \bibnamefont
  {Schaller}},\ and\ \bibinfo {author} {\bibfnamefont {D.~E.}\ \bibnamefont
  {Freedman}},\ }\href {https://doi.org/10.1021/jacs.0c06909} {\bibfield
  {journal} {\bibinfo  {journal} {Journal of the American Chemical Society}\
  }\textbf {\bibinfo {volume} {142}},\ \bibinfo {pages} {14826} (\bibinfo
  {year} {2020})}\BibitemShut {NoStop}%
\bibitem [{\citenamefont {Bayliss}\ \emph {et~al.}(2020)\citenamefont
  {Bayliss}, \citenamefont {Laorenza}, \citenamefont {Mintun}, \citenamefont
  {Kovos}, \citenamefont {Freedman},\ and\ \citenamefont
  {Awschalom}}]{Bayliss2020}%
  \BibitemOpen
  \bibfield  {author} {\bibinfo {author} {\bibfnamefont {S.~L.}\ \bibnamefont
  {Bayliss}}, \bibinfo {author} {\bibfnamefont {D.~W.}\ \bibnamefont
  {Laorenza}}, \bibinfo {author} {\bibfnamefont {P.~J.}\ \bibnamefont
  {Mintun}}, \bibinfo {author} {\bibfnamefont {B.~D.}\ \bibnamefont {Kovos}},
  \bibinfo {author} {\bibfnamefont {D.~E.}\ \bibnamefont {Freedman}},\ and\
  \bibinfo {author} {\bibfnamefont {D.~D.}\ \bibnamefont {Awschalom}},\ }\href
  {https://doi.org/10.1126/science.abb9352} {\bibfield  {journal} {\bibinfo
  {journal} {Science}\ }\textbf {\bibinfo {volume} {370}},\ \bibinfo {pages}
  {1309} (\bibinfo {year} {2020})}\BibitemShut {NoStop}%
\bibitem [{\citenamefont {Fataftah}\ \emph {et~al.}(2020)\citenamefont
  {Fataftah}, \citenamefont {Bayliss}, \citenamefont {Laorenza}, \citenamefont
  {Wang}, \citenamefont {Phelan}, \citenamefont {Wilson}, \citenamefont
  {Mintun}, \citenamefont {Kovos}, \citenamefont {Wasielewski}, \citenamefont
  {Han}, \citenamefont {Sherwin}, \citenamefont {Awschalom},\ and\
  \citenamefont {Freedman}}]{Fataftah2020}%
  \BibitemOpen
  \bibfield  {author} {\bibinfo {author} {\bibfnamefont {M.~S.}\ \bibnamefont
  {Fataftah}}, \bibinfo {author} {\bibfnamefont {S.~L.}\ \bibnamefont
  {Bayliss}}, \bibinfo {author} {\bibfnamefont {D.~W.}\ \bibnamefont
  {Laorenza}}, \bibinfo {author} {\bibfnamefont {X.}~\bibnamefont {Wang}},
  \bibinfo {author} {\bibfnamefont {B.~T.}\ \bibnamefont {Phelan}}, \bibinfo
  {author} {\bibfnamefont {C.~B.}\ \bibnamefont {Wilson}}, \bibinfo {author}
  {\bibfnamefont {P.~J.}\ \bibnamefont {Mintun}}, \bibinfo {author}
  {\bibfnamefont {B.~D.}\ \bibnamefont {Kovos}}, \bibinfo {author}
  {\bibfnamefont {M.~R.}\ \bibnamefont {Wasielewski}}, \bibinfo {author}
  {\bibfnamefont {S.}~\bibnamefont {Han}}, \bibinfo {author} {\bibfnamefont
  {M.~S.}\ \bibnamefont {Sherwin}}, \bibinfo {author} {\bibfnamefont {D.~D.}\
  \bibnamefont {Awschalom}},\ and\ \bibinfo {author} {\bibfnamefont {D.~E.}\
  \bibnamefont {Freedman}},\ }\href {https://doi.org/10.1021/jacs.0c08986}
  {\bibfield  {journal} {\bibinfo  {journal} {Journal of the American Chemical
  Society}\ }\textbf {\bibinfo {volume} {142}},\ \bibinfo {pages} {20400}
  (\bibinfo {year} {2020})}\BibitemShut {NoStop}%
\bibitem [{\citenamefont {Mirzoyan}\ \emph {et~al.}(2021)\citenamefont
  {Mirzoyan}, \citenamefont {Kazmierczak},\ and\ \citenamefont
  {Hadt}}]{Mirzoyan2021}%
  \BibitemOpen
  \bibfield  {author} {\bibinfo {author} {\bibfnamefont {R.}~\bibnamefont
  {Mirzoyan}}, \bibinfo {author} {\bibfnamefont {N.~P.}\ \bibnamefont
  {Kazmierczak}},\ and\ \bibinfo {author} {\bibfnamefont {R.~G.}\ \bibnamefont
  {Hadt}},\ }\href {https://doi.org/10.1002/chem.202100845} {\bibfield
  {journal} {\bibinfo  {journal} {Chemistry – A European Journal}\ }\textbf
  {\bibinfo {volume} {27}},\ \bibinfo {pages} {9482} (\bibinfo {year}
  {2021})}\BibitemShut {NoStop}%
\bibitem [{\citenamefont {Kazmierczak}\ \emph {et~al.}(2021)\citenamefont
  {Kazmierczak}, \citenamefont {Mirzoyan},\ and\ \citenamefont
  {Hadt}}]{Kazmierczak2021}%
  \BibitemOpen
  \bibfield  {author} {\bibinfo {author} {\bibfnamefont {N.~P.}\ \bibnamefont
  {Kazmierczak}}, \bibinfo {author} {\bibfnamefont {R.}~\bibnamefont
  {Mirzoyan}},\ and\ \bibinfo {author} {\bibfnamefont {R.~G.}\ \bibnamefont
  {Hadt}},\ }\href {https://doi.org/10.1021/jacs.1c04605} {\bibfield  {journal}
  {\bibinfo  {journal} {Journal of the American Chemical Society}\ }\textbf
  {\bibinfo {volume} {143}},\ \bibinfo {pages} {17305} (\bibinfo {year}
  {2021})}\BibitemShut {NoStop}%
\bibitem [{\citenamefont {Laorenza}\ \emph {et~al.}(2021)\citenamefont
  {Laorenza}, \citenamefont {Kairalapova}, \citenamefont {Bayliss},
  \citenamefont {Goldzak}, \citenamefont {Greene}, \citenamefont {Weiss},
  \citenamefont {Deb}, \citenamefont {Mintun}, \citenamefont {Collins},
  \citenamefont {Awschalom}, \citenamefont {Berkelbach},\ and\ \citenamefont
  {Freedman}}]{Laorenza2021}%
  \BibitemOpen
  \bibfield  {author} {\bibinfo {author} {\bibfnamefont {D.~W.}\ \bibnamefont
  {Laorenza}}, \bibinfo {author} {\bibfnamefont {A.}~\bibnamefont
  {Kairalapova}}, \bibinfo {author} {\bibfnamefont {S.~L.}\ \bibnamefont
  {Bayliss}}, \bibinfo {author} {\bibfnamefont {T.}~\bibnamefont {Goldzak}},
  \bibinfo {author} {\bibfnamefont {S.~M.}\ \bibnamefont {Greene}}, \bibinfo
  {author} {\bibfnamefont {L.~R.}\ \bibnamefont {Weiss}}, \bibinfo {author}
  {\bibfnamefont {P.}~\bibnamefont {Deb}}, \bibinfo {author} {\bibfnamefont
  {P.~J.}\ \bibnamefont {Mintun}}, \bibinfo {author} {\bibfnamefont {K.~A.}\
  \bibnamefont {Collins}}, \bibinfo {author} {\bibfnamefont {D.~D.}\
  \bibnamefont {Awschalom}}, \bibinfo {author} {\bibfnamefont {T.~C.}\
  \bibnamefont {Berkelbach}},\ and\ \bibinfo {author} {\bibfnamefont {D.~E.}\
  \bibnamefont {Freedman}},\ }\href {https://doi.org/10.1021/jacs.1c10145}
  {\bibfield  {journal} {\bibinfo  {journal} {Journal of the American Chemical
  Society}\ }\textbf {\bibinfo {volume} {143}},\ \bibinfo {pages} {21350}
  (\bibinfo {year} {2021})}\BibitemShut {NoStop}%
\bibitem [{\citenamefont {Amdur}\ \emph {et~al.}(2022)\citenamefont {Amdur},
  \citenamefont {Mullin}, \citenamefont {Waters}, \citenamefont {Puggioni},
  \citenamefont {Wojnar}, \citenamefont {Gu}, \citenamefont {Sun},
  \citenamefont {Oyala}, \citenamefont {Rondinelli},\ and\ \citenamefont
  {Freedman}}]{Amdur2022}%
  \BibitemOpen
  \bibfield  {author} {\bibinfo {author} {\bibfnamefont {M.~J.}\ \bibnamefont
  {Amdur}}, \bibinfo {author} {\bibfnamefont {K.~R.}\ \bibnamefont {Mullin}},
  \bibinfo {author} {\bibfnamefont {M.~J.}\ \bibnamefont {Waters}}, \bibinfo
  {author} {\bibfnamefont {D.}~\bibnamefont {Puggioni}}, \bibinfo {author}
  {\bibfnamefont {M.~K.}\ \bibnamefont {Wojnar}}, \bibinfo {author}
  {\bibfnamefont {M.}~\bibnamefont {Gu}}, \bibinfo {author} {\bibfnamefont
  {L.}~\bibnamefont {Sun}}, \bibinfo {author} {\bibfnamefont {P.~H.}\
  \bibnamefont {Oyala}}, \bibinfo {author} {\bibfnamefont {J.~M.}\ \bibnamefont
  {Rondinelli}},\ and\ \bibinfo {author} {\bibfnamefont {D.~E.}\ \bibnamefont
  {Freedman}},\ }\href {https://doi.org/10.1039/D1SC06130E} {\bibfield
  {journal} {\bibinfo  {journal} {Chemical Science}\ }\textbf {\bibinfo
  {volume} {13}},\ \bibinfo {pages} {7034} (\bibinfo {year}
  {2022})}\BibitemShut {NoStop}%
\bibitem [{\citenamefont {Bayliss}\ \emph {et~al.}(2022)\citenamefont
  {Bayliss}, \citenamefont {Deb}, \citenamefont {Laorenza}, \citenamefont
  {Onizhuk}, \citenamefont {Galli}, \citenamefont {Freedman},\ and\
  \citenamefont {Awschalom}}]{Bayliss2022}%
  \BibitemOpen
  \bibfield  {author} {\bibinfo {author} {\bibfnamefont {S.}~\bibnamefont
  {Bayliss}}, \bibinfo {author} {\bibfnamefont {P.}~\bibnamefont {Deb}},
  \bibinfo {author} {\bibfnamefont {D.}~\bibnamefont {Laorenza}}, \bibinfo
  {author} {\bibfnamefont {M.}~\bibnamefont {Onizhuk}}, \bibinfo {author}
  {\bibfnamefont {G.}~\bibnamefont {Galli}}, \bibinfo {author} {\bibfnamefont
  {D.}~\bibnamefont {Freedman}},\ and\ \bibinfo {author} {\bibfnamefont
  {D.}~\bibnamefont {Awschalom}},\ }\href
  {https://doi.org/10.1103/PhysRevX.12.031028} {\bibfield  {journal} {\bibinfo
  {journal} {Physical Review X}\ }\textbf {\bibinfo {volume} {12}},\ \bibinfo
  {pages} {031028} (\bibinfo {year} {2022})}\BibitemShut {NoStop}%
\bibitem [{\citenamefont {Goh}\ \emph {et~al.}(2022)\citenamefont {Goh},
  \citenamefont {Pandharkar},\ and\ \citenamefont {Gagliardi}}]{Goh2022}%
  \BibitemOpen
  \bibfield  {author} {\bibinfo {author} {\bibfnamefont {T.}~\bibnamefont
  {Goh}}, \bibinfo {author} {\bibfnamefont {R.}~\bibnamefont {Pandharkar}},\
  and\ \bibinfo {author} {\bibfnamefont {L.}~\bibnamefont {Gagliardi}},\ }\href
  {https://doi.org/10.1021/acs.jpca.2c04730} {\bibfield  {journal} {\bibinfo
  {journal} {The Journal of Physical Chemistry A}\ }\textbf {\bibinfo {volume}
  {126}},\ \bibinfo {pages} {6329} (\bibinfo {year} {2022})}\BibitemShut
  {NoStop}%
\bibitem [{\citenamefont {Kazmierczak}\ and\ \citenamefont
  {Hadt}(2022)}]{Kazmierczak2022}%
  \BibitemOpen
  \bibfield  {author} {\bibinfo {author} {\bibfnamefont {N.~P.}\ \bibnamefont
  {Kazmierczak}}\ and\ \bibinfo {author} {\bibfnamefont {R.~G.}\ \bibnamefont
  {Hadt}},\ }\href {https://doi.org/10.1021/jacs.2c08729} {\bibfield  {journal}
  {\bibinfo  {journal} {Journal of the American Chemical Society}\ }\textbf
  {\bibinfo {volume} {144}},\ \bibinfo {pages} {20804} (\bibinfo {year}
  {2022})}\BibitemShut {NoStop}%
\bibitem [{\citenamefont {Kazmierczak}\ \emph {et~al.}(2023)\citenamefont
  {Kazmierczak}, \citenamefont {Luedecke}, \citenamefont {Gallmeier},\ and\
  \citenamefont {Hadt}}]{Kazmierczak2023}%
  \BibitemOpen
  \bibfield  {author} {\bibinfo {author} {\bibfnamefont {N.~P.}\ \bibnamefont
  {Kazmierczak}}, \bibinfo {author} {\bibfnamefont {K.~M.}\ \bibnamefont
  {Luedecke}}, \bibinfo {author} {\bibfnamefont {E.~T.}\ \bibnamefont
  {Gallmeier}},\ and\ \bibinfo {author} {\bibfnamefont {R.~G.}\ \bibnamefont
  {Hadt}},\ }\href {https://doi.org/10.1021/acs.jpclett.3c01964} {\bibfield
  {journal} {\bibinfo  {journal} {The Journal of Physical Chemistry Letters}\
  }\textbf {\bibinfo {volume} {14}},\ \bibinfo {pages} {7658} (\bibinfo {year}
  {2023})}\BibitemShut {NoStop}%
\bibitem [{\citenamefont {Mullin}\ \emph {et~al.}(2023)\citenamefont {Mullin},
  \citenamefont {Laorenza}, \citenamefont {Freedman},\ and\ \citenamefont
  {Rondinelli}}]{Mullin2023}%
  \BibitemOpen
  \bibfield  {author} {\bibinfo {author} {\bibfnamefont {K.~R.}\ \bibnamefont
  {Mullin}}, \bibinfo {author} {\bibfnamefont {D.~W.}\ \bibnamefont
  {Laorenza}}, \bibinfo {author} {\bibfnamefont {D.~E.}\ \bibnamefont
  {Freedman}},\ and\ \bibinfo {author} {\bibfnamefont {J.~M.}\ \bibnamefont
  {Rondinelli}},\ }\href {https://doi.org/10.1103/PhysRevResearch.5.L042023}
  {\bibfield  {journal} {\bibinfo  {journal} {Physical Review Research}\
  }\textbf {\bibinfo {volume} {5}},\ \bibinfo {pages} {L042023} (\bibinfo
  {year} {2023})}\BibitemShut {NoStop}%
\bibitem [{\citenamefont {Smyser}\ and\ \citenamefont
  {Eaves}(2020)}]{Smyser2020}%
  \BibitemOpen
  \bibfield  {author} {\bibinfo {author} {\bibfnamefont {K.~E.}\ \bibnamefont
  {Smyser}}\ and\ \bibinfo {author} {\bibfnamefont {J.~D.}\ \bibnamefont
  {Eaves}},\ }\href {https://doi.org/10.1038/s41598-020-75459-x} {\bibfield
  {journal} {\bibinfo  {journal} {Scientific Reports}\ }\textbf {\bibinfo
  {volume} {10}},\ \bibinfo {pages} {18480} (\bibinfo {year}
  {2020})}\BibitemShut {NoStop}%
\bibitem [{\citenamefont {Dill}\ \emph {et~al.}(2023)\citenamefont {Dill},
  \citenamefont {Smyser}, \citenamefont {Rugg}, \citenamefont {Damrauer},\ and\
  \citenamefont {Eaves}}]{Dill2023}%
  \BibitemOpen
  \bibfield  {author} {\bibinfo {author} {\bibfnamefont {R.~D.}\ \bibnamefont
  {Dill}}, \bibinfo {author} {\bibfnamefont {K.~E.}\ \bibnamefont {Smyser}},
  \bibinfo {author} {\bibfnamefont {B.~K.}\ \bibnamefont {Rugg}}, \bibinfo
  {author} {\bibfnamefont {N.~H.}\ \bibnamefont {Damrauer}},\ and\ \bibinfo
  {author} {\bibfnamefont {J.~D.}\ \bibnamefont {Eaves}},\ }\href
  {https://doi.org/10.1038/s41467-023-36529-6} {\bibfield  {journal} {\bibinfo
  {journal} {Nature Communications}\ }\textbf {\bibinfo {volume} {14}},\
  \bibinfo {pages} {1180} (\bibinfo {year} {2023})}\BibitemShut {NoStop}%
\bibitem [{\citenamefont {Gorgon}\ \emph {et~al.}(2023)\citenamefont {Gorgon},
  \citenamefont {Lv}, \citenamefont {Grüne}, \citenamefont {Drummond},
  \citenamefont {Myers}, \citenamefont {Londi}, \citenamefont {Ricci},
  \citenamefont {Valverde}, \citenamefont {Tonnelé}, \citenamefont {Murto},
  \citenamefont {Romanov}, \citenamefont {Casanova}, \citenamefont {Dyakonov},
  \citenamefont {Sperlich}, \citenamefont {Beljonne}, \citenamefont {Olivier},
  \citenamefont {Li}, \citenamefont {Friend},\ and\ \citenamefont
  {Evans}}]{Gorgon2023}%
  \BibitemOpen
  \bibfield  {author} {\bibinfo {author} {\bibfnamefont {S.}~\bibnamefont
  {Gorgon}}, \bibinfo {author} {\bibfnamefont {K.}~\bibnamefont {Lv}}, \bibinfo
  {author} {\bibfnamefont {J.}~\bibnamefont {Grüne}}, \bibinfo {author}
  {\bibfnamefont {B.~H.}\ \bibnamefont {Drummond}}, \bibinfo {author}
  {\bibfnamefont {W.~K.}\ \bibnamefont {Myers}}, \bibinfo {author}
  {\bibfnamefont {G.}~\bibnamefont {Londi}}, \bibinfo {author} {\bibfnamefont
  {G.}~\bibnamefont {Ricci}}, \bibinfo {author} {\bibfnamefont
  {D.}~\bibnamefont {Valverde}}, \bibinfo {author} {\bibfnamefont
  {C.}~\bibnamefont {Tonnelé}}, \bibinfo {author} {\bibfnamefont
  {P.}~\bibnamefont {Murto}}, \bibinfo {author} {\bibfnamefont {A.~S.}\
  \bibnamefont {Romanov}}, \bibinfo {author} {\bibfnamefont {D.}~\bibnamefont
  {Casanova}}, \bibinfo {author} {\bibfnamefont {V.}~\bibnamefont {Dyakonov}},
  \bibinfo {author} {\bibfnamefont {A.}~\bibnamefont {Sperlich}}, \bibinfo
  {author} {\bibfnamefont {D.}~\bibnamefont {Beljonne}}, \bibinfo {author}
  {\bibfnamefont {Y.}~\bibnamefont {Olivier}}, \bibinfo {author} {\bibfnamefont
  {F.}~\bibnamefont {Li}}, \bibinfo {author} {\bibfnamefont {R.~H.}\
  \bibnamefont {Friend}},\ and\ \bibinfo {author} {\bibfnamefont {E.~W.}\
  \bibnamefont {Evans}},\ }\href {https://doi.org/10.1038/s41586-023-06222-1}
  {\bibfield  {journal} {\bibinfo  {journal} {Nature}\ }\textbf {\bibinfo
  {volume} {620}},\ \bibinfo {pages} {538} (\bibinfo {year}
  {2023})}\BibitemShut {NoStop}%
\bibitem [{\citenamefont {Palmer}\ \emph {et~al.}(2024)\citenamefont {Palmer},
  \citenamefont {Williams}, \citenamefont {Young}, \citenamefont {Peinkofer},
  \citenamefont {Phelan}, \citenamefont {Krzyaniak},\ and\ \citenamefont
  {Wasielewski}}]{Palmer2024}%
  \BibitemOpen
  \bibfield  {author} {\bibinfo {author} {\bibfnamefont {J.~R.}\ \bibnamefont
  {Palmer}}, \bibinfo {author} {\bibfnamefont {M.~L.}\ \bibnamefont
  {Williams}}, \bibinfo {author} {\bibfnamefont {R.~M.}\ \bibnamefont {Young}},
  \bibinfo {author} {\bibfnamefont {K.~R.}\ \bibnamefont {Peinkofer}}, \bibinfo
  {author} {\bibfnamefont {B.~T.}\ \bibnamefont {Phelan}}, \bibinfo {author}
  {\bibfnamefont {M.~D.}\ \bibnamefont {Krzyaniak}},\ and\ \bibinfo {author}
  {\bibfnamefont {M.~R.}\ \bibnamefont {Wasielewski}},\ }\href
  {https://doi.org/10.1021/jacs.3c12277} {\bibfield  {journal} {\bibinfo
  {journal} {Journal of the American Chemical Society}\ }\textbf {\bibinfo
  {volume} {146}},\ \bibinfo {pages} {1089} (\bibinfo {year}
  {2024})}\BibitemShut {NoStop}%
\bibitem [{\citenamefont {Mena}\ \emph {et~al.}(2024)\citenamefont {Mena},
  \citenamefont {Mann}, \citenamefont {Cowley-Semple}, \citenamefont {Bryan},
  \citenamefont {Heutz}, \citenamefont {McCamey}, \citenamefont {Attwood},\
  and\ \citenamefont {Bayliss}}]{Mena2024}%
  \BibitemOpen
  \bibfield  {author} {\bibinfo {author} {\bibfnamefont {A.}~\bibnamefont
  {Mena}}, \bibinfo {author} {\bibfnamefont {S.~K.}\ \bibnamefont {Mann}},
  \bibinfo {author} {\bibfnamefont {A.}~\bibnamefont {Cowley-Semple}}, \bibinfo
  {author} {\bibfnamefont {E.}~\bibnamefont {Bryan}}, \bibinfo {author}
  {\bibfnamefont {S.}~\bibnamefont {Heutz}}, \bibinfo {author} {\bibfnamefont
  {D.~R.}\ \bibnamefont {McCamey}}, \bibinfo {author} {\bibfnamefont
  {M.}~\bibnamefont {Attwood}},\ and\ \bibinfo {author} {\bibfnamefont {S.~L.}\
  \bibnamefont {Bayliss}},\ }\href@noop {} {\bibinfo {title} {Room-temperature
  optically detected coherent control of molecular spins}} (\bibinfo {year}
  {2024}),\ \Eprint {https://arxiv.org/abs/2402.07572} {arXiv:2402.07572
  [quant-ph]} \BibitemShut {NoStop}%
\bibitem [{\citenamefont {Singh}\ \emph {et~al.}(2024)\citenamefont {Singh},
  \citenamefont {D'Souza}, \citenamefont {Zhong}, \citenamefont {Druga},
  \citenamefont {Oshiro}, \citenamefont {Blankenship}, \citenamefont {Reimer},
  \citenamefont {Breeze},\ and\ \citenamefont {Ajoy}}]{Singh2024}%
  \BibitemOpen
  \bibfield  {author} {\bibinfo {author} {\bibfnamefont {H.}~\bibnamefont
  {Singh}}, \bibinfo {author} {\bibfnamefont {N.}~\bibnamefont {D'Souza}},
  \bibinfo {author} {\bibfnamefont {K.}~\bibnamefont {Zhong}}, \bibinfo
  {author} {\bibfnamefont {E.}~\bibnamefont {Druga}}, \bibinfo {author}
  {\bibfnamefont {J.}~\bibnamefont {Oshiro}}, \bibinfo {author} {\bibfnamefont
  {B.}~\bibnamefont {Blankenship}}, \bibinfo {author} {\bibfnamefont {J.~A.}\
  \bibnamefont {Reimer}}, \bibinfo {author} {\bibfnamefont {J.~D.}\
  \bibnamefont {Breeze}},\ and\ \bibinfo {author} {\bibfnamefont
  {A.}~\bibnamefont {Ajoy}},\ }\href@noop {} {\bibinfo {title}
  {Room-temperature quantum sensing with photoexcited triplet electrons in
  organic crystals}} (\bibinfo {year} {2024}),\ \Eprint
  {https://arxiv.org/abs/2402.13898} {arXiv:2402.13898 [quant-ph]} \BibitemShut
  {NoStop}%
\bibitem [{\citenamefont {Doherty}\ \emph {et~al.}(2013)\citenamefont
  {Doherty}, \citenamefont {Manson}, \citenamefont {Delaney}, \citenamefont
  {Jelezko}, \citenamefont {Wrachtrup},\ and\ \citenamefont
  {Hollenberg}}]{Doherty2013}%
  \BibitemOpen
  \bibfield  {author} {\bibinfo {author} {\bibfnamefont {M.~W.}\ \bibnamefont
  {Doherty}}, \bibinfo {author} {\bibfnamefont {N.~B.}\ \bibnamefont {Manson}},
  \bibinfo {author} {\bibfnamefont {P.}~\bibnamefont {Delaney}}, \bibinfo
  {author} {\bibfnamefont {F.}~\bibnamefont {Jelezko}}, \bibinfo {author}
  {\bibfnamefont {J.}~\bibnamefont {Wrachtrup}},\ and\ \bibinfo {author}
  {\bibfnamefont {L.~C.~L.}\ \bibnamefont {Hollenberg}},\ }\href
  {https://doi.org/10.1016/j.physrep.2013.02.001} {\bibfield  {journal}
  {\bibinfo  {journal} {Physics Reports}\ }\bibinfo {series} {The
  nitrogen-vacancy colour centre in diamond},\ \textbf {\bibinfo {volume}
  {528}},\ \bibinfo {pages} {1} (\bibinfo {year} {2013})}\BibitemShut {NoStop}%
\bibitem [{\citenamefont {Hattori}\ \emph {et~al.}(2019)\citenamefont
  {Hattori}, \citenamefont {Michail}, \citenamefont {Schmiedel}, \citenamefont
  {Moos}, \citenamefont {Holzapfel}, \citenamefont {Krummenacher},
  \citenamefont {Braunschweig}, \citenamefont {Müller}, \citenamefont
  {Pflaum},\ and\ \citenamefont {Lambert}}]{Hattori2019}%
  \BibitemOpen
  \bibfield  {author} {\bibinfo {author} {\bibfnamefont {Y.}~\bibnamefont
  {Hattori}}, \bibinfo {author} {\bibfnamefont {E.}~\bibnamefont {Michail}},
  \bibinfo {author} {\bibfnamefont {A.}~\bibnamefont {Schmiedel}}, \bibinfo
  {author} {\bibfnamefont {M.}~\bibnamefont {Moos}}, \bibinfo {author}
  {\bibfnamefont {M.}~\bibnamefont {Holzapfel}}, \bibinfo {author}
  {\bibfnamefont {I.}~\bibnamefont {Krummenacher}}, \bibinfo {author}
  {\bibfnamefont {H.}~\bibnamefont {Braunschweig}}, \bibinfo {author}
  {\bibfnamefont {U.}~\bibnamefont {Müller}}, \bibinfo {author} {\bibfnamefont
  {J.}~\bibnamefont {Pflaum}},\ and\ \bibinfo {author} {\bibfnamefont
  {C.}~\bibnamefont {Lambert}},\ }\href
  {https://doi.org/10.1002/chem.201903007} {\bibfield  {journal} {\bibinfo
  {journal} {Chemistry – A European Journal}\ }\textbf {\bibinfo {volume}
  {25}},\ \bibinfo {pages} {15463} (\bibinfo {year} {2019})}\BibitemShut
  {NoStop}%
\bibitem [{\citenamefont {Kimura}\ \emph {et~al.}(2021)\citenamefont {Kimura},
  \citenamefont {Uejima}, \citenamefont {Ota}, \citenamefont {Sato},
  \citenamefont {Kusaka}, \citenamefont {Matsuda}, \citenamefont {Nishihara},\
  and\ \citenamefont {Kusamoto}}]{Kimura2021}%
  \BibitemOpen
  \bibfield  {author} {\bibinfo {author} {\bibfnamefont {S.}~\bibnamefont
  {Kimura}}, \bibinfo {author} {\bibfnamefont {M.}~\bibnamefont {Uejima}},
  \bibinfo {author} {\bibfnamefont {W.}~\bibnamefont {Ota}}, \bibinfo {author}
  {\bibfnamefont {T.}~\bibnamefont {Sato}}, \bibinfo {author} {\bibfnamefont
  {S.}~\bibnamefont {Kusaka}}, \bibinfo {author} {\bibfnamefont
  {R.}~\bibnamefont {Matsuda}}, \bibinfo {author} {\bibfnamefont
  {H.}~\bibnamefont {Nishihara}},\ and\ \bibinfo {author} {\bibfnamefont
  {T.}~\bibnamefont {Kusamoto}},\ }\href {https://doi.org/10.1021/jacs.0c13310}
  {\bibfield  {journal} {\bibinfo  {journal} {Journal of the American Chemical
  Society}\ }\textbf {\bibinfo {volume} {143}},\ \bibinfo {pages} {4329}
  (\bibinfo {year} {2021})}\BibitemShut {NoStop}%
\bibitem [{\citenamefont {Wonink}\ \emph {et~al.}(2021)\citenamefont {Wonink},
  \citenamefont {Corbet}, \citenamefont {Kulago}, \citenamefont {Boursalian},
  \citenamefont {de~Bruin}, \citenamefont {Otten}, \citenamefont {Browne},\
  and\ \citenamefont {Feringa}}]{Wonink2021}%
  \BibitemOpen
  \bibfield  {author} {\bibinfo {author} {\bibfnamefont {M.~B.~S.}\
  \bibnamefont {Wonink}}, \bibinfo {author} {\bibfnamefont {B.~P.}\
  \bibnamefont {Corbet}}, \bibinfo {author} {\bibfnamefont {A.~A.}\
  \bibnamefont {Kulago}}, \bibinfo {author} {\bibfnamefont {G.~B.}\
  \bibnamefont {Boursalian}}, \bibinfo {author} {\bibfnamefont
  {B.}~\bibnamefont {de~Bruin}}, \bibinfo {author} {\bibfnamefont
  {E.}~\bibnamefont {Otten}}, \bibinfo {author} {\bibfnamefont {W.~R.}\
  \bibnamefont {Browne}},\ and\ \bibinfo {author} {\bibfnamefont {B.~L.}\
  \bibnamefont {Feringa}},\ }\href {https://doi.org/10.1021/jacs.1c05938}
  {\bibfield  {journal} {\bibinfo  {journal} {Journal of the American Chemical
  Society}\ }\textbf {\bibinfo {volume} {143}},\ \bibinfo {pages} {18020}
  (\bibinfo {year} {2021})}\BibitemShut {NoStop}%
\bibitem [{\citenamefont {Feng}\ \emph {et~al.}(2021)\citenamefont {Feng},
  \citenamefont {Chong}, \citenamefont {Tang}, \citenamefont {Fang},
  \citenamefont {Zhao}, \citenamefont {Jiang},\ and\ \citenamefont
  {Wang}}]{Feng2021}%
  \BibitemOpen
  \bibfield  {author} {\bibinfo {author} {\bibfnamefont {Z.}~\bibnamefont
  {Feng}}, \bibinfo {author} {\bibfnamefont {Y.}~\bibnamefont {Chong}},
  \bibinfo {author} {\bibfnamefont {S.}~\bibnamefont {Tang}}, \bibinfo {author}
  {\bibfnamefont {Y.}~\bibnamefont {Fang}}, \bibinfo {author} {\bibfnamefont
  {Y.}~\bibnamefont {Zhao}}, \bibinfo {author} {\bibfnamefont {J.}~\bibnamefont
  {Jiang}},\ and\ \bibinfo {author} {\bibfnamefont {X.}~\bibnamefont {Wang}},\
  }\href {https://doi.org/10.1039/D1SC04486A} {\bibfield  {journal} {\bibinfo
  {journal} {Chemical Science}\ }\textbf {\bibinfo {volume} {12}},\ \bibinfo
  {pages} {15151} (\bibinfo {year} {2021})}\BibitemShut {NoStop}%
\bibitem [{\citenamefont {Huang}\ \emph {et~al.}(2022)\citenamefont {Huang},
  \citenamefont {Kang}, \citenamefont {Zhang}, \citenamefont {Zhao},
  \citenamefont {Shi},\ and\ \citenamefont {Yang}}]{Huang2022}%
  \BibitemOpen
  \bibfield  {author} {\bibinfo {author} {\bibfnamefont {B.}~\bibnamefont
  {Huang}}, \bibinfo {author} {\bibfnamefont {H.}~\bibnamefont {Kang}},
  \bibinfo {author} {\bibfnamefont {C.-W.}\ \bibnamefont {Zhang}}, \bibinfo
  {author} {\bibfnamefont {X.-L.}\ \bibnamefont {Zhao}}, \bibinfo {author}
  {\bibfnamefont {X.}~\bibnamefont {Shi}},\ and\ \bibinfo {author}
  {\bibfnamefont {H.-B.}\ \bibnamefont {Yang}},\ }\bibfield  {journal}
  {\bibinfo  {journal} {Communications Chemistry}\ }\textbf {\bibinfo {volume}
  {5}},\ \href {https://doi.org/10.1038/s42004-022-00747-8}
  {10.1038/s42004-022-00747-8} (\bibinfo {year} {2022})\BibitemShut {NoStop}%
\bibitem [{\citenamefont {Abdurahman}\ \emph
  {et~al.}(2023{\natexlab{a}})\citenamefont {Abdurahman}, \citenamefont {Wang},
  \citenamefont {Zhao}, \citenamefont {Li}, \citenamefont {Shen},\ and\
  \citenamefont {Peng}}]{Abdurahman2023}%
  \BibitemOpen
  \bibfield  {author} {\bibinfo {author} {\bibfnamefont {A.}~\bibnamefont
  {Abdurahman}}, \bibinfo {author} {\bibfnamefont {J.}~\bibnamefont {Wang}},
  \bibinfo {author} {\bibfnamefont {Y.}~\bibnamefont {Zhao}}, \bibinfo {author}
  {\bibfnamefont {P.}~\bibnamefont {Li}}, \bibinfo {author} {\bibfnamefont
  {L.}~\bibnamefont {Shen}},\ and\ \bibinfo {author} {\bibfnamefont
  {Q.}~\bibnamefont {Peng}},\ }\href {https://doi.org/10.1002/anie.202300772}
  {\bibfield  {journal} {\bibinfo  {journal} {Angewandte Chemie International
  Edition}\ }\textbf {\bibinfo {volume} {62}},\ \bibinfo {pages} {e202300772}
  (\bibinfo {year} {2023}{\natexlab{a}})}\BibitemShut {NoStop}%
\bibitem [{\citenamefont {Matsuoka}\ \emph {et~al.}(2023)\citenamefont
  {Matsuoka}, \citenamefont {Kimura}, \citenamefont {Miura}, \citenamefont
  {Ikoma},\ and\ \citenamefont {Kusamoto}}]{Matsuoka2023}%
  \BibitemOpen
  \bibfield  {author} {\bibinfo {author} {\bibfnamefont {R.}~\bibnamefont
  {Matsuoka}}, \bibinfo {author} {\bibfnamefont {S.}~\bibnamefont {Kimura}},
  \bibinfo {author} {\bibfnamefont {T.}~\bibnamefont {Miura}}, \bibinfo
  {author} {\bibfnamefont {T.}~\bibnamefont {Ikoma}},\ and\ \bibinfo {author}
  {\bibfnamefont {T.}~\bibnamefont {Kusamoto}},\ }\href
  {https://doi.org/10.1021/jacs.3c01076} {\bibfield  {journal} {\bibinfo
  {journal} {Journal of the American Chemical Society}\ }\textbf {\bibinfo
  {volume} {145}},\ \bibinfo {pages} {13615} (\bibinfo {year}
  {2023})}\BibitemShut {NoStop}%
\bibitem [{\citenamefont {Abdurahman}\ \emph
  {et~al.}(2023{\natexlab{b}})\citenamefont {Abdurahman}, \citenamefont {Shen},
  \citenamefont {Wang}, \citenamefont {Niu}, \citenamefont {Li}, \citenamefont
  {Qiming}, \citenamefont {Wang},\ and\ \citenamefont {Lu}}]{Abdurahman2023-2}%
  \BibitemOpen
  \bibfield  {author} {\bibinfo {author} {\bibfnamefont {A.}~\bibnamefont
  {Abdurahman}}, \bibinfo {author} {\bibfnamefont {L.}~\bibnamefont {Shen}},
  \bibinfo {author} {\bibfnamefont {J.}~\bibnamefont {Wang}}, \bibinfo {author}
  {\bibfnamefont {M.}~\bibnamefont {Niu}}, \bibinfo {author} {\bibfnamefont
  {P.}~\bibnamefont {Li}}, \bibinfo {author} {\bibfnamefont {P.}~\bibnamefont
  {Qiming}}, \bibinfo {author} {\bibfnamefont {J.}~\bibnamefont {Wang}},\ and\
  \bibinfo {author} {\bibfnamefont {G.}~\bibnamefont {Lu}},\ }\href
  {https://doi.org/10.21203/rs.3.rs-3179748/v1} {\bibinfo {title} {A {Highly}
  {Efficient} {Open}-shell {Singlet} {Luminescent} {Diradical} with {Strong}
  {Magnetoluminescence} {Properties}}} (\bibinfo {year}
  {2023}{\natexlab{b}})\BibitemShut {NoStop}%
\bibitem [{\citenamefont {Abe}(2013)}]{Abe2013}%
  \BibitemOpen
  \bibfield  {author} {\bibinfo {author} {\bibfnamefont {M.}~\bibnamefont
  {Abe}},\ }\href {https://doi.org/10.1021/cr400056a} {\bibfield  {journal}
  {\bibinfo  {journal} {Chemical Reviews}\ }\textbf {\bibinfo {volume} {113}},\
  \bibinfo {pages} {7011} (\bibinfo {year} {2013})}\BibitemShut {NoStop}%
\bibitem [{\citenamefont {Casado}(2017)}]{Casado2017}%
  \BibitemOpen
  \bibfield  {author} {\bibinfo {author} {\bibfnamefont {J.}~\bibnamefont
  {Casado}},\ }in\ \href {https://doi.org/10.1007/978-3-319-93302-3_5} {\emph
  {\bibinfo {booktitle} {Physical Organic Chemistry of Quinodimethanes}}},\
  \bibinfo {series and number} {Topics in {Current} {Chemistry}
  {Collections}},\ \bibinfo {editor} {edited by\ \bibinfo {editor}
  {\bibfnamefont {Y.}~\bibnamefont {Tobe}}\ and\ \bibinfo {editor}
  {\bibfnamefont {T.}~\bibnamefont {Kubo}}}\ (\bibinfo  {publisher}
  {Springer},\ \bibinfo {address} {Cham},\ \bibinfo {year} {2017})\ pp.\
  \bibinfo {pages} {209--248}\BibitemShut {NoStop}%
\bibitem [{\citenamefont {Stuyver}\ \emph {et~al.}(2019)\citenamefont
  {Stuyver}, \citenamefont {Chen}, \citenamefont {Zeng}, \citenamefont
  {Geerlings}, \citenamefont {De~Proft},\ and\ \citenamefont
  {Hoffmann}}]{Stuyver2019}%
  \BibitemOpen
  \bibfield  {author} {\bibinfo {author} {\bibfnamefont {T.}~\bibnamefont
  {Stuyver}}, \bibinfo {author} {\bibfnamefont {B.}~\bibnamefont {Chen}},
  \bibinfo {author} {\bibfnamefont {T.}~\bibnamefont {Zeng}}, \bibinfo {author}
  {\bibfnamefont {P.}~\bibnamefont {Geerlings}}, \bibinfo {author}
  {\bibfnamefont {F.}~\bibnamefont {De~Proft}},\ and\ \bibinfo {author}
  {\bibfnamefont {R.}~\bibnamefont {Hoffmann}},\ }\href
  {https://doi.org/10.1021/acs.chemrev.9b00260} {\bibfield  {journal} {\bibinfo
   {journal} {Chemical Reviews}\ }\textbf {\bibinfo {volume} {119}},\ \bibinfo
  {pages} {11291} (\bibinfo {year} {2019})}\BibitemShut {NoStop}%
\bibitem [{\citenamefont {Murto}\ and\ \citenamefont
  {Bronstein}(2022)}]{Murto2022}%
  \BibitemOpen
  \bibfield  {author} {\bibinfo {author} {\bibfnamefont {P.}~\bibnamefont
  {Murto}}\ and\ \bibinfo {author} {\bibfnamefont {H.}~\bibnamefont
  {Bronstein}},\ }\href {https://doi.org/10.1039/D1TC05268C} {\bibfield
  {journal} {\bibinfo  {journal} {Journal of Materials Chemistry C}\ }\textbf
  {\bibinfo {volume} {10}},\ \bibinfo {pages} {7368} (\bibinfo {year}
  {2022})}\BibitemShut {NoStop}%
\bibitem [{\citenamefont {Prajapati}\ \emph {et~al.}(2023)\citenamefont
  {Prajapati}, \citenamefont {Ambhore}, \citenamefont {Dang}, \citenamefont
  {Chmielewski}, \citenamefont {Lis}, \citenamefont {Gómez-García},
  \citenamefont {Zimmerman},\ and\ \citenamefont {Stępień}}]{Prajapati2023}%
  \BibitemOpen
  \bibfield  {author} {\bibinfo {author} {\bibfnamefont {B.}~\bibnamefont
  {Prajapati}}, \bibinfo {author} {\bibfnamefont {M.~D.}\ \bibnamefont
  {Ambhore}}, \bibinfo {author} {\bibfnamefont {D.-K.}\ \bibnamefont {Dang}},
  \bibinfo {author} {\bibfnamefont {P.~J.}\ \bibnamefont {Chmielewski}},
  \bibinfo {author} {\bibfnamefont {T.}~\bibnamefont {Lis}}, \bibinfo {author}
  {\bibfnamefont {C.~J.}\ \bibnamefont {Gómez-García}}, \bibinfo {author}
  {\bibfnamefont {P.~M.}\ \bibnamefont {Zimmerman}},\ and\ \bibinfo {author}
  {\bibfnamefont {M.}~\bibnamefont {Stępień}},\ }\href
  {https://doi.org/10.1038/s41557-023-01341-8} {\bibfield  {journal} {\bibinfo
  {journal} {Nature Chemistry}\ }\textbf {\bibinfo {volume} {15}},\ \bibinfo
  {pages} {1541} (\bibinfo {year} {2023})}\BibitemShut {NoStop}%
\bibitem [{\citenamefont {Khvorost}\ \emph {et~al.}(2024)\citenamefont
  {Khvorost}, \citenamefont {Wojcik}, \citenamefont {Chang}, \citenamefont
  {Calvillo}, \citenamefont {Dickerson}, \citenamefont {Krylov},\ and\
  \citenamefont {Alexandrova}}]{Khvorost2024}%
  \BibitemOpen
  \bibfield  {author} {\bibinfo {author} {\bibfnamefont {T.}~\bibnamefont
  {Khvorost}}, \bibinfo {author} {\bibfnamefont {P.}~\bibnamefont {Wojcik}},
  \bibinfo {author} {\bibfnamefont {C.}~\bibnamefont {Chang}}, \bibinfo
  {author} {\bibfnamefont {M.}~\bibnamefont {Calvillo}}, \bibinfo {author}
  {\bibfnamefont {C.}~\bibnamefont {Dickerson}}, \bibinfo {author}
  {\bibfnamefont {A.}~\bibnamefont {Krylov}},\ and\ \bibinfo {author}
  {\bibfnamefont {A.}~\bibnamefont {Alexandrova}}\ }\href
  {https://doi.org/10.26434/chemrxiv-2024-fmrnw-v2}
  {10.26434/chemrxiv-2024-fmrnw-v2} (\bibinfo {year} {2024})\BibitemShut
  {NoStop}%
\bibitem [{\citenamefont {Dickerson}\ \emph {et~al.}(2021)\citenamefont
  {Dickerson}, \citenamefont {Guo}, \citenamefont {Zhu}, \citenamefont
  {Hudson}, \citenamefont {Caram}, \citenamefont {Campbell},\ and\
  \citenamefont {Alexandrova}}]{Dickerson2021}%
  \BibitemOpen
  \bibfield  {author} {\bibinfo {author} {\bibfnamefont {C.~E.}\ \bibnamefont
  {Dickerson}}, \bibinfo {author} {\bibfnamefont {H.}~\bibnamefont {Guo}},
  \bibinfo {author} {\bibfnamefont {G.-Z.}\ \bibnamefont {Zhu}}, \bibinfo
  {author} {\bibfnamefont {E.~R.}\ \bibnamefont {Hudson}}, \bibinfo {author}
  {\bibfnamefont {J.~R.}\ \bibnamefont {Caram}}, \bibinfo {author}
  {\bibfnamefont {W.~C.}\ \bibnamefont {Campbell}},\ and\ \bibinfo {author}
  {\bibfnamefont {A.~N.}\ \bibnamefont {Alexandrova}},\ }\href
  {https://doi.org/10.1021/acs.jpclett.1c00733} {\bibfield  {journal} {\bibinfo
   {journal} {The Journal of Physical Chemistry Letters}\ }\textbf {\bibinfo
  {volume} {12}},\ \bibinfo {pages} {3989} (\bibinfo {year}
  {2021})}\BibitemShut {NoStop}%
\bibitem [{\citenamefont {Zhu}\ \emph {et~al.}(2022)\citenamefont {Zhu},
  \citenamefont {Mitra}, \citenamefont {Augenbraun}, \citenamefont {Dickerson},
  \citenamefont {Frim}, \citenamefont {Lao}, \citenamefont {Lasner},
  \citenamefont {Alexandrova}, \citenamefont {Campbell}, \citenamefont {Caram},
  \citenamefont {Doyle},\ and\ \citenamefont {Hudson}}]{Zhu2022}%
  \BibitemOpen
  \bibfield  {author} {\bibinfo {author} {\bibfnamefont {G.-Z.}\ \bibnamefont
  {Zhu}}, \bibinfo {author} {\bibfnamefont {D.}~\bibnamefont {Mitra}}, \bibinfo
  {author} {\bibfnamefont {B.~L.}\ \bibnamefont {Augenbraun}}, \bibinfo
  {author} {\bibfnamefont {C.~E.}\ \bibnamefont {Dickerson}}, \bibinfo {author}
  {\bibfnamefont {M.~J.}\ \bibnamefont {Frim}}, \bibinfo {author}
  {\bibfnamefont {G.}~\bibnamefont {Lao}}, \bibinfo {author} {\bibfnamefont
  {Z.~D.}\ \bibnamefont {Lasner}}, \bibinfo {author} {\bibfnamefont {A.~N.}\
  \bibnamefont {Alexandrova}}, \bibinfo {author} {\bibfnamefont {W.~C.}\
  \bibnamefont {Campbell}}, \bibinfo {author} {\bibfnamefont {J.~R.}\
  \bibnamefont {Caram}}, \bibinfo {author} {\bibfnamefont {J.~M.}\ \bibnamefont
  {Doyle}},\ and\ \bibinfo {author} {\bibfnamefont {E.~R.}\ \bibnamefont
  {Hudson}},\ }\href {https://doi.org/10.1038/s41557-022-00998-x} {\bibfield
  {journal} {\bibinfo  {journal} {Nature Chemistry}\ }\textbf {\bibinfo
  {volume} {14}},\ \bibinfo {pages} {995} (\bibinfo {year} {2022})}\BibitemShut
  {NoStop}%
\bibitem [{\citenamefont {Gately}\ \emph {et~al.}(2023)\citenamefont {Gately},
  \citenamefont {Boto}, \citenamefont {Tauber}, \citenamefont {Casanova},\ and\
  \citenamefont {Bardeen}}]{Gately2023}%
  \BibitemOpen
  \bibfield  {author} {\bibinfo {author} {\bibfnamefont {T.~J.}\ \bibnamefont
  {Gately}}, \bibinfo {author} {\bibfnamefont {R.~A.}\ \bibnamefont {Boto}},
  \bibinfo {author} {\bibfnamefont {M.~J.}\ \bibnamefont {Tauber}}, \bibinfo
  {author} {\bibfnamefont {D.}~\bibnamefont {Casanova}},\ and\ \bibinfo
  {author} {\bibfnamefont {C.~J.}\ \bibnamefont {Bardeen}},\ }\href
  {https://doi.org/10.1021/acs.jpcc.2c08231} {\bibfield  {journal} {\bibinfo
  {journal} {The Journal of Physical Chemistry C}\ }\textbf {\bibinfo {volume}
  {127}},\ \bibinfo {pages} {4816} (\bibinfo {year} {2023})}\BibitemShut
  {NoStop}%
\bibitem [{\citenamefont {Penfold}\ \emph {et~al.}(2018)\citenamefont
  {Penfold}, \citenamefont {Gindensperger}, \citenamefont {Daniel},\ and\
  \citenamefont {Marian}}]{Penfold2018}%
  \BibitemOpen
  \bibfield  {author} {\bibinfo {author} {\bibfnamefont {T.~J.}\ \bibnamefont
  {Penfold}}, \bibinfo {author} {\bibfnamefont {E.}~\bibnamefont
  {Gindensperger}}, \bibinfo {author} {\bibfnamefont {C.}~\bibnamefont
  {Daniel}},\ and\ \bibinfo {author} {\bibfnamefont {C.~M.}\ \bibnamefont
  {Marian}},\ }\href {https://doi.org/10.1021/acs.chemrev.7b00617} {\bibfield
  {journal} {\bibinfo  {journal} {Chemical Reviews}\ }\textbf {\bibinfo
  {volume} {118}},\ \bibinfo {pages} {6975} (\bibinfo {year}
  {2018})}\BibitemShut {NoStop}%
\bibitem [{\citenamefont {El‐Sayed}(1963)}]{El-Sayed1963}%
  \BibitemOpen
  \bibfield  {author} {\bibinfo {author} {\bibfnamefont {M.~A.}\ \bibnamefont
  {El‐Sayed}},\ }\href {https://doi.org/10.1063/1.1733610} {\bibfield
  {journal} {\bibinfo  {journal} {The Journal of Chemical Physics}\ }\textbf
  {\bibinfo {volume} {38}},\ \bibinfo {pages} {2834} (\bibinfo {year}
  {1963})}\BibitemShut {NoStop}%
\bibitem [{\citenamefont {Hong}\ and\ \citenamefont {Meng}(2001)}]{Hong2001}%
  \BibitemOpen
  \bibfield  {author} {\bibinfo {author} {\bibfnamefont {T.-M.}\ \bibnamefont
  {Hong}}\ and\ \bibinfo {author} {\bibfnamefont {H.-F.}\ \bibnamefont
  {Meng}},\ }\href {https://doi.org/10.1103/PhysRevB.63.075206} {\bibfield
  {journal} {\bibinfo  {journal} {Physical Review B}\ }\textbf {\bibinfo
  {volume} {63}},\ \bibinfo {pages} {075206} (\bibinfo {year}
  {2001})}\BibitemShut {NoStop}%
\bibitem [{\citenamefont {Rybicki}\ and\ \citenamefont
  {Wohlgenannt}(2009)}]{Rybicki2009}%
  \BibitemOpen
  \bibfield  {author} {\bibinfo {author} {\bibfnamefont {J.}~\bibnamefont
  {Rybicki}}\ and\ \bibinfo {author} {\bibfnamefont {M.}~\bibnamefont
  {Wohlgenannt}},\ }\href {https://doi.org/10.1103/PhysRevB.79.153202}
  {\bibfield  {journal} {\bibinfo  {journal} {Physical Review B}\ }\textbf
  {\bibinfo {volume} {79}},\ \bibinfo {pages} {153202} (\bibinfo {year}
  {2009})}\BibitemShut {NoStop}%
\bibitem [{\citenamefont {Barford}\ \emph {et~al.}(2010)\citenamefont
  {Barford}, \citenamefont {Bursill},\ and\ \citenamefont
  {Makhov}}]{Barford2010}%
  \BibitemOpen
  \bibfield  {author} {\bibinfo {author} {\bibfnamefont {W.}~\bibnamefont
  {Barford}}, \bibinfo {author} {\bibfnamefont {R.~J.}\ \bibnamefont
  {Bursill}},\ and\ \bibinfo {author} {\bibfnamefont {D.~V.}\ \bibnamefont
  {Makhov}},\ }\href {https://doi.org/10.1103/PhysRevB.81.035206} {\bibfield
  {journal} {\bibinfo  {journal} {Physical Review B}\ }\textbf {\bibinfo
  {volume} {81}},\ \bibinfo {pages} {035206} (\bibinfo {year}
  {2010})}\BibitemShut {NoStop}%
\bibitem [{\citenamefont {Yu}(2012)}]{Yu2012}%
  \BibitemOpen
  \bibfield  {author} {\bibinfo {author} {\bibfnamefont {Z.~G.}\ \bibnamefont
  {Yu}},\ }\href {https://doi.org/10.1103/PhysRevB.85.115201} {\bibfield
  {journal} {\bibinfo  {journal} {Physical Review B}\ }\textbf {\bibinfo
  {volume} {85}},\ \bibinfo {pages} {115201} (\bibinfo {year}
  {2012})}\BibitemShut {NoStop}%
\bibitem [{\citenamefont {Pariser}\ and\ \citenamefont
  {Parr}(1953)}]{Pariser1953}%
  \BibitemOpen
  \bibfield  {author} {\bibinfo {author} {\bibfnamefont {R.}~\bibnamefont
  {Pariser}}\ and\ \bibinfo {author} {\bibfnamefont {R.~G.}\ \bibnamefont
  {Parr}},\ }\href {https://doi.org/10.1063/1.1699030} {\bibfield  {journal}
  {\bibinfo  {journal} {The Journal of Chemical Physics}\ }\textbf {\bibinfo
  {volume} {21}},\ \bibinfo {pages} {767} (\bibinfo {year} {1953})}\BibitemShut
  {NoStop}%
\bibitem [{\citenamefont {Pople}(1953)}]{Pople1953}%
  \BibitemOpen
  \bibfield  {author} {\bibinfo {author} {\bibfnamefont {J.~A.}\ \bibnamefont
  {Pople}},\ }\href@noop {} {\bibfield  {journal} {\bibinfo  {journal}
  {Transactions of the Faraday Society}\ }\textbf {\bibinfo {volume} {49}},\
  \bibinfo {pages} {1375} (\bibinfo {year} {1953})}\BibitemShut {NoStop}%
\bibitem [{\citenamefont {Linderberg}\ and\ \citenamefont
  {öhrn}(1968)}]{Linderberg1968}%
  \BibitemOpen
  \bibfield  {author} {\bibinfo {author} {\bibfnamefont {J.}~\bibnamefont
  {Linderberg}}\ and\ \bibinfo {author} {\bibfnamefont {Y.}~\bibnamefont
  {öhrn}},\ }\href {https://doi.org/10.1063/1.1670129} {\bibfield  {journal}
  {\bibinfo  {journal} {The Journal of Chemical Physics}\ }\textbf {\bibinfo
  {volume} {49}},\ \bibinfo {pages} {716} (\bibinfo {year} {1968})}\BibitemShut
  {NoStop}%
\bibitem [{\citenamefont {Longuet‐Higgins}(1950)}]{Longuet-Higgins1950}%
  \BibitemOpen
  \bibfield  {author} {\bibinfo {author} {\bibfnamefont {H.~C.}\ \bibnamefont
  {Longuet‐Higgins}},\ }\href {https://doi.org/10.1063/1.1747618} {\bibfield
  {journal} {\bibinfo  {journal} {The Journal of Chemical Physics}\ }\textbf
  {\bibinfo {volume} {18}},\ \bibinfo {pages} {265} (\bibinfo {year}
  {1950})}\BibitemShut {NoStop}%
\bibitem [{\citenamefont {Dewar}\ and\ \citenamefont
  {Longuet-Higgins}(1954)}]{Dewar1954}%
  \BibitemOpen
  \bibfield  {author} {\bibinfo {author} {\bibfnamefont {M.~J.~S.}\
  \bibnamefont {Dewar}}\ and\ \bibinfo {author} {\bibfnamefont {H.~C.}\
  \bibnamefont {Longuet-Higgins}},\ }\href
  {https://doi.org/10.1088/0370-1298/67/9/307} {\bibfield  {journal} {\bibinfo
  {journal} {Proceedings of the Physical Society. Section A}\ }\textbf
  {\bibinfo {volume} {67}},\ \bibinfo {pages} {795} (\bibinfo {year}
  {1954})}\BibitemShut {NoStop}%
\bibitem [{\citenamefont {Longuet-Higgins}\ and\ \citenamefont
  {Pople}(1955)}]{Longuet-Higgins1955}%
  \BibitemOpen
  \bibfield  {author} {\bibinfo {author} {\bibfnamefont {H.~C.}\ \bibnamefont
  {Longuet-Higgins}}\ and\ \bibinfo {author} {\bibfnamefont {J.~A.}\
  \bibnamefont {Pople}},\ }\href@noop {} {\bibfield  {journal} {\bibinfo
  {journal} {Proceedings of the Physical Society. Section A}\ }\textbf
  {\bibinfo {volume} {68}},\ \bibinfo {pages} {591} (\bibinfo {year}
  {1955})}\BibitemShut {NoStop}%
\bibitem [{\citenamefont {Hele}(2021)}]{Hele2021}%
  \BibitemOpen
  \bibfield  {author} {\bibinfo {author} {\bibfnamefont {T.~J.~H.}\
  \bibnamefont {Hele}},\ }in\ \href {https://doi.org/10.1117/12.2593712} {\emph
  {\bibinfo {booktitle} {Physical Chemistry of Semiconductor Materials and
  Interfaces XX}}},\ Vol.\ \bibinfo {volume} {11799},\ \bibinfo {editor}
  {edited by\ \bibinfo {editor} {\bibfnamefont {A.~J.}\ \bibnamefont {Musser}}\
  and\ \bibinfo {editor} {\bibfnamefont {D.}~\bibnamefont {Baran}}},\ \bibinfo
  {organization} {International Society for Optics and Photonics}\ (\bibinfo
  {publisher} {SPIE},\ \bibinfo {year} {2021})\ p.\ \bibinfo {pages}
  {117991A}\BibitemShut {NoStop}%
\bibitem [{\citenamefont {Szabo}\ and\ \citenamefont
  {Ostlund}(1989)}]{Szabo1989}%
  \BibitemOpen
  \bibfield  {author} {\bibinfo {author} {\bibfnamefont {A.}~\bibnamefont
  {Szabo}}\ and\ \bibinfo {author} {\bibfnamefont {N.~S.}\ \bibnamefont
  {Ostlund}},\ }\href@noop {} {\emph {\bibinfo {title} {{Modern Quantum
  Chemistry: Introduction to Advanced Electronic Structure Theory}}}}\
  (\bibinfo  {publisher} {Dover Publications},\ \bibinfo {year}
  {1989})\BibitemShut {NoStop}%
\bibitem [{\citenamefont {Pariser}(1956)}]{Pariser1956}%
  \BibitemOpen
  \bibfield  {author} {\bibinfo {author} {\bibfnamefont {R.}~\bibnamefont
  {Pariser}},\ }\href {https://doi.org/10.1063/1.1742461} {\bibfield  {journal}
  {\bibinfo  {journal} {The Journal of Chemical Physics}\ }\textbf {\bibinfo
  {volume} {24}},\ \bibinfo {pages} {250} (\bibinfo {year} {1956})}\BibitemShut
  {NoStop}%
\bibitem [{\citenamefont {Hele}\ \emph {et~al.}(2019)\citenamefont {Hele},
  \citenamefont {Fuemmeler}, \citenamefont {Sanders}, \citenamefont
  {Kumarasamy}, \citenamefont {Sfeir}, \citenamefont {Campos},\ and\
  \citenamefont {Ananth}}]{Hele2019}%
  \BibitemOpen
  \bibfield  {author} {\bibinfo {author} {\bibfnamefont {T.~J.~H.}\
  \bibnamefont {Hele}}, \bibinfo {author} {\bibfnamefont {E.~G.}\ \bibnamefont
  {Fuemmeler}}, \bibinfo {author} {\bibfnamefont {S.~N.}\ \bibnamefont
  {Sanders}}, \bibinfo {author} {\bibfnamefont {E.}~\bibnamefont {Kumarasamy}},
  \bibinfo {author} {\bibfnamefont {M.~Y.}\ \bibnamefont {Sfeir}}, \bibinfo
  {author} {\bibfnamefont {L.~M.}\ \bibnamefont {Campos}},\ and\ \bibinfo
  {author} {\bibfnamefont {N.}~\bibnamefont {Ananth}},\ }\href
  {https://doi.org/10.1021/acs.jpca.8b12222} {\bibfield  {journal} {\bibinfo
  {journal} {The Journal of Physical Chemistry A}\ }\textbf {\bibinfo {volume}
  {123}},\ \bibinfo {pages} {2527} (\bibinfo {year} {2019})}\BibitemShut
  {NoStop}%
\bibitem [{\citenamefont {Abdurahman}\ \emph {et~al.}(2020)\citenamefont
  {Abdurahman}, \citenamefont {Hele}, \citenamefont {Gu}, \citenamefont
  {Zhang}, \citenamefont {Peng}, \citenamefont {Zhang}, \citenamefont {Friend},
  \citenamefont {Li},\ and\ \citenamefont {Evans}}]{Abdurahman2020}%
  \BibitemOpen
  \bibfield  {author} {\bibinfo {author} {\bibfnamefont {A.}~\bibnamefont
  {Abdurahman}}, \bibinfo {author} {\bibfnamefont {T.~J.~H.}\ \bibnamefont
  {Hele}}, \bibinfo {author} {\bibfnamefont {Q.}~\bibnamefont {Gu}}, \bibinfo
  {author} {\bibfnamefont {J.}~\bibnamefont {Zhang}}, \bibinfo {author}
  {\bibfnamefont {Q.}~\bibnamefont {Peng}}, \bibinfo {author} {\bibfnamefont
  {M.}~\bibnamefont {Zhang}}, \bibinfo {author} {\bibfnamefont {R.~H.}\
  \bibnamefont {Friend}}, \bibinfo {author} {\bibfnamefont {F.}~\bibnamefont
  {Li}},\ and\ \bibinfo {author} {\bibfnamefont {E.~W.}\ \bibnamefont
  {Evans}},\ }\href {https://doi.org/10.1038/s41563-020-0705-9} {\bibfield
  {journal} {\bibinfo  {journal} {Nature Materials}\ }\textbf {\bibinfo
  {volume} {19}},\ \bibinfo {pages} {1224} (\bibinfo {year}
  {2020})}\BibitemShut {NoStop}%
\bibitem [{\citenamefont {Salem}\ and\ \citenamefont
  {Rowland}(1972)}]{Salem1972}%
  \BibitemOpen
  \bibfield  {author} {\bibinfo {author} {\bibfnamefont {L.}~\bibnamefont
  {Salem}}\ and\ \bibinfo {author} {\bibfnamefont {C.}~\bibnamefont
  {Rowland}},\ }\href {https://doi.org/10.1002/anie.197200921} {\bibfield
  {journal} {\bibinfo  {journal} {Angewandte Chemie International Edition in
  English}\ }\textbf {\bibinfo {volume} {11}},\ \bibinfo {pages} {92} (\bibinfo
  {year} {1972})}\BibitemShut {NoStop}%
\bibitem [{\citenamefont {Atherton}(1993)}]{Atherton1993}%
  \BibitemOpen
  \bibfield  {author} {\bibinfo {author} {\bibfnamefont {N.~M.}\ \bibnamefont
  {Atherton}},\ }\bibinfo {title} {{Principles of Electron Spin Resonance}}\
  (\bibinfo  {publisher} {Ellis Horwood},\ \bibinfo {year} {1993})\ Chap.\
  \bibinfo {chapter} {High-spin systems}, pp.\ \bibinfo {pages}
  {224--263}\BibitemShut {NoStop}%
\bibitem [{\citenamefont {Englman}\ and\ \citenamefont
  {Jortner}(1970)}]{Englman1970}%
  \BibitemOpen
  \bibfield  {author} {\bibinfo {author} {\bibfnamefont {R.}~\bibnamefont
  {Englman}}\ and\ \bibinfo {author} {\bibfnamefont {J.}~\bibnamefont
  {Jortner}},\ }\href {https://doi.org/10.1080/00268977000100171} {\bibfield
  {journal} {\bibinfo  {journal} {Molecular Physics}\ }\textbf {\bibinfo
  {volume} {18}},\ \bibinfo {pages} {145} (\bibinfo {year} {1970})}\BibitemShut
  {NoStop}%
\bibitem [{\citenamefont {Gamero}\ \emph {et~al.}(2006)\citenamefont {Gamero},
  \citenamefont {Velasco}, \citenamefont {Latorre}, \citenamefont
  {López-Calahorra}, \citenamefont {Brillas},\ and\ \citenamefont
  {Juliá}}]{Gamero2006}%
  \BibitemOpen
  \bibfield  {author} {\bibinfo {author} {\bibfnamefont {V.}~\bibnamefont
  {Gamero}}, \bibinfo {author} {\bibfnamefont {D.}~\bibnamefont {Velasco}},
  \bibinfo {author} {\bibfnamefont {S.}~\bibnamefont {Latorre}}, \bibinfo
  {author} {\bibfnamefont {F.}~\bibnamefont {López-Calahorra}}, \bibinfo
  {author} {\bibfnamefont {E.}~\bibnamefont {Brillas}},\ and\ \bibinfo {author}
  {\bibfnamefont {L.}~\bibnamefont {Juliá}},\ }\href
  {https://doi.org/10.1016/j.tetlet.2006.02.022} {\bibfield  {journal}
  {\bibinfo  {journal} {Tetrahedron Letters}\ }\textbf {\bibinfo {volume}
  {47}},\ \bibinfo {pages} {2305} (\bibinfo {year} {2006})}\BibitemShut
  {NoStop}%
\bibitem [{Note1()}]{Note1}%
  \BibitemOpen
  \bibinfo {note} {These values were calculated using the Q-Chem package
  (version 6.0.2) \protect \citep {Epifanovsky2021}. For the same computation
  on the Gaussian16 software \protect \citep {g16}, the respective values were
  $-0.005\protect \text { eV}$ and $+0.001\protect \text { eV}$ at $\theta
  =90^{\protect \text {o}}$, and $-0.009\protect \text { eV}$ and
  $+0.001\protect \text { eV}$ at $\theta =110^{\protect \text
  {o}}$.}\BibitemShut {Stop}%
\bibitem [{\citenamefont {Laurent}\ and\ \citenamefont
  {Jacquemin}(2013)}]{Laurent2013}%
  \BibitemOpen
  \bibfield  {author} {\bibinfo {author} {\bibfnamefont {A.~D.}\ \bibnamefont
  {Laurent}}\ and\ \bibinfo {author} {\bibfnamefont {D.}~\bibnamefont
  {Jacquemin}},\ }\href {https://doi.org/10.1002/qua.24438} {\bibfield
  {journal} {\bibinfo  {journal} {International Journal of Quantum Chemistry}\
  }\textbf {\bibinfo {volume} {113}},\ \bibinfo {pages} {2019} (\bibinfo {year}
  {2013})}\BibitemShut {NoStop}%
\bibitem [{\citenamefont {Nielsen}\ and\ \citenamefont
  {Chuang}(2010)}]{Nielsen2010}%
  \BibitemOpen
  \bibfield  {author} {\bibinfo {author} {\bibfnamefont {M.~A.}\ \bibnamefont
  {Nielsen}}\ and\ \bibinfo {author} {\bibfnamefont {I.~L.}\ \bibnamefont
  {Chuang}},\ }\bibinfo {title} {{Quantum Computation and Quantum
  Information}}\ (\bibinfo  {publisher} {Cambridge University Press},\ \bibinfo
  {year} {2010})\ Chap.\ \bibinfo {chapter} {Quantum circuits}, pp.\ \bibinfo
  {pages} {171--215},\ \bibinfo {edition} {{10th Anniversary}}\ ed.\BibitemShut
  {Stop}%
\bibitem [{\citenamefont {Epifanovsky}\ \emph {et~al.}(2021)\citenamefont
  {Epifanovsky}, \citenamefont {Gilbert}, \citenamefont {Feng}, \citenamefont
  {Lee}, \citenamefont {Mao}, \citenamefont {Mardirossian}, \citenamefont
  {Pokhilko}, \citenamefont {White}, \citenamefont {Coons}, \citenamefont
  {Dempwolff}, \citenamefont {Gan}, \citenamefont {Hait}, \citenamefont {Horn},
  \citenamefont {Jacobson}, \citenamefont {Kaliman}, \citenamefont {Kussmann},
  \citenamefont {Lange}, \citenamefont {Lao}, \citenamefont {Levine},
  \citenamefont {Liu}, \citenamefont {McKenzie}, \citenamefont {Morrison},
  \citenamefont {Nanda}, \citenamefont {Plasser}, \citenamefont {Rehn},
  \citenamefont {Vidal}, \citenamefont {You}, \citenamefont {Zhu},
  \citenamefont {Alam}, \citenamefont {Albrecht}, \citenamefont {Aldossary},
  \citenamefont {Alguire}, \citenamefont {Andersen}, \citenamefont {Athavale},
  \citenamefont {Barton}, \citenamefont {Begam}, \citenamefont {Behn},
  \citenamefont {Bellonzi}, \citenamefont {Bernard}, \citenamefont {Berquist},
  \citenamefont {Burton}, \citenamefont {Carreras}, \citenamefont
  {Carter-Fenk}, \citenamefont {Chakraborty}, \citenamefont {Chien},
  \citenamefont {Closser}, \citenamefont {Cofer-Shabica}, \citenamefont
  {Dasgupta}, \citenamefont {de~Wergifosse}, \citenamefont {Deng},
  \citenamefont {Diedenhofen}, \citenamefont {Do}, \citenamefont {Ehlert},
  \citenamefont {Fang}, \citenamefont {Fatehi}, \citenamefont {Feng},
  \citenamefont {Friedhoff}, \citenamefont {Gayvert}, \citenamefont {Ge},
  \citenamefont {Gidofalvi}, \citenamefont {Goldey}, \citenamefont {Gomes},
  \citenamefont {Gonz{\'a}lez-Espinoza}, \citenamefont {Gulania}, \citenamefont
  {Gunina}, \citenamefont {Hanson-Heine}, \citenamefont {Harbach},
  \citenamefont {Hauser}, \citenamefont {Herbst}, \citenamefont
  {Hern{\'a}ndez~Vera}, \citenamefont {Hodecker}, \citenamefont {Holden},
  \citenamefont {Houck}, \citenamefont {Huang}, \citenamefont {Hui},
  \citenamefont {Huynh}, \citenamefont {Ivanov}, \citenamefont {J{\'a}sz},
  \citenamefont {Ji}, \citenamefont {Jiang}, \citenamefont {Kaduk},
  \citenamefont {K{\"a}hler}, \citenamefont {Khistyaev}, \citenamefont {Kim},
  \citenamefont {Kis}, \citenamefont {Klunzinger}, \citenamefont
  {Koczor-Benda}, \citenamefont {Koh}, \citenamefont {Kosenkov}, \citenamefont
  {Koulias}, \citenamefont {Kowalczyk}, \citenamefont {Krauter}, \citenamefont
  {Kue}, \citenamefont {Kunitsa}, \citenamefont {Kus}, \citenamefont
  {Ladj{\'a}nszki}, \citenamefont {Landau}, \citenamefont {Lawler},
  \citenamefont {Lefrancois}, \citenamefont {Lehtola}, \citenamefont {Li},
  \citenamefont {Li}, \citenamefont {Liang}, \citenamefont {Liebenthal},
  \citenamefont {Lin}, \citenamefont {Lin}, \citenamefont {Liu}, \citenamefont
  {Liu}, \citenamefont {Loipersberger}, \citenamefont {Luenser}, \citenamefont
  {Manjanath}, \citenamefont {Manohar}, \citenamefont {Mansoor}, \citenamefont
  {Manzer}, \citenamefont {Mao}, \citenamefont {Marenich}, \citenamefont
  {Markovich}, \citenamefont {Mason}, \citenamefont {Maurer}, \citenamefont
  {McLaughlin}, \citenamefont {Menger}, \citenamefont {Mewes}, \citenamefont
  {Mewes}, \citenamefont {Morgante}, \citenamefont {Mullinax}, \citenamefont
  {Oosterbaan}, \citenamefont {Paran}, \citenamefont {Paul}, \citenamefont
  {Paul}, \citenamefont {Pavo{\v s}evi{\'c}}, \citenamefont {Pei},
  \citenamefont {Prager}, \citenamefont {Proynov}, \citenamefont {R{\'a}k},
  \citenamefont {Ramos-Cordoba}, \citenamefont {Rana}, \citenamefont {Rask},
  \citenamefont {Rettig}, \citenamefont {Richard}, \citenamefont {Rob},
  \citenamefont {Rossomme}, \citenamefont {Scheele}, \citenamefont {Scheurer},
  \citenamefont {Schneider}, \citenamefont {Sergueev}, \citenamefont {Sharada},
  \citenamefont {Skomorowski}, \citenamefont {Small}, \citenamefont {Stein},
  \citenamefont {Su}, \citenamefont {Sundstrom}, \citenamefont {Tao},
  \citenamefont {Thirman}, \citenamefont {Tornai}, \citenamefont {Tsuchimochi},
  \citenamefont {Tubman}, \citenamefont {Veccham}, \citenamefont {Vydrov},
  \citenamefont {Wenzel}, \citenamefont {Witte}, \citenamefont {Yamada},
  \citenamefont {Yao}, \citenamefont {Yeganeh}, \citenamefont {Yost},
  \citenamefont {Zech}, \citenamefont {Zhang}, \citenamefont {Zhang},
  \citenamefont {Zhang}, \citenamefont {Zuev}, \citenamefont {Aspuru-Guzik},
  \citenamefont {Bell}, \citenamefont {Besley}, \citenamefont {Bravaya},
  \citenamefont {Brooks}, \citenamefont {Casanova}, \citenamefont {Chai},
  \citenamefont {Coriani}, \citenamefont {Cramer}, \citenamefont {Cserey},
  \citenamefont {DePrince}, \citenamefont {DiStasio}, \citenamefont {Dreuw},
  \citenamefont {Dunietz}, \citenamefont {Furlani}, \citenamefont {Goddard},
  \citenamefont {Hammes-Schiffer}, \citenamefont {Head-Gordon}, \citenamefont
  {Hehre}, \citenamefont {Hsu}, \citenamefont {Jagau}, \citenamefont {Jung},
  \citenamefont {Klamt}, \citenamefont {Kong}, \citenamefont {Lambrecht},
  \citenamefont {Liang}, \citenamefont {Mayhall}, \citenamefont {McCurdy},
  \citenamefont {Neaton}, \citenamefont {Ochsenfeld}, \citenamefont {Parkhill},
  \citenamefont {Peverati}, \citenamefont {Rassolov}, \citenamefont {Shao},
  \citenamefont {Slipchenko}, \citenamefont {Stauch}, \citenamefont {Steele},
  \citenamefont {Subotnik}, \citenamefont {Thom}, \citenamefont {Tkatchenko},
  \citenamefont {Truhlar}, \citenamefont {Van~Voorhis}, \citenamefont
  {Wesolowski}, \citenamefont {Whaley}, \citenamefont {Woodcock}, \citenamefont
  {Zimmerman}, \citenamefont {Faraji}, \citenamefont {Gill}, \citenamefont
  {Head-Gordon}, \citenamefont {Herbert},\ and\ \citenamefont
  {Krylov}}]{Epifanovsky2021}%
  \BibitemOpen
  \bibfield  {author} {\bibinfo {author} {\bibfnamefont {E.}~\bibnamefont
  {Epifanovsky}}, \bibinfo {author} {\bibfnamefont {A.~T.~B.}\ \bibnamefont
  {Gilbert}}, \bibinfo {author} {\bibfnamefont {X.}~\bibnamefont {Feng}},
  \bibinfo {author} {\bibfnamefont {J.}~\bibnamefont {Lee}}, \bibinfo {author}
  {\bibfnamefont {Y.}~\bibnamefont {Mao}}, \bibinfo {author} {\bibfnamefont
  {N.}~\bibnamefont {Mardirossian}}, \bibinfo {author} {\bibfnamefont
  {P.}~\bibnamefont {Pokhilko}}, \bibinfo {author} {\bibfnamefont {A.~F.}\
  \bibnamefont {White}}, \bibinfo {author} {\bibfnamefont {M.~P.}\ \bibnamefont
  {Coons}}, \bibinfo {author} {\bibfnamefont {A.~L.}\ \bibnamefont
  {Dempwolff}}, \bibinfo {author} {\bibfnamefont {Z.}~\bibnamefont {Gan}},
  \bibinfo {author} {\bibfnamefont {D.}~\bibnamefont {Hait}}, \bibinfo {author}
  {\bibfnamefont {P.~R.}\ \bibnamefont {Horn}}, \bibinfo {author}
  {\bibfnamefont {L.~D.}\ \bibnamefont {Jacobson}}, \bibinfo {author}
  {\bibfnamefont {I.}~\bibnamefont {Kaliman}}, \bibinfo {author} {\bibfnamefont
  {J.}~\bibnamefont {Kussmann}}, \bibinfo {author} {\bibfnamefont {A.~W.}\
  \bibnamefont {Lange}}, \bibinfo {author} {\bibfnamefont {K.~U.}\ \bibnamefont
  {Lao}}, \bibinfo {author} {\bibfnamefont {D.~S.}\ \bibnamefont {Levine}},
  \bibinfo {author} {\bibfnamefont {J.}~\bibnamefont {Liu}}, \bibinfo {author}
  {\bibfnamefont {S.~C.}\ \bibnamefont {McKenzie}}, \bibinfo {author}
  {\bibfnamefont {A.~F.}\ \bibnamefont {Morrison}}, \bibinfo {author}
  {\bibfnamefont {K.~D.}\ \bibnamefont {Nanda}}, \bibinfo {author}
  {\bibfnamefont {F.}~\bibnamefont {Plasser}}, \bibinfo {author} {\bibfnamefont
  {D.~R.}\ \bibnamefont {Rehn}}, \bibinfo {author} {\bibfnamefont {M.~L.}\
  \bibnamefont {Vidal}}, \bibinfo {author} {\bibfnamefont {Z.-Q.}\ \bibnamefont
  {You}}, \bibinfo {author} {\bibfnamefont {Y.}~\bibnamefont {Zhu}}, \bibinfo
  {author} {\bibfnamefont {B.}~\bibnamefont {Alam}}, \bibinfo {author}
  {\bibfnamefont {B.~J.}\ \bibnamefont {Albrecht}}, \bibinfo {author}
  {\bibfnamefont {A.}~\bibnamefont {Aldossary}}, \bibinfo {author}
  {\bibfnamefont {E.}~\bibnamefont {Alguire}}, \bibinfo {author} {\bibfnamefont
  {J.~H.}\ \bibnamefont {Andersen}}, \bibinfo {author} {\bibfnamefont
  {V.}~\bibnamefont {Athavale}}, \bibinfo {author} {\bibfnamefont
  {D.}~\bibnamefont {Barton}}, \bibinfo {author} {\bibfnamefont
  {K.}~\bibnamefont {Begam}}, \bibinfo {author} {\bibfnamefont
  {A.}~\bibnamefont {Behn}}, \bibinfo {author} {\bibfnamefont {N.}~\bibnamefont
  {Bellonzi}}, \bibinfo {author} {\bibfnamefont {Y.~A.}\ \bibnamefont
  {Bernard}}, \bibinfo {author} {\bibfnamefont {E.~J.}\ \bibnamefont
  {Berquist}}, \bibinfo {author} {\bibfnamefont {H.~G.~A.}\ \bibnamefont
  {Burton}}, \bibinfo {author} {\bibfnamefont {A.}~\bibnamefont {Carreras}},
  \bibinfo {author} {\bibfnamefont {K.}~\bibnamefont {Carter-Fenk}}, \bibinfo
  {author} {\bibfnamefont {R.}~\bibnamefont {Chakraborty}}, \bibinfo {author}
  {\bibfnamefont {A.~D.}\ \bibnamefont {Chien}}, \bibinfo {author}
  {\bibfnamefont {K.~D.}\ \bibnamefont {Closser}}, \bibinfo {author}
  {\bibfnamefont {V.}~\bibnamefont {Cofer-Shabica}}, \bibinfo {author}
  {\bibfnamefont {S.}~\bibnamefont {Dasgupta}}, \bibinfo {author}
  {\bibfnamefont {M.}~\bibnamefont {de~Wergifosse}}, \bibinfo {author}
  {\bibfnamefont {J.}~\bibnamefont {Deng}}, \bibinfo {author} {\bibfnamefont
  {M.}~\bibnamefont {Diedenhofen}}, \bibinfo {author} {\bibfnamefont
  {H.}~\bibnamefont {Do}}, \bibinfo {author} {\bibfnamefont {S.}~\bibnamefont
  {Ehlert}}, \bibinfo {author} {\bibfnamefont {P.-T.}\ \bibnamefont {Fang}},
  \bibinfo {author} {\bibfnamefont {S.}~\bibnamefont {Fatehi}}, \bibinfo
  {author} {\bibfnamefont {Q.}~\bibnamefont {Feng}}, \bibinfo {author}
  {\bibfnamefont {T.}~\bibnamefont {Friedhoff}}, \bibinfo {author}
  {\bibfnamefont {J.}~\bibnamefont {Gayvert}}, \bibinfo {author} {\bibfnamefont
  {Q.}~\bibnamefont {Ge}}, \bibinfo {author} {\bibfnamefont {G.}~\bibnamefont
  {Gidofalvi}}, \bibinfo {author} {\bibfnamefont {M.}~\bibnamefont {Goldey}},
  \bibinfo {author} {\bibfnamefont {J.}~\bibnamefont {Gomes}}, \bibinfo
  {author} {\bibfnamefont {C.~E.}\ \bibnamefont {Gonz{\'a}lez-Espinoza}},
  \bibinfo {author} {\bibfnamefont {S.}~\bibnamefont {Gulania}}, \bibinfo
  {author} {\bibfnamefont {A.~O.}\ \bibnamefont {Gunina}}, \bibinfo {author}
  {\bibfnamefont {M.~W.~D.}\ \bibnamefont {Hanson-Heine}}, \bibinfo {author}
  {\bibfnamefont {P.~H.~P.}\ \bibnamefont {Harbach}}, \bibinfo {author}
  {\bibfnamefont {A.}~\bibnamefont {Hauser}}, \bibinfo {author} {\bibfnamefont
  {M.~F.}\ \bibnamefont {Herbst}}, \bibinfo {author} {\bibfnamefont
  {M.}~\bibnamefont {Hern{\'a}ndez~Vera}}, \bibinfo {author} {\bibfnamefont
  {M.}~\bibnamefont {Hodecker}}, \bibinfo {author} {\bibfnamefont {Z.~C.}\
  \bibnamefont {Holden}}, \bibinfo {author} {\bibfnamefont {S.}~\bibnamefont
  {Houck}}, \bibinfo {author} {\bibfnamefont {X.}~\bibnamefont {Huang}},
  \bibinfo {author} {\bibfnamefont {K.}~\bibnamefont {Hui}}, \bibinfo {author}
  {\bibfnamefont {B.~C.}\ \bibnamefont {Huynh}}, \bibinfo {author}
  {\bibfnamefont {M.}~\bibnamefont {Ivanov}}, \bibinfo {author} {\bibfnamefont
  {{\'A}.}~\bibnamefont {J{\'a}sz}}, \bibinfo {author} {\bibfnamefont
  {H.}~\bibnamefont {Ji}}, \bibinfo {author} {\bibfnamefont {H.}~\bibnamefont
  {Jiang}}, \bibinfo {author} {\bibfnamefont {B.}~\bibnamefont {Kaduk}},
  \bibinfo {author} {\bibfnamefont {S.}~\bibnamefont {K{\"a}hler}}, \bibinfo
  {author} {\bibfnamefont {K.}~\bibnamefont {Khistyaev}}, \bibinfo {author}
  {\bibfnamefont {J.}~\bibnamefont {Kim}}, \bibinfo {author} {\bibfnamefont
  {G.}~\bibnamefont {Kis}}, \bibinfo {author} {\bibfnamefont {P.}~\bibnamefont
  {Klunzinger}}, \bibinfo {author} {\bibfnamefont {Z.}~\bibnamefont
  {Koczor-Benda}}, \bibinfo {author} {\bibfnamefont {J.~H.}\ \bibnamefont
  {Koh}}, \bibinfo {author} {\bibfnamefont {D.}~\bibnamefont {Kosenkov}},
  \bibinfo {author} {\bibfnamefont {L.}~\bibnamefont {Koulias}}, \bibinfo
  {author} {\bibfnamefont {T.}~\bibnamefont {Kowalczyk}}, \bibinfo {author}
  {\bibfnamefont {C.~M.}\ \bibnamefont {Krauter}}, \bibinfo {author}
  {\bibfnamefont {K.}~\bibnamefont {Kue}}, \bibinfo {author} {\bibfnamefont
  {A.}~\bibnamefont {Kunitsa}}, \bibinfo {author} {\bibfnamefont
  {T.}~\bibnamefont {Kus}}, \bibinfo {author} {\bibfnamefont {I.}~\bibnamefont
  {Ladj{\'a}nszki}}, \bibinfo {author} {\bibfnamefont {A.}~\bibnamefont
  {Landau}}, \bibinfo {author} {\bibfnamefont {K.~V.}\ \bibnamefont {Lawler}},
  \bibinfo {author} {\bibfnamefont {D.}~\bibnamefont {Lefrancois}}, \bibinfo
  {author} {\bibfnamefont {S.}~\bibnamefont {Lehtola}}, \bibinfo {author}
  {\bibfnamefont {R.~R.}\ \bibnamefont {Li}}, \bibinfo {author} {\bibfnamefont
  {Y.-P.}\ \bibnamefont {Li}}, \bibinfo {author} {\bibfnamefont
  {J.}~\bibnamefont {Liang}}, \bibinfo {author} {\bibfnamefont
  {M.}~\bibnamefont {Liebenthal}}, \bibinfo {author} {\bibfnamefont {H.-H.}\
  \bibnamefont {Lin}}, \bibinfo {author} {\bibfnamefont {Y.-S.}\ \bibnamefont
  {Lin}}, \bibinfo {author} {\bibfnamefont {F.}~\bibnamefont {Liu}}, \bibinfo
  {author} {\bibfnamefont {K.-Y.}\ \bibnamefont {Liu}}, \bibinfo {author}
  {\bibfnamefont {M.}~\bibnamefont {Loipersberger}}, \bibinfo {author}
  {\bibfnamefont {A.}~\bibnamefont {Luenser}}, \bibinfo {author} {\bibfnamefont
  {A.}~\bibnamefont {Manjanath}}, \bibinfo {author} {\bibfnamefont
  {P.}~\bibnamefont {Manohar}}, \bibinfo {author} {\bibfnamefont
  {E.}~\bibnamefont {Mansoor}}, \bibinfo {author} {\bibfnamefont {S.~F.}\
  \bibnamefont {Manzer}}, \bibinfo {author} {\bibfnamefont {S.-P.}\
  \bibnamefont {Mao}}, \bibinfo {author} {\bibfnamefont {A.~V.}\ \bibnamefont
  {Marenich}}, \bibinfo {author} {\bibfnamefont {T.}~\bibnamefont {Markovich}},
  \bibinfo {author} {\bibfnamefont {S.}~\bibnamefont {Mason}}, \bibinfo
  {author} {\bibfnamefont {S.~A.}\ \bibnamefont {Maurer}}, \bibinfo {author}
  {\bibfnamefont {P.~F.}\ \bibnamefont {McLaughlin}}, \bibinfo {author}
  {\bibfnamefont {M.~F. S.~J.}\ \bibnamefont {Menger}}, \bibinfo {author}
  {\bibfnamefont {J.-M.}\ \bibnamefont {Mewes}}, \bibinfo {author}
  {\bibfnamefont {S.~A.}\ \bibnamefont {Mewes}}, \bibinfo {author}
  {\bibfnamefont {P.}~\bibnamefont {Morgante}}, \bibinfo {author}
  {\bibfnamefont {J.~W.}\ \bibnamefont {Mullinax}}, \bibinfo {author}
  {\bibfnamefont {K.~J.}\ \bibnamefont {Oosterbaan}}, \bibinfo {author}
  {\bibfnamefont {G.}~\bibnamefont {Paran}}, \bibinfo {author} {\bibfnamefont
  {A.~C.}\ \bibnamefont {Paul}}, \bibinfo {author} {\bibfnamefont {S.~K.}\
  \bibnamefont {Paul}}, \bibinfo {author} {\bibfnamefont {F.}~\bibnamefont
  {Pavo{\v s}evi{\'c}}}, \bibinfo {author} {\bibfnamefont {Z.}~\bibnamefont
  {Pei}}, \bibinfo {author} {\bibfnamefont {S.}~\bibnamefont {Prager}},
  \bibinfo {author} {\bibfnamefont {E.~I.}\ \bibnamefont {Proynov}}, \bibinfo
  {author} {\bibfnamefont {{\'A}.}~\bibnamefont {R{\'a}k}}, \bibinfo {author}
  {\bibfnamefont {E.}~\bibnamefont {Ramos-Cordoba}}, \bibinfo {author}
  {\bibfnamefont {B.}~\bibnamefont {Rana}}, \bibinfo {author} {\bibfnamefont
  {A.~E.}\ \bibnamefont {Rask}}, \bibinfo {author} {\bibfnamefont
  {A.}~\bibnamefont {Rettig}}, \bibinfo {author} {\bibfnamefont {R.~M.}\
  \bibnamefont {Richard}}, \bibinfo {author} {\bibfnamefont {F.}~\bibnamefont
  {Rob}}, \bibinfo {author} {\bibfnamefont {E.}~\bibnamefont {Rossomme}},
  \bibinfo {author} {\bibfnamefont {T.}~\bibnamefont {Scheele}}, \bibinfo
  {author} {\bibfnamefont {M.}~\bibnamefont {Scheurer}}, \bibinfo {author}
  {\bibfnamefont {M.}~\bibnamefont {Schneider}}, \bibinfo {author}
  {\bibfnamefont {N.}~\bibnamefont {Sergueev}}, \bibinfo {author}
  {\bibfnamefont {S.~M.}\ \bibnamefont {Sharada}}, \bibinfo {author}
  {\bibfnamefont {W.}~\bibnamefont {Skomorowski}}, \bibinfo {author}
  {\bibfnamefont {D.~W.}\ \bibnamefont {Small}}, \bibinfo {author}
  {\bibfnamefont {C.~J.}\ \bibnamefont {Stein}}, \bibinfo {author}
  {\bibfnamefont {Y.-C.}\ \bibnamefont {Su}}, \bibinfo {author} {\bibfnamefont
  {E.~J.}\ \bibnamefont {Sundstrom}}, \bibinfo {author} {\bibfnamefont
  {Z.}~\bibnamefont {Tao}}, \bibinfo {author} {\bibfnamefont {J.}~\bibnamefont
  {Thirman}}, \bibinfo {author} {\bibfnamefont {G.~J.}\ \bibnamefont {Tornai}},
  \bibinfo {author} {\bibfnamefont {T.}~\bibnamefont {Tsuchimochi}}, \bibinfo
  {author} {\bibfnamefont {N.~M.}\ \bibnamefont {Tubman}}, \bibinfo {author}
  {\bibfnamefont {S.~P.}\ \bibnamefont {Veccham}}, \bibinfo {author}
  {\bibfnamefont {O.}~\bibnamefont {Vydrov}}, \bibinfo {author} {\bibfnamefont
  {J.}~\bibnamefont {Wenzel}}, \bibinfo {author} {\bibfnamefont
  {J.}~\bibnamefont {Witte}}, \bibinfo {author} {\bibfnamefont
  {A.}~\bibnamefont {Yamada}}, \bibinfo {author} {\bibfnamefont
  {K.}~\bibnamefont {Yao}}, \bibinfo {author} {\bibfnamefont {S.}~\bibnamefont
  {Yeganeh}}, \bibinfo {author} {\bibfnamefont {S.~R.}\ \bibnamefont {Yost}},
  \bibinfo {author} {\bibfnamefont {A.}~\bibnamefont {Zech}}, \bibinfo {author}
  {\bibfnamefont {I.~Y.}\ \bibnamefont {Zhang}}, \bibinfo {author}
  {\bibfnamefont {X.}~\bibnamefont {Zhang}}, \bibinfo {author} {\bibfnamefont
  {Y.}~\bibnamefont {Zhang}}, \bibinfo {author} {\bibfnamefont
  {D.}~\bibnamefont {Zuev}}, \bibinfo {author} {\bibfnamefont {A.}~\bibnamefont
  {Aspuru-Guzik}}, \bibinfo {author} {\bibfnamefont {A.~T.}\ \bibnamefont
  {Bell}}, \bibinfo {author} {\bibfnamefont {N.~A.}\ \bibnamefont {Besley}},
  \bibinfo {author} {\bibfnamefont {K.~B.}\ \bibnamefont {Bravaya}}, \bibinfo
  {author} {\bibfnamefont {B.~R.}\ \bibnamefont {Brooks}}, \bibinfo {author}
  {\bibfnamefont {D.}~\bibnamefont {Casanova}}, \bibinfo {author}
  {\bibfnamefont {J.-D.}\ \bibnamefont {Chai}}, \bibinfo {author}
  {\bibfnamefont {S.}~\bibnamefont {Coriani}}, \bibinfo {author} {\bibfnamefont
  {C.~J.}\ \bibnamefont {Cramer}}, \bibinfo {author} {\bibfnamefont
  {G.}~\bibnamefont {Cserey}}, \bibinfo {author} {\bibfnamefont {A.~E.}\
  \bibnamefont {DePrince}, \bibfnamefont {3rd}}, \bibinfo {author}
  {\bibfnamefont {R.~A.}\ \bibnamefont {DiStasio}, \bibfnamefont {Jr}},
  \bibinfo {author} {\bibfnamefont {A.}~\bibnamefont {Dreuw}}, \bibinfo
  {author} {\bibfnamefont {B.~D.}\ \bibnamefont {Dunietz}}, \bibinfo {author}
  {\bibfnamefont {T.~R.}\ \bibnamefont {Furlani}}, \bibinfo {author}
  {\bibfnamefont {W.~A.}\ \bibnamefont {Goddard}, \bibfnamefont {3rd}},
  \bibinfo {author} {\bibfnamefont {S.}~\bibnamefont {Hammes-Schiffer}},
  \bibinfo {author} {\bibfnamefont {T.}~\bibnamefont {Head-Gordon}}, \bibinfo
  {author} {\bibfnamefont {W.~J.}\ \bibnamefont {Hehre}}, \bibinfo {author}
  {\bibfnamefont {C.-P.}\ \bibnamefont {Hsu}}, \bibinfo {author} {\bibfnamefont
  {T.-C.}\ \bibnamefont {Jagau}}, \bibinfo {author} {\bibfnamefont
  {Y.}~\bibnamefont {Jung}}, \bibinfo {author} {\bibfnamefont {A.}~\bibnamefont
  {Klamt}}, \bibinfo {author} {\bibfnamefont {J.}~\bibnamefont {Kong}},
  \bibinfo {author} {\bibfnamefont {D.~S.}\ \bibnamefont {Lambrecht}}, \bibinfo
  {author} {\bibfnamefont {W.}~\bibnamefont {Liang}}, \bibinfo {author}
  {\bibfnamefont {N.~J.}\ \bibnamefont {Mayhall}}, \bibinfo {author}
  {\bibfnamefont {C.~W.}\ \bibnamefont {McCurdy}}, \bibinfo {author}
  {\bibfnamefont {J.~B.}\ \bibnamefont {Neaton}}, \bibinfo {author}
  {\bibfnamefont {C.}~\bibnamefont {Ochsenfeld}}, \bibinfo {author}
  {\bibfnamefont {J.~A.}\ \bibnamefont {Parkhill}}, \bibinfo {author}
  {\bibfnamefont {R.}~\bibnamefont {Peverati}}, \bibinfo {author}
  {\bibfnamefont {V.~A.}\ \bibnamefont {Rassolov}}, \bibinfo {author}
  {\bibfnamefont {Y.}~\bibnamefont {Shao}}, \bibinfo {author} {\bibfnamefont
  {L.~V.}\ \bibnamefont {Slipchenko}}, \bibinfo {author} {\bibfnamefont
  {T.}~\bibnamefont {Stauch}}, \bibinfo {author} {\bibfnamefont {R.~P.}\
  \bibnamefont {Steele}}, \bibinfo {author} {\bibfnamefont {J.~E.}\
  \bibnamefont {Subotnik}}, \bibinfo {author} {\bibfnamefont {A.~J.~W.}\
  \bibnamefont {Thom}}, \bibinfo {author} {\bibfnamefont {A.}~\bibnamefont
  {Tkatchenko}}, \bibinfo {author} {\bibfnamefont {D.~G.}\ \bibnamefont
  {Truhlar}}, \bibinfo {author} {\bibfnamefont {T.}~\bibnamefont
  {Van~Voorhis}}, \bibinfo {author} {\bibfnamefont {T.~A.}\ \bibnamefont
  {Wesolowski}}, \bibinfo {author} {\bibfnamefont {K.~B.}\ \bibnamefont
  {Whaley}}, \bibinfo {author} {\bibfnamefont {H.~L.}\ \bibnamefont {Woodcock},
  \bibfnamefont {3rd}}, \bibinfo {author} {\bibfnamefont {P.~M.}\ \bibnamefont
  {Zimmerman}}, \bibinfo {author} {\bibfnamefont {S.}~\bibnamefont {Faraji}},
  \bibinfo {author} {\bibfnamefont {P.~M.~W.}\ \bibnamefont {Gill}}, \bibinfo
  {author} {\bibfnamefont {M.}~\bibnamefont {Head-Gordon}}, \bibinfo {author}
  {\bibfnamefont {J.~M.}\ \bibnamefont {Herbert}},\ and\ \bibinfo {author}
  {\bibfnamefont {A.~I.}\ \bibnamefont {Krylov}},\ }\href
  {https://doi.org/10.1063/5.0055522} {\bibfield  {journal} {\bibinfo
  {journal} {The Journal of Chemical Physics}\ }\textbf {\bibinfo {volume}
  {155}},\ \bibinfo {pages} {084801} (\bibinfo {year} {2021})}\BibitemShut
  {NoStop}%
\bibitem [{\citenamefont {Yamaguchi}\ \emph {et~al.}(1988)\citenamefont
  {Yamaguchi}, \citenamefont {Jensen}, \citenamefont {Dorigo},\ and\
  \citenamefont {Houk}}]{Yamaguchi1988}%
  \BibitemOpen
  \bibfield  {author} {\bibinfo {author} {\bibfnamefont {K.}~\bibnamefont
  {Yamaguchi}}, \bibinfo {author} {\bibfnamefont {F.}~\bibnamefont {Jensen}},
  \bibinfo {author} {\bibfnamefont {A.}~\bibnamefont {Dorigo}},\ and\ \bibinfo
  {author} {\bibfnamefont {K.~N.}\ \bibnamefont {Houk}},\ }\href
  {https://doi.org/10.1016/0009-2614(88)80378-6} {\bibfield  {journal}
  {\bibinfo  {journal} {Chemical Physics Letters}\ }\textbf {\bibinfo {volume}
  {149}},\ \bibinfo {pages} {537} (\bibinfo {year} {1988})}\BibitemShut
  {NoStop}%
\bibitem [{\citenamefont {Neese}(2022)}]{Neese2022}%
  \BibitemOpen
  \bibfield  {author} {\bibinfo {author} {\bibfnamefont {F.}~\bibnamefont
  {Neese}},\ }\href {https://doi.org/10.1002/wcms.1606} {\bibfield  {journal}
  {\bibinfo  {journal} {WIREs Computational Molecular Science}\ }\textbf
  {\bibinfo {volume} {12}},\ \bibinfo {pages} {e1606} (\bibinfo {year}
  {2022})}\BibitemShut {NoStop}%
\end{thebibliography}
\end{document}